\documentclass[prd,aps,preprint,showpacs,nofootinbib,tightenlines,amsmath,amssymb]{revtex4-1}
\usepackage{epsfig,graphicx}
\usepackage{bm}
\usepackage{times}
\usepackage{braket}
\usepackage{color}
\usepackage{slashed}
\usepackage{hyperref}
\usepackage{ulem}

\graphicspath{{./figure/}}

\definecolor{nicered}{rgb}{0.7,0.1,0.1}
\definecolor{nicegreen}{rgb}{0.1,0.5,0.1}
\hypersetup{colorlinks,citecolor=nicegreen,linkcolor=nicered}

\def\sglinespb{\noalign{\hrule\vskip 0.05truecm}}

\newcommand{\beq}{\begin{eqnarray}}
\newcommand{\eeq}{\end{eqnarray}}
\newcommand{\non}{\nonumber\\ }
\newcommand{\psl}{ p \hspace{-2truemm}/ }

\newcommand{\nsl}{ n \hspace{-2.2truemm}/ }

\newcommand{\epsl}{\epsilon \hspace{-1.8truemm}/\,  }

\newcommand{\calm}{ {\cal M} }
\newcommand{\calh}{ {\cal H} }
\newcommand{\calo}{ {\cal O} }

\newcommand{\acp}{{\cal A}_{\rm CP}}

\begin{document}
\title{  Charmless two-body $B$ meson decays in   perturbative QCD factorization approach}
\author{Jian Chai$^{a}$}
\author{Shan Cheng$^{a,b}$}\email{scheng@hnu.edu.cn}
\author{Yao-hui Ju$^{a}$}
\affiliation{{\footnotesize
$^a$ School of Physics and Electronics, Hunan University, Changsha 410082, China, \\
$^b$  School for Theoretical Physics, Hunan University, Changsha 410082, China,}}
\author{Da-Cheng Yan$^c$}\email{yandac@126.com}
\affiliation{{\footnotesize
$^c$ School of Mathematics and Physics, Changzhou University, Changzhou 213164, China,}}
\author{Cai-Dian L\"u$^{d,e}$}\email{lucd@ihep.ac.cn}
\affiliation{{\footnotesize
$^d$ Institute of High Energy Physics, CAS, Beijing 100049, China, \\
$^e$ School of Physics, University of Chinese Academy of Sciences, Beijing 100049, China,}}
\author{Zhen-Jun Xiao$^f$}\email{xiaozhenjun@njnu.edu.cn}
\affiliation{{\footnotesize
$^f$ Department of Physics and Institute of Theoretical Physics, Nanjing Normal University, Nanjing 210023, China.}}

\date{\today}

\begin{abstract}
The perturbative QCD (PQCD) approach based on $k_T$ factorization has made a great achievement for the QCD calculation of the hadronic B decays. 
Regulating the endpoint divergence by the transverse momentum of quarks in the propagators, 
one can do the perturbation calculation for kinds of diagrams including the annihilation type diagrams. 
In this paper, we review the current status of PQCD factorization calculation of two-body charmless $B\to PP, PV, VV$ decays 
up to the next-to-leading order (NLO) QCD corrections.
two new power suppressed terms in decaying amplitudes are also taken into account.
By using the universal input (non-perturbative) parameters, we collected the branching ratios 
and ${\bf CP}$ asymmetry parameters for all the charmless two body $B$ decays, 
calculated in the PQCD approach up to the NLO. 
The results are compared with the ones from QCD factorization approach, soft-collinear effective theory approach and the current experimental measurements.
For most considered B meson decays, the PQCD results for branching ratios agree well with other approaches and the experimental data. 
The PQCD predictions for the ${\bf CP}$ asymmetry parameters for many of the decay channels do not agree with other approaches, 
but have a better agreement with the experimental data. 
The longstanding $K \pi $ puzzle about the pattern of the direct CP asymmetries of the penguin-dominated $B \to K \pi $ decays can be understood 
after the inclusion of the NLO contributions in PQCD.
The NLO corrections and power suppressed terms play an important role in the color suppressed and pure annihilation type $B$  decay modes.
These rare decays are more sensitive to different types of corrections, 
providing opportunity to examine the factorization approach with the more precise experimental measurements. 
\end{abstract}

\maketitle


\section{Introduction}

Two-body non-leptonic $B$ meson decays play an essential role in particle physics to help us understand the quantum chromodynamics (QCD)
and the CP violation in the Standard Model (SM) \cite{Kou:2018nap,CEPCStudyGroup:2018ghi,Cerri:2018ypt,Gambino:2020jvv}.
With the precision measurements, it also provides a window to investigate the possible new physics beyond SM
\cite{Li:2018lxi,Aebischer:2019mlg,Jager:2019bgk,Charles:2020dfl}.

Since the running of Large Hadron Collider at CERN, the LHCb collaboration has provided plenty of more precise B decay measurements than the B factories. 
The High Luminosity phase of the Large Hadron Collider will improve the measurements of the B decay channels
with the integrated luminosity increasing from $23 \, \rm{fb^{-1}}$ in Phase 1 to $300 \, \rm{fb^{-1}}$ in Phase 2 \cite{Cerri:2018ypt}.
After the successful running of B factory, the upgrade of Super-KEKB is expected to provide 50 times more data than the B factory \cite{Kou:2018nap}, 
which will provide another independent precise measurement.
Belle-II is expected to improve the measurements of most of the charmless two-body $B$ decays.
For example, the measurement of the direct CP violation in $B \to K^\ast \pi, \, K \rho, \, K^\ast \rho$ channels
could be improved greatly to achieve the same accuracy as that for the $K\pi$ channels.
Another type of precision measurements in Belle-II focus on decays to two vector meson final states. 
Besides the branching ratios, more physical observables, such as the perpendicular polarization fraction ($f_\perp$), 
the relative phase ($\varphi_\parallel, \, \varphi_\perp, \, \delta_0$) and the helicity CP asymmetry parameters ($\acp^0, \, \acp^\perp$) in these decays are waiting to be measured. 

At the theoretical side, high precision calculations of the two-body charmless $B$ decays are moving forward on a variety of fronts within different approaches. 
One example is the light cone sum rules (LCSRs) approach, which is one of the traditional method to calculate the heavy-to-light transition form factors \cite{Straub:2015ica,Wang:2015vgv,Gao:2019lta,Lu:2018cfc,Gubernari:2018wyi,Cheng:2020vwr,Cheng:2019tgh}.
The form factors are the key inputs for many of the factorization approaches on hadronic B decays. 
The high order power corrections in this method have been performed in recent years. 
The systematic power corrections study of the radiative leptonic decay $B \to \gamma l \nu_l$ \cite{Wang:2016qii,Beneke:2018wjp} 
shed light on the sub-leading twist B meson wave function information\cite{Braun:2017liq}, which are also key inputs for hadronic B meson decay study. 
Besides the form factors, the light cone sum rules approach is also applied to the two-body non-leptonic $B \to \pi\pi$ decays
\cite{Khodjamirian:2000mi,Khodjamirian:2002pk,Khodjamirian:2003eq,Khodjamirian:2005wn}.

The QCD calculation of non-leptonic B decays has a long history. 
The first try is the naive factorization approach \cite{Bauer:1984zv,Bauer:1986bm}, 
which assumes that the two body non-leptonic decay amplitude is a production of transition form factor and the meson decay constant. 
The so called generalized factorization approach is the first to include the perturbative QCD corrections of effective operators 
and the chiral enhanced penguin contribution in hadronic B decays \cite{Ali:1998eb,Ali:1998gb}. 
The factorization is approved order by order later in the QCD factorization approach (QCDF) \cite{Beneke:1999br,Beneke:2001ev}, 
allowing people to calculation the high order QCD corrections leaving the non-perturbative parameters to be determined by experiments \cite{Huber:2021cgk}. 
The vertex corrections to the tree amplitudes \cite{Beneke:2009ek,Huber:2016xod} and the penguin amplitudes \cite{Bell:2015koa,Bell:2020qus} 
have been worked out at next-to-next-to-leading-order (NNLO) recently.
These calculations, together with the next-to-leading-order (NLO) calculations of the spectator scattering (NNLO in $\alpha_s$)
\cite{Beneke:2005vv,Beneke:2006mk,Jain:2007dy}, compose the full corrections to the hard kernels at NNLO.
With introducing different fields in different energy regions,
an improved factorization approach named as soft-collinear effective theory (SCET)  is established by two-step matching
\cite{Bauer:2000yr,Chay:2003ju,Becher:2014oda,Beneke:2015wfa}.
SCET has simple kinematic but more complicated dynamics with several typical scales, which results in a more apparent and efficient factorization formalism.

Although the NNLO calculation has been performed in the QCDF, the dominant contribution for hadronic B decays comes from the transition form factors, 
which is not calculable in the QCDF. 
In the QCDF/SCET, people neglect the transverse momentum of the valence quark to simplify the perturbative calculation, 
but result in an endpoint divergence in the Feynman diagram of form factor calculation. 
This divergence also occurs in the annihilation type diagrams, which is not a physical divergence. 
Recently, the LO weak annihilation diagrams are shown calculable without encounting the end-point divergence 
by taking into account the hard-collinear gluon exchange effect \cite{Lu:2022fgz}. 
In the perturbative QCD (PQCD) factorization approach \cite{Li:1994cka,Li:1994iu}, 
the transversal momenta $k_T$ are picked up for each external light quark lines to regularize the end-point singularity. 
The additional scale $k_T$ apparently will result in extra logarithms in the QCD calculation, which may spoil the perturbative expansion. 
The resummation techniques are carried out for the large logarithms,
resulting in a Sudakov exponent which highly suppress the dynamics in small $k_T$ and small $x$, 
here $x$ is the longitudinal momentum fraction of parton in meson.
Two-body charmless $B$ decays are firstly calculated at leading order in PQCD approach for the $PP$, 
with $P$ denoting a pseudo-scalar meson \cite{Keum:2000wi,Lu:2000em,Xiao:2006mg},
$PV$ \cite{Melic:1998em,Lu:2000hj}, with $V$ denoting a vector meson and $VV$ \cite{Li:2004ti,Li:2005hg,Zou:2015iwa} final states, 
which made a right prediction of the first direct CP asymmetry measurement in B decays.  
A recent global analysis of all $B\to PP$, $PV$ decays in the leading order PQCD approach is also available \cite{Hua:2020usv}.

With the success of the leading order PQCD results and the more and more precise experimental measurements, 
the NLO corrections in PQCD are required to improve the accuracy of the approach.
There are two types of NLO corrections to two-body charmless $B$ decays in the framework of $k_T$ factorization,
one is associated to the four fermion effective operators,
the other one is accompanied to the transition matrix elements sandwiched between $B$ meson and two light mesons.
The first type of corrections include the vertex correction, quark loop correction and also the chromo-magnetic penguin correction to operator $O_{8g}$. 
These corrections are first considered in PQCD to investigate the $B \to \pi K, \phi K$ puzzles \cite{Li:2005kt,Mishima:2003wm}.
One part of these NLO correlations are included in the effective Wilson coefficients, while the other parts provide the independent decay amplitudes for certain channels. 
The second type of NLO QCD corrections carry the dynamics from $m_B$ scale to the hadronic scale,
which are manifested by means of the heavy-to-light transition form factors \cite{Li:2012nk,Cheng:2014fwa},
the timelike form factors of final light mesons \cite{Hu:2012cp,Cheng:2015qra,Cheng:2014rka,Zhang:2015mxa,Hua:2018kho},
the Glauber effects in the hard scattering spectator and annihilation amplitudes \cite{Li:2014haa,Liu:2015sra,Liu:2015upa},
and also the other possible corrections which have not been studies in detail \cite{Cheng:2020fcx}.
Besides the QCD correction at NLO, the sub-leading power corrections from high twist LCDAs
are studied recently for the pion form factors and the radiative leptonic decay $B \to \gamma l \nu_l$ in the PQCD approach
\cite{Cheng:2019ruz,Shen:2018abs,Shen:2019vdc}.

The effects from the NLO corrections mentioned above are partly examined case by case in some charmless two-body $B$ decay channels
\cite{Xiao:2011tx,Bai:2013tsa,Fan:2012kn,Zhang:2014bsa,Zhang:2009zg,Li:2006jv,Zhang:2008by,Rui:2011dr,Liu:2005mm,Zhang:2009zzn}. 
The theoretical precision is explicitly improved with less theoretical uncertainty.
The agreements between the PQCD predictions with the experimental measurements are improved effectively for the branching ratios and other physical observables.
In this review, we summarize all the 78 channels of charmless $B \to PP, PV, VV$ decays in the PQCD approach with the updated input parameters.
With the inclusion of all current known NLO QCD corrections and power corrections, 
we discuss some longstanding ``puzzle'' channels, particularly for the $B \to K \pi, K\rho, K^\ast \pi, K^\ast\rho$ decays.

The paper is organized as follows. In the next section, the three scale factorization approach is introduced
in terms of the effective Hamilton of $b$ quark decay and the definition of meson wave functions.
In section \ref{sec:amp-B}, we exhibit the PQCD calculation of charmless hadronic $B$ decay matrix elements at LO.
We then discuss the NLO corrections in section \ref{PQCD-NLO}.
The phenomenological results for all two-body charmless decays $B \to PP$, $PV$ and $VV$ are discussed 
in sections \ref{sec:b2pp}. The summary is given in the last section.

\section{Theoretical framework}\label{sec:framework}

All the charmless B meson decays are weak decays in the SM, which are induced by the charged current. Since the $W$ boson mass is much larger than the $b$ quark mass, people usually integrate out the heavy $W$ boson and top quark to result in an effective Hamiltonian of four quark operators with QCD corrections. 
The relevant effective Hamiltonian of $b \to q U \bar{U}$ decays with $U \in \{ u, c\}$ and $q \in \{ d , s \}$ is
\beq
\calh_{{\rm eff}}(\Delta b = - 1) &=& \frac{{\rm G_F}}{\sqrt{2}} \Big[
V_{U q}^\ast \, V_{Ub} \left( C_1(\mu) \, O_1(\mu) + C_2(\mu) \, O_2(\mu) \right) \non
&~& \hspace{0.4cm} - V_{tq}^\ast \, V_{tb} \, \sum_{i=3}^{10} \, C_{i}(\mu) \, O_{i}(\mu) - V_{tq}^\ast \, V_{tb} 
C_{8g}(\mu) \, O_{8g}(\mu)  \Big] ,
\label{eq:Hamiltonian-b}
\eeq
where $V_{ij}$ are the CKM matrix elements.  
With the chiral representation of the fermion fields $\left(\bar{q}_1 q_2\right)_{{\rm V-A}} = \bar{q}_1 \gamma_\mu (1 - \gamma_5) q_2$ 
and $\left(\bar{q}_1 q_2\right)_{{\rm V+A}} = \bar{q}_1 \gamma^\mu (1 + \gamma_5) q_2$, 
the local operators involved in nonleptonic $B$ decay processes are  
\beq
&&O_1 = \left( \bar{q}_\alpha \, U_\beta \right)_{{\rm V-A}} \left( \bar{U}_\beta \, b_\alpha \right)_{{\rm V-A}} \,, 
\quad\quad\quad\quad\quad\quad
O_2 = \left( \bar{q}_\alpha \, U_\alpha \right)_{{\rm V-A}} \left( \bar{U}_\beta \, b_\beta \right)_{{\rm V-A}} \,, 
\label{eq:operator-tree}\\
&&O_3 = \left( \bar{q}_\alpha \, b_\alpha \right)_{{\rm V-A}} \sum_{q^\prime} \left( \bar{q}^\prime_\beta \, q^\prime_\beta \right)_{{\rm V-A}} \,, 
\quad\quad\quad\quad \;\;\;
O_4 = \left( \bar{q}_\alpha \, b_\beta \right)_{{\rm V-A}} \sum_{q^\prime} \left( \bar{q}^\prime_\beta \, q^\prime_\alpha \right)_{{\rm V-A}} \,, \non
&&O_5 = \left( \bar{q}_\alpha \, b_\alpha \right)_{{\rm V-A}} \sum_{q^\prime} \left( \bar{q}^\prime_\beta \, q^\prime_\beta \right)_{{\rm V+A}} \,, 
\quad\quad\quad\quad \;\;\;
O_6 = \left( \bar{q}_\alpha \, b_\beta \right)_{{\rm V-A}} \sum_{q^\prime} \left( \bar{q}^\prime_\beta \, q^\prime_\alpha \right)_{{\rm V+A}} \,, 
\label{eq:operator-QCD}\\
&&O_7 = \frac{3}{2} \left( \bar{q}_\alpha \, b_\alpha \right)_{{\rm V-A}} 
\sum_{q^\prime} e_{q^\prime} \left( \bar{q}^\prime_\beta \, q^\prime_\beta \right)_{{\rm V+A}} \,, \quad\quad\;\;
O_8 = \frac{3}{2} \left( \bar{q}_\alpha \, b_\beta \right)_{{\rm V-A}} 
\sum_{q^\prime} e_{q^\prime} \left( \bar{q}^\prime_\beta \, q^\prime_\alpha \right)_{{\rm V+A}} \,, \non
&&O_9 = \frac{3}{2} \left( \bar{q}_\alpha \, b_\alpha \right)_{{\rm V-A}} 
\sum_{q^\prime} e_{q^\prime} \left( \bar{q}^\prime_\beta \, q^\prime_\beta \right)_{{\rm V-A}} \,, \quad\quad \;\,
O_{10} = \frac{3}{2} \left( \bar{q}_\alpha \, b_\beta \right)_{{\rm V-A}} 
\sum_{q^\prime} e_{q^\prime} \left( \bar{q}^\prime_\beta \, q^\prime_\alpha \right)_{{\rm V-A}} \,,
\label{eq:operator-QED}\\
&&O_{8 g} = \frac{g_s}{8 \pi^2} \, m_b \, \bar{q}_\alpha \, \sigma^{\mu\nu} \left( 1 + \gamma_5 \right) T_{\alpha\beta}^a \, G_{\mu\nu}^a \, b_\beta \,. \label{eq:operator-8g}
\eeq
All these effective operators are grouped as    the current-current (tree) operators $O_{1,2}$, 
the QCD (electroweak) penguin operators $O_{3-6}$ ($O_{7-10}$) and the chromomagnetic operator  $O_{8 g}$.

The Wilson coefficients $C_{1-10}$ and $C_{8g}$  are obtained by matching the effective Hamiltonian with the full theory of weak decays
\cite{Buchalla:1995vs,Ma:1979px,Clements:1982mk,Inami:1980fz}, including the NLO QCD corrections. 
The explicit  renormalization scale dependence of the Wilson coefficients should be canceled by the matrix elements of effective operators. 
For this reason, in the leading order calculation of the perturbative QCD factorization approach, 
we usually use the leading order Wilson coefficients, although the NLO corrections are already on the market \cite{Lu:2000em}.

\subsection{The three scale factorization frame}\label{PQCD}

Since the masses of charmless mesons are all negligible comparing to the large $B$ meson mass, 
the two final state mesons are at the collinear state with large momentum  at the rest frame of the $B$ meson. It is convenient to work at the light cone coordinate.
If we define one of the outgoing light meson direction is ``-'', then its momentum in the light cone coordinate is 
\beq
p_3 & =&  \frac{1}{\sqrt{2}} \left( 0, m_B, \bf{0}_T \right)  .
\eeq 
The valence quark (anti-quark) in this final state meson is also collinear, whose momentum is $x_3p_3$ and $\bar x_3p_3=(1-x_3)p_3$, respectively.  
Here $x_3$ is the momentum fraction carried by the quark defined in the parton model.
All charmless two body B decays are characterized by the $b$ quark weak decay through the four quark operators defined in Eqs. (\ref{eq:operator-tree}-\ref{eq:operator-QED}). 
The large $b$ quark mass makes sure that all the three final state light quarks from the four-quark operator are energetic (collinear), 
and hence a hard gluon is needed to kick the spectator quark, which is soft in the $B$ meson, to make it collinear to form the final light mesons.  

\begin{figure}[t]
\begin{center}
\vspace{-1cm} 
\includegraphics[scale=0.9]{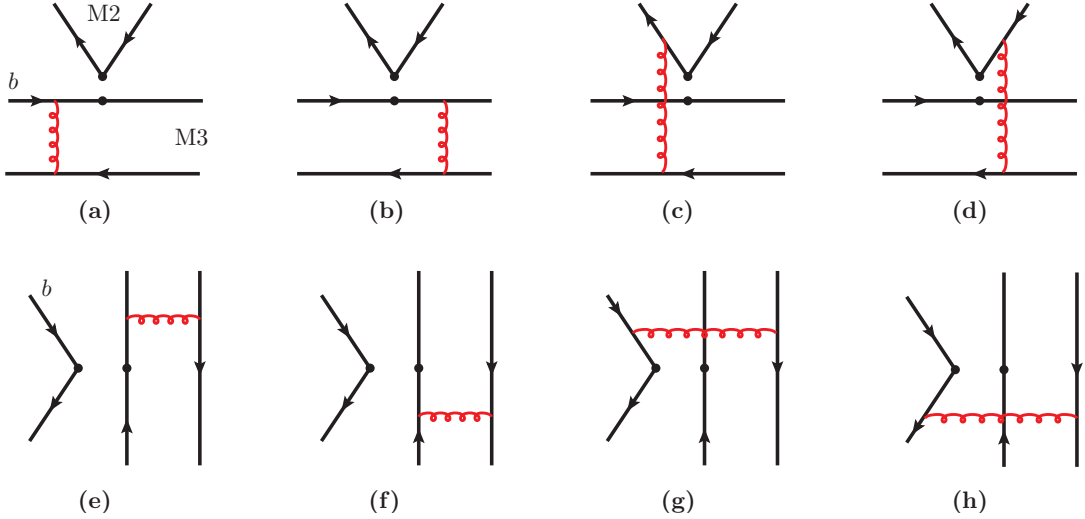}
\caption{Leading order Feynman diagrams of the two-body hadronic $\bar{B}^0/B^-$ decays. 
The two black dots denote the vertexes of effective four quark operator.} 
\label{fig:PQCD}
\vspace{-0.5cm} 
\end{center} 
\end{figure}

The hard gluon connecting the spectator quark with the four quark operators making the four Feynman diagrams for the nonleptonic two-body $B$ decays 
in the framework of PQCD approach at leading order, which is depicted in Fig.~\ref{fig:PQCD} (a,b,c,d). 
In fact, the first two diagrams in Fig.~\ref{fig:PQCD} (a,b) are also the leading Feynman diagrams in the QCDF approach and the SCET framework 
at the numerical calculations. 
The mainly difference between these approaches is the treatment of Fig.~\ref{fig:PQCD} (a,b). 
For example, in the calculation of the second diagram, the gluon propagator together with the quark propagator will be proportional to $x_1^2x_2$, 
which appear in the denominator of the decay amplitude. 
The range of a parton momentum fraction $x$,  is not experimentally controllable, and runs from 0 to 1. 
Hence, the end point region with $x \to 0$ is unavoidable. 
The leading twist distribution amplitude is proportional to $x$, therefore the leading digram in Fig.~\ref{fig:PQCD} (a,b) diverge at the endpoint region. 
In QCDF,  people argue that these two diagrams can be treated as the generalized factorization approach, 
which are products of transition form factor and meson decay constant.  
As a result, the most important diagrams in the numerical calculation of QCDF is not perturbatively calculable.  

In fact, at the endpoint region, when $x\to 0$, the transverse momentum of the valence quark is not negligible any more. 
In the PQCD approach, we keep the transverse momentum of quarks at the denominator of decay amplitude to kill the endpoint divergence \cite{Keum:2000wi,Lu:2000em}. 
While in the numerator of the decay amplitude, we still neglect the transverse momentum, comparing with the large longitudinal momentum.  
After this treatment, the gauge invariance of the decay amplitude is still held, while the endpoint singularity is avoided. 
The introduction of transversal momentum enriches the study of hadron distribution amplitudes,
where the light-cone aligned definition is corrected to the transversal momentum dependent  definition with a general Wilson link. 
The collinear factorization will break down, and the $k_T$ factorization should be adopted. 
It has been shown that infrared divergences appearing in loop corrections to exclusive processes 
can be absorbed into hadron LCDAs in the $k_T$ factorization without breaking the gauge invariance \cite{Nandi:2007qx}. 
The transverse momentum of quark is an extra scale in the QCD calculation of the decay amplitude, 
which will result in extra logarithms to spoil the perturbative expansion. 
These double logarithms should be resummed using the renormalization group equation to repair the perturbative expansion.  
The resummation is done to the leading logarithms or next-to-leading logarithms to result in a $k_T$ Sudakov factor \cite{Li:1992nu},
which exhibit the highly suppression for the large  distance (small $k_T$ when $x \sim 0$).
Integrating over the transversal momentum, the decay amplitude is still proportional to logarithm term $\ln^2 x$. 
To improve the perturbative expansion in the PQCD approach, these logarithms are also resummed by the so called threshold resummation  
\cite{Li:1994cka-Li:1994iu,Li:1996gi,Li:1998is,Feng:2007qv}.
As a remarkable result, 
the two-stage application of resummation repairs the self-consistency between the perturbative strong coupling $\alpha_s(t)$ and the hard logarithm $\ln (x_1x_2Q/t^2)$.
Here $t$ is the factorization scale usually chosen at the largest virtuality in the hard scattering, 
i.e, the characterized factorization scale in $B \to \pi l \nu$ decay is $t \sim 8 \Lambda_{QCD} m_b$.
The typical infrare divergence, including the soft divergence and the collinear divergence in the PQCD approach 
are treated the same way as the soft-collinear effective theory \cite{Bauer:2000yr}, which we will not discuss in detail.  

There is also another possibility that the light quark of the $B$ meson is one of the quarks in the four quark operators, 
with a light quark anti-quark pair generated by a hard gluon. 
These kinds of Feynman diagrams are usually called annihilation type diagrams, which are shown in  Fig.~\ref{fig:PQCD} (e,f,g,h). 
Since the $B$ meson is a pseudoscalar meson, its decay to two massless quarks in weak interaction is helicity suppressed. 
These diagrams are neglected in the generalized factorization approach \cite{Ali:1998eb}. 
Similar as the Fig.~\ref{fig:PQCD} (a,b), there is also endpoint singularity in the calculation of annihilation type diagrams. 
Their contributions in QCDF are parametrized as free parameters to be fitted from experiments.
Again, we can do the perturbative QCD calculation of these diagrams in PQCD by including the transverse momentum of valence quark in meson.

For the emission topological diagrams in Fig.~\ref{fig:PQCD} (a,b), 
the decaying amplitudes are real functions because this transition is happened in the space-like region.
While in the annihilation topology, the decaying amplitudes are complex functions with the internal propagators varying on shell, 
which can be re-expressed by the use of the identity
\beq
\frac{1}{k_T^2 - xm_B^2 - i \epsilon} = {\cal P} \left( \frac{1}{k_T^2 - xm_B^2} \right) + i \pi \delta(k_T^2 - xm_B^2).
\label{eq:pr}
\eeq
Here ${\cal P} (f)$ is the Cauchy's principle value obtained by evenly approaching the singular point from both sides such that the diverging pieces will cancel each other. 
This is not in conflict with the hard mechanism of the amplitudes which states a large off-shellness.
In fact, on-shell configuration indeed happens for the internal propagators
even though the soft mechanism is highly suppressed by the Sudakov exponents,
i.e, the factorizable annihilation amplitudes in $B \to \pi\pi$ decay is proportional to timelike pion form factor,
where the energy dependent imaginary part continues from the resonance region to the $\calo(m_B^2)$ energy region.
The strong phase in Eq. (\ref{eq:pr})   could induce large ${\bf CP}$ violation in the two-body hadronic $B$ meson decays, 
which is essential to explain the large direct CP asymmetry of $B^0\to K^+\pi^-$ decay \cite{Keum:2000wi} and $B^0\to \pi^+\pi^-$ decay \cite{Lu:2000em}.

Besides the annihilation amplitudes, there are other sources of strong phase in the PQCD approach.
The first candidate is the Sudakov exponent which is related to the center of mass scattering angle and the angular distribution of scattering hadrons.
This contribution maybe important in the baryon decays due to the angle distribution, while in the $B$ meson decays it is negligible.
The second candidate comes from the NLO corrections to the spectator emission amplitudes with Glauber gluon \cite{Li:2014haa,Liu:2015sra,Liu:2015upa}.
This effect only supplements a sizable phase to pion final state and modifies the interaction manners between different topological amplitudes. The strong phase from the quark loop correction, with an on shell charm quark, is the leading strong phase in QCDF approach, but a kind of NLO corrections in $\alpha_s$ expansion. 
We mark that  all these sources of strong phases reflect either the soft or the Glauber gluon corrections to the wave function at high orders,
while the on shell configuration of the annihilation amplitudes is  {mechanized} by the hard gluon exchange at leading order.

The ultraviolet  divergence in the high order calculations are performed by the standard renormalization method, which we do not go through much on it.
The large ultraviolet  logarithms $\ln (m_W/t)$ and $\ln (t/\Lambda_{{\rm QCD}})$ are summed by the standard renormalization-group method to give the two-stage evolutions,
where the interaction happened in energy interval $t \sim \calo [m_b, m_W]$ is described by the effective Wilson coefficients and the interaction happened below the energy scale $t$ is demonstrated by the hadronic transition matrix element.
The renormalization scale dependences cancel in these two evolutions and results in a scale independent amplitude, 
which shows a reliable prediction for the charmed $B$ decays \cite{Chang:1996dw,Yeh:1997rq,Li:1997un}.
The standard formula of PQCD approach to handle an exclusive scattering/weak decay
is written in the combination of the hard scattering mechanism and the transversal momentum dependent  wave functions,
with the guidance of factorization theorem to detach the physical amplitude according to the acting intervals between interactions.
 
By employing the resummation techniques to sum all the double and single logarithms between the $W$ boson mass, $b$ quark mass and the transverse momentum, the three scale factorization schemes allow us to calculate the two-body $B$ decays perturbatively. 
The decay amplitude can therefore be written in the convolution of hard kernels and meson LCDAs
\beq
&~&\calm(B \to M_2M_3) = H_{r}(M_W, t ) \otimes H(t,\mu) \otimes \phi(x,P^+,b,\mu) \non
&=& \sum_{i} C_i(M_W,t) \otimes H_i(t,b) \otimes \phi(x,b,1/b)   \cdot
\exp \Big[ -s(P^+,b) - \int_{1/b}^{t} \frac{d\overline{\mu}}{\overline{\mu}} \gamma_\phi(\alpha_s(\overline{\mu}))\Big].
\label{eq:PQCD}
\eeq
Here $H_r$ ($C_i$) is the hard kernel (Wilson coefficients) carrying the first-stage evolution stretched from the $m_W$ down to $t$ scale, 
$H(t,\mu)$ is the perturbative calculable  hard part shown in Fig.\ref{fig:PQCD}.
$H_i \otimes \phi $ takes along the second-stage evolution from the hard scale $t$ down to $\Lambda_{QCD}$,
in which $\phi$ is the LCDAs defined with certain twist in the meson wave functions.
 $s(P^+,b)$ in the exponent is the so called   Sudakov   factor resulting in from the resummation of double logarithms.
$\gamma_\phi$ is the anomalous dimension of the wave function emerged from the resummation of the single logarithms in quark self-energy correction.
 
\subsection{Meson distribution amplitudes  }\label{DAs}

\subsubsection{Distribution amplitude of B meson}\label{B-DAs}

$B$ meson distribution amplitude is defined under the heavy quark effective theory by the dynamical twist expansion
\cite{Grozin:1996pq,Li:1999kna,Beneke:2000wa,Kawamura:2001jm,Geyer:2005fb}.
 The leading LCDA of  $B$ meson   in the momentum space is
\beq
&~&\int d^4z_1 \, e^{i \bar{k}_1 \cdot z_1} \big\langle 0 \vert \bar{d}_\sigma(z_1) \, b_{ \beta}(0)  \vert \bar{B}^0(p_1) \big\rangle \non
&=& \frac{i f_B}{\sqrt {2 N_c}} \left\{ (\psl_1 + m_B) \gamma_5 \left[ \frac{\nsl_+}{\sqrt{2}} \varphi_+(\bar{x}_1, b_1) +
\left( \frac{\nsl_-}{\sqrt{2}} - k_1^+ \gamma_\perp^\nu \frac{\partial}{\partial \bf{k}^\nu_{1 T}}\right) \varphi_-(\bar{x}_1, b_1) \right] \right\}_{\beta\sigma} \, \non
&=& \frac{-i }{\sqrt {2 N_c}} \left\{ (\psl_1 + m_B) \gamma_5 \left[ \varphi_B(x_1, b_1) - \frac{\nsl_+ - \nsl_-}{\sqrt{2}} \bar{\varphi}_B(x_1, b_1) \right] \right\} _{\beta\sigma} \,. \label{eq:B-wf-2p-2}
\eeq
Here $x_1 = k_1^-/p_1^-$ denotes the momentum fraction of antiquark moving along the minus direction on the light cone,
the underlying integral $\varphi_{\pm} (x_1, b_1)= \int d k_1^+ d^2 {\bf k}_{\rm 1T} \, e^{i {\bf k}_{\rm 1T} \cdot {\bf b}_1} \, \varphi_{\pm}(k_1)$ is implemented.
To obtain the approximate expression in the 2rd line, we have omitted the transversal projection term. 
Several comments are in order to explain the definition:
\begin{itemize}
\item[(1)]
The definition of matrix elements in Eq. (\ref{eq:B-wf-2p-2}) is valid only in the factorizable scatterings
where the hard interaction happens locally, which is independent with the nonlocal matrix elements.

\item[(2)]
By definition, it is easy to show that
\beq
\varphi_B(x_1, b_1) &=& \frac{1}{2} \left[ \varphi_+(x_1, b_1) + \varphi_-(x_1, b_1) \right] \,, \non
\bar{\varphi}_B(x_1, b_2) &=& \frac{1}{2} \left[ \varphi_-(x_1, b_1) - \varphi_+(x_1, b_1) \right] \,. \label{eq::B-wf-2p-rel}
\eeq
The contribution from the LCDA $\bar{\varphi}_B(x_1) = \bar{\varphi}_B(x_1, 0)$ is argued to be suppressed by
$\mathcal{O}(\ln \frac{\bar{\Lambda}}{m_B})$ in contrast to $\varphi_B(x_1)$ \cite{Kurimoto:2001zj},
with the relation $\varphi_-(x) = \int_x^\infty dx' \, \varphi_+(x')/x'$ and the hadronic scale $\bar{\Lambda} \simeq m_B - m_b$.
In the symmetry limit of   $ \varphi_+$ and $ \varphi_-$, we have $\varphi_B = \varphi_+$ and $\bar{\varphi}_B = 0$.
  This approximation would be employed in our calculation with the accuracy up to $\mathcal{O}(\bar{\Lambda}/m_b)$. 
We notice that the symmetry broken between $\varphi_+$ and $ \varphi_-$ have been considered recently in 
two-body hadronic $B$ decays and the result is positive to the approximation \cite{Yang:2020xal}.

\item[(3)]
The LCDA is usually parameterized in the exponential model
\beq
\varphi_B(x_1,b_1) = N_B \,x_1^2 \, (1-x_1)^2 \,
{\rm exp} \left[ - \frac{x_1^2m_B^2}{2 \omega_B^2} - \frac{(\omega_B b_1)^2}{2} \right] ,
\label{eq:B-DA}
\eeq
where the parameter $N_B$ is determined by  the distribution amplitude normalization   as
\beq
\int_0^1 dx_1 \, \varphi_B(x_1, b_1=0) = \frac{f_B}{2 \sqrt{2N_c}} \,.
\label{eq:B-DA-norm}
\eeq
 
\end{itemize}

\subsubsection{ LCDAs of light pseudoscalar meson} \label{P-DAs}

LCDAs are rigorously defined by the matrix element sandwiched with the quark bilinears with light-cone separation,
and then switch to the actual momenta and the near lightlike distance $x$ for the practice of phenomenas.
 
In this paper, we would not consider the three-particle LCDAs of light mesons whose contributions can be expected to be
power suppressed in the $B$ decays,
although three-particle LCDAs relate to the high twist LCDA with two-particle assignment by the equation of motion \cite{Ball:1998sk}.
The three-particle LCDA contributions are carefully examined in the $\pi$ and $K$ electromagnetic and transition form factor \cite{Cheng:2019ruz,Shen:2019vdc},
which shows at least one order of magnitude smaller than the two-particle contribution in the large energies region $Q^2 \geqslant 10 \, {\rm GeV}^2$.
In the momentum space, the vacuum to pion matrix element with possible currents can be written in the    twits  expansion as the following form, up to the   twist-3 accuracy,  
\beq
&~& \int d^4z \, e^{-i \bar{k} \cdot z} \,\big\langle \pi^+(p) \big\vert \bar{u}_\delta(0) \, d_\alpha(z) \big\vert 0 \big\rangle \non
&=& \frac{-i }{\sqrt {2N_c}} \, \left\{ \gamma_5 \left[ \, \psl \, \varphi_\pi^{\rm a}( x, b) +  m_0^\pi \, \varphi_\pi^{\rm p}(x, b) -m_0^\pi \left(\nsl_+\nsl_- - 1\right) \, \varphi_\pi^{\rm t}(x, b) \right] \right\}_{\alpha\delta} .
\label{eq:DAs-pi-P-2}
\eeq
$f_\pi$ is the decay constant, $m_0^\pi \equiv m_\pi^2/(m_u+m_d)$ is the chiral mass,
$\varphi_\pi^a$ and $ \varphi^{{\rm p}, \sigma}_\pi$ are the LCDAs at dynamical leading twist and twist-3, respectively.
They have the similar normalization $\int_0^1 dx \, \varphi_\pi^{\rm a}(x) = f_{\pi}/2\sqrt{2N_c}$.

LCDAs are usually formulated by the conformal partial expansion and expressed in terms of the Gegenbauer polynomials $C_n^{j/2}$.
The leading twist LCDA of pseudoscalar meson is
\beq
\varphi_P^{\rm a}(x, \mu) = \frac{f_{\pi}}{2\sqrt{2N_c}}6x\bar{x} \sum_{n=0} \, a^P_n(\mu)  \, C_n^{3/2}(2x-1) \,.
\label{eq:P-DA-t2}
\eeq
Two-particle twist-3 LCDAs are related to the three-particle LCDA and the leading twist LCDA by the QCD equation of motion.
The parameter $\rho^P = (m_{q_1}+m_{q_2})/m_0^P$ is then introduced to reflect the quark mass terms in the equation of motion.
Up to the accuracy with conformal spin at next-to-leading order and second Gegenbauer moment, the LCDAs of pseudoscalar meson are 
\beq
\varphi^{\rm p}_P(x, \mu) &=& \frac{f_{\pi}}{2\sqrt{2N_c}} \big[1 + 3 \rho^P \Big( 1-3a_1^P+6a_2^P\Big)(1+\ln x)
- \frac{\rho^P}{2} \Big( 3-27a_1^P+54a_2^P\Big) \, C_1^{1/2}(2x-1) \non
&+& 3 \Big(10 \eta_{3P} - \rho^P (a_1^P - 5 a_2^P) \Big) \, C_2^{1/2}(2x-1)
+ \Big(10 \eta_{3P} \lambda_{3P} - \frac{9}{2}\rho^Pa_2^P \Big) \, C_3^{1/2}(2x-1) \non
&-& 3 \eta_{3P} \omega_{3P} \, C_4^{1/2}(2x-1) ] \, , \label{eq:P-DA-t3-P}\\  
\varphi_P^{\rm t}(x, \mu) &=&   \frac{1}{6} \, \frac{d}{dx }\varphi_P^\sigma(x, \mu),\non
\varphi_P^\sigma(x, \mu) &=& \frac{f_{\pi}}{2\sqrt{2N_c}} 6x(1-x) \Big\{ 1 + \frac{\rho^P}{2} \Big(2 - 15 a_1^P + 30 a_2^P\Big)
+ \rho^P \Big(3a_1^P - \frac{15}{2}a_2^P\Big) \, C_1^{3/2}(2x-1) \non
&+& \frac{1}{2} \Big( \eta_{3P}(10-\omega_{3P}) + 3\rho^P a_2^P \Big) \, C_2^{3/2}(2x-1)
+ \eta_{3P} \lambda_{3P} \, C_3^{3/2}(2x-1) \non
&+& 3 \rho^P \Big( 1- 3a_1^P + 6 a_2^P \Big) \, \ln x \Big\} \,.
\label{eq:P-DA-t3-T}
\eeq
In the above expression, contributions from the three-particle configuration and the two-particle configuration by equation of motion are separated clearly. 
The three-particle parameters $f_{3P}$, $\lambda_{3P}$, $\omega_{3P}$ are defined by the matrix element of local twist-3 operators, 
and their evolutions have the mixing terms with the quark mass \cite{Ball:2006wn}.
In our consideration we would only take into account the mass of strange quark, neglecting the $u, d$ quark masses, 
what's more, we would not include the terms proportional to parameters $f_{3P}, \lambda_{3P}, \omega_{3P}$ within the present accuracy.

\subsubsection{LCDAs of light vector meson }\label{V-DAs}

The longitudinal and  transverse decay constants of the vector mesons are defined as
\beq
&&\big\langle \rho^+(p, \epsilon^\lambda) \big\vert \bar{u}(0) \gamma_\tau d(0) \big\vert 0 \big\rangle
= - i f^\parallel_\rho m_\rho \epsilon^\lambda_\tau \,, \non
&&\big\langle \rho^+(p, \epsilon^\lambda) \big\vert \bar{u}(0) \sigma_{\tau,\tau^\prime} d(0) \big\vert 0 \big\rangle
= - i f^\perp_\rho \left( \epsilon^\lambda_\tau p_{\tau^\prime} - \epsilon^\lambda_{\tau^\prime} p_\tau \right) \,. 
\label{eq:rho-decay-c}
\eeq
In the convenient momentum space in practice, the matrix elements of vacuum to vector meson up to twist-3 are arranged for longitudinal and transverse polarization, 
respectively, 
\beq
&~&\int d^4 z \, e^{- i \bar{k} \cdot z} \, \big\langle \rho^+(p, \epsilon^\parallel) \big\vert \bar{u}_{\delta}(0) d_\alpha(z) \big\vert 0 \big\rangle \non
&=&\frac{- i }{\sqrt {2N_c}} \left\{m_\rho \epsl^\parallel \, \varphi_{\rho}^\parallel( x) + \epsl^\parallel \psl \, \varphi_{\rho}^{t,\parallel}( x)
- m_\rho  \,  \bar{\psi}_{\rho}^{s,\parallel}( x) \right\}_{\alpha\delta} \,, 
\eeq
\beq
&~&\int d^4z \, e^{- i \bar{k} \cdot z} \, \big\langle \rho^+(p, \epsilon^\perp) \big\vert \bar{u}_{\delta}(0) d_\alpha(z) \big\vert 0 \big\rangle \non
&=&\frac{- i }{\sqrt {2N_c}} \left\{ \epsl^\perp \psl \, \varphi_{\rho}^\perp( x) +   m_\rho \epsl^\perp\, \varphi_{\rho}^{t,\perp}( x)  
- \frac{i \, m_\rho}{(p \cdot n_-)} \varepsilon_{\tau\tau'\kappa\kappa'}
\gamma_5 \gamma^\tau \epsilon^{\perp \tau'} p^\kappa n_-^{\kappa'} \, \bar{\psi}_{\rho}^{s,\perp}(x) \right\}_{\alpha\delta} \,.
\label{eq:DA-rho-p-2}
\eeq  
The normalizations of LCDAs $\varphi_\rho = \{ \varphi^{\parallel(\perp)}_\rho, \varphi_{\rho}^{t,\parallel(\perp)} \}$ are the following
\beq
\int_0^1 dx \, \varphi^{\parallel(\perp)}_\rho(x) =\frac{f^{\parallel(\perp)}_{\rho}}{2\sqrt{2N_c}}  \,, \quad
\int_0^1 dx \, \varphi_{\rho}^{t,\parallel(\perp)}(x) = \frac{1}{2\sqrt{2N_c}} \left(f^{\perp(\parallel)}_{\rho} - f^{\parallel(\perp)}_{\rho} \frac{m_u+m_d}{m_\rho} \right)\,.\label{eq:rho-DA-norm}
\eeq

The LCDAs of light vector mesons are more complicated than the pseudoscalar ones due to the polarizations,
which are quoted as \cite{Ball:1998sk,Ball:1998ff,Ball:2007rt}
\beq
\varphi^\parallel_{V}(x, \mu) &=& \frac{3 f_\rho^\parallel}{\sqrt{6}} x\bar{x} \sum_{n=0} \, a^{V, \parallel}_n(\mu)  \, C_n^{3/2}(2x-1) \,,
\label{eq:V-DA-L-t2} \\
\varphi^{t,\parallel}_{V}(x, \mu) &=& \frac{3 f_\rho^\perp}{2\sqrt{6}} (2x-1) \left\{C_1^{1/2}(2x-1) +  \, a_{V, 1}^\perp \, C_2^{1/2}(2x-1)
+ \, a_{V, 2}^\perp \, C_3^{1/2}(2x-1) \right\} \,,
\label{eq:V-DA-L-t31} \\
\psi^{s,\parallel}_{V}(x, \mu) &=& \frac{3 f_\rho^\perp}{2\sqrt{6}} x\bar{x} \left\{1 + \frac{a_{V,1}^\perp}{3} \, C_1^{3/2}(2x-1) +  \frac{a_{V, 2}^\perp}{6} \, C_2^{3/2}(2x-1) \right\} \,.
\label{eq:V-DA-L-t32} \\
\varphi^\perp_{V}(x, \mu) &=& \frac {3 f_\rho^\perp}{\sqrt{6}} x\bar{x} \sum_{n=0} \, a^{V, \perp}_n(\mu)  \, C_n^{3/2}(2x-1) \,,
\label{eq:V-DA-T-t2} \\
\varphi^{t,\perp}_{V}(x, \mu) &=& \frac{3 f_\rho^\parallel}{8 \sqrt{6}} \big\{  [1 + (2x-1)^2] + 2\, a_{V,1}^{\parallel} \, (2x-1)^3 +8 \, a_{V, 2}^\parallel \, C_2^{1/2}(2x-1) \non
&& + \frac{6}{7} \, a_{V, 2}^\parallel \, C_4^{1/2}(2x-1) \big\} \,,
\label{eq:V-DA-T-t31} \\
\psi^{s,\perp}_{V}(x, \mu) &=& \frac{3 f_\rho^\parallel}{4 \sqrt{6}} x\bar{x} \left\{1 + \frac{a_{V,1}^\parallel}{3} \, C_1^{3/2}(2x-1) +  \frac{a_{V, 2}^\parallel}{6} \, C_2^{3/2}(2x-1) \right\} \,.
\label{eq:V-DA-T-t32}
\eeq
We notice the relation ${\bar \psi}^{s,\perp(\parallel)}_{V}(x) = \frac{d}{dx} \psi^{s,\perp(\parallel)}_{V}(x)$.
For the isospin-half light mesons ($K$ and $K^\ast$), the definition of LCDAs is similar with the isospin-vector mesons ($\pi$ and $\rho$)
by substituting the non-perturbative parameters such as $m_{K^{(\ast)}}, f_{K^{(\ast)}}, m_0^{K^{(\ast)}}, a_n^{K^{(\ast)}}, \rho^{K^{(\ast)}}$.

\subsubsection{ \texorpdfstring{$\eta$-$\eta^\prime$}{} mixing }\label{eta-DAs}

For the isospin-singlet light mesons, $\eta$ and $\eta^\prime$, the mixing \cite{Ambrosino:2006gk,Cheng:2008ss,Tsai:2011dp} should be considered.
We consider the $\eta_q-\eta_s$ mixing scheme \cite{Feldmann:1998vh,Feldmann:1998sh},
where the physical states are expressed as a linear combination of orthogonal quark-flavor basis $\eta_q = (\bar{u}u + \bar{d}d)/\sqrt{2}$ and $\eta_s = \bar{s}s$
via the octet-singlet basis $\eta_1 = (\bar{u}u + \bar{d}d + \bar{s}s)/\sqrt{3}$ and $\eta_8 = (\bar{u}u + \bar{d}d - 2\bar{s}s)/\sqrt{6}$ in the way
\beq
\left(
\begin{array}{c}  \big\vert \eta \rangle      \\   \big\vert \eta^\prime \rangle \end{array} \right) = U(\theta) \, \left( \begin{array}{c}   \big\vert \eta_8 \rangle      \\   \big\vert \eta_1 \rangle \end{array} \right) \,
= U(\phi) \, \left( \begin{array}{c}   \big\vert \eta_q \rangle      \\   \big\vert \eta_s \rangle \end{array}
\right) \, = \left( \begin{array}{ccc} \cos \phi  & - \sin \phi    \\ \sin \phi   & \cos \phi \end{array} \right) \, \left( \begin{array}{c}
  \big\vert \eta_q \rangle      \\   \big\vert \eta_s \rangle \end{array} \right) \, .
\label{eq:eta-etap-mix}
\eeq
The decay constants of physical states
\beq
\big\langle 0 \big\vert \, \bar{q} \gamma_\tau \gamma_5 q \, \big\vert \eta (\eta') \big\rangle = i f_{\eta (\eta')}^{q} p_\tau \,, \,\,\,\,\,\,
\big\langle 0 \big\vert \, \bar{s} \gamma_\tau \gamma_5 s \, \big\vert \eta (\eta') \big\rangle = i f_{\eta (\eta')}^{s} p_\tau \,,
\label{eq:decay-cons-physical}
\eeq
are obtained from that of quark flavor basis
\beq
&&\big\langle 0 \big\vert \, \bar{q} \gamma_\tau \gamma_5 q \, \big\vert \eta_q (p) \big\rangle = \frac{i}{\sqrt{2}} \, p_\tau \, f_q \,,  \,\,\,\,\,\,
\big\langle 0 \big\vert \, \bar{s} \gamma_\tau \gamma_5 s\, \big\vert \eta_s (p) \big\rangle = i \, p_\tau \, f_s \,
\label{eq:decay-cons-flavor}
\eeq
by the same rotation, which are written in terms of mass independent superpositions of $f_q$ and $f_s$
\beq
\left( \begin{array}{c c }    f_\eta^q & f_\eta^s      \\    f_{\eta^\prime}^q & f_{\eta^\prime}^s
\end{array} \right) = U(\phi) \, \left( \begin{array}{c c}   f_q & 0    \\   0 & f_s \end{array} \right) \,. \label{eq:eta-etap-decay-cons}
\eeq
Considering the well-known anomaly of the axial vector currents
\beq
&&\partial^\tau \left(J_{\tau, 5}^{q} \right) =
\sqrt{2} \left( m_u \bar{u} \gamma_5 u + m_d \bar{d} \gamma_5 d + \frac{\alpha_s}{4\pi} G \tilde{G} \right) \,, \non
&&\partial^\tau \left(J_{\tau, 5}^{s} \right) = 2 m_s \bar{s} \gamma_5 s + \frac{\alpha_s}{4\pi} G \tilde{G} \,, 
\label{eq:av-anomaly}
\eeq
where $J_{\tau, 5}^{q} = \frac{1}{\sqrt{2}} (\bar{u} \gamma_\tau \gamma_5 u + \bar{d} \gamma_\tau \gamma_5 d)$, 
$J_{\tau, 5}^{s} = \bar{s} \gamma_\tau \gamma_5 s$, $m_i$ is the current quark mass, 
and $G$ and $\tilde{G}$ are the gluon field strength tensor and its dual, respectively.
Similar to those defined in Eq.(\ref{eq:decay-cons-physical}),
the matrix elements of the axial vector current are given by the product of the decay constants of mesons and the square of meson mass as follow,
\beq
\left\langle 0 \left\vert \partial^\tau \left( \begin{array}{c c}J_{\tau, 5}^{q}  & 0   \\
0  & J_{\tau, 5}^{s}\end{array} \right)\right\vert \left(\begin{array}{c c} \eta  & 0   \\ 0  & \eta^\prime \end{array} \right) \right\rangle = \left(\begin{array}{c c }
   m_{\eta}^2 & 0      \\    0 & m_{\eta^\prime}^2 \end{array}
\right) \left(\begin{array}{c c }
   f_\eta^q & f_\eta^s      \\
   f_{\eta^\prime}^q & f_{\eta^\prime}^s
\end{array} \right) = \mathcal{M}^2 U(\phi)
\left( \begin{array}{c c} f_q & 0   \\ 0  & f_s \end{array} \right) \,, \label{eq:av-matrix-element}
\eeq
which resolves out the mass matrix in the quark flavor basis
\beq
\mathcal{M}_{qs}^2 = U^\dag(\phi) \mathcal{M}^2 U(\phi) =
\left(
\begin{array}{c c}
m_{qq}^2+\frac{\sqrt{2}}{f_q}\big\langle 0 \big\vert \frac{\alpha_s}{4\pi} G \tilde{G} \big\vert \eta_q \big\rangle &
\frac{1}{f_s}\big\langle 0 \big\vert \frac{\alpha_s}{4\pi} G \tilde{G} \big\vert \eta_q \big\rangle     \\
\frac{\sqrt{2}}{f_q}\big\langle 0 \big\vert \frac{\alpha_s}{4\pi} G \tilde{G} \big\vert \eta_s \big\rangle  &
m_{ss}^2+\frac{1}{f_s}\big\langle 0 \big\vert \frac{\alpha_s}{4\pi} G \tilde{G} \big\vert \eta_s \big\rangle
\end{array}
\right) \,
\label{eq:eta-etap-mass}
\eeq
with the quark mass contributions
\beq
m_{qq}^2 &\equiv& i \frac{\sqrt{2}}{f_q} \big\langle 0 \big\vert m_u \bar{u} \gamma_5 u + m_d \bar{d} \gamma_5 d \big\vert \eta_q \big\rangle\non
&=& m_\eta^2 \cos^2 \phi + m^2_{\eta^\prime} \sin^2\phi- \frac{\sqrt{2}f_s}{f_q} \left( m^2_{\eta^\prime} - m^2_\eta \right) \cos\phi \sin\phi \,, \non
m_{ss}^2 &\equiv& \frac{2}{f_s} \big\langle 0 \big\vert m_s \bar{s} \gamma_5 s \big\vert \eta_s \big\rangle
= m_\eta^2 \sin^2 \phi + m^2_{\eta^\prime} \cos^2\phi - \frac{f_q}{\sqrt{2}f_s} \left( m^2_{\eta^\prime} - m^2_\eta \right) \cos\phi \sin\phi\,.
\label{eq:eta-etap-quark-mass}
\eeq
The chiral mass entered into the high twist LCDAs of quark flavour $\eta_{i}$ state is $m_0^i \equiv m_{ii}^2/(2 m_i)$ with $i=q,s$.
The $\eta_q$ and $\eta_s$ components of $\eta, \eta^\prime$ mesons obey the similar twist expansion as in the pion and kaon mesons.

\subsection{Input parameters}\label{paras}

The main uncertainty of the PQCD approach comes from higher order QCD corrections and the nonperturbative parameters of meson LCDAs.
The high order QCD corrections characterized by  the variation of factorization scale is usually minimized by setting the factorization scale 
as the largest virtuality in the hard scattering processes.
We adopt the two-loop expression for the strong coupling constant with the $\beta_{1,2}$ functions \cite{Workman:2022ynf}
\beq
\alpha_s(\mu) = \frac{\pi}{2\,\beta_1\,\textrm{log}(\mu/\Lambda^{(n_f)})}
\left[ 1 - \frac{\beta_2}{\beta_1^2} \, \frac{\textrm{log}(2\,\textrm{log}(\mu/\Lambda^{(n_f)}))}{2\,\textrm{log}(\mu/\Lambda^{(n_f)})} \right] \,,
\label{eq:alphas-2loop}
\eeq
where the active flavor number is chosen as $n_f(\mu) = 3, 4, 5$ when the involved scale $\mu$ locates in $[0,  \overline{m}_c )$, 
$[\overline{m}_c,  \overline{m}_b)$ and $[ \overline{m}_b,  \overline{m}_t)$, respectively, 
by taking into account the quark masses in the $\overline{{\rm MS}}$ scheme \cite{Workman:2022ynf}
\beq
\overline{m}_c(\overline{m}_c) = 1.28 \, \textrm{GeV}\,, \,\,\,\,\,\, \overline{m}_b(\overline{m}_b)= 4.18 \,
\mathrm{GeV}\,, \,\,\,\,\,\, \overline{m}_t(\overline{m}_t) = 165 \, \mathrm{GeV} \,.
\label{eq:quark-MSbar-mass}
\eeq
The QCD scale $\Lambda^{(n_f)}$ is determined by the experimental value of $\alpha_{s}(m_Z) = 0.1182$.

The definition of $B$ meson wave function in Eqs. (\ref{eq:B-wf-2p-2},\ref{eq:B-DA}) relies on the three independent parameters,
which are the mass $m_B$, the decay constant $f_B$ and the first inverse moment $\omega_B$.
We take $m_{B} = 5.28 \, {\rm GeV}$ from PDG \cite{Workman:2022ynf} and
adopt $f_B = 190.0 \pm 1.3\, {\rm MeV}$ from    the lattice QCD calculation \cite{Aoki:2019cca}.
For the inverse moment $\omega_B$,   there are a lot of studies in the literatures  \cite{Braun:2012kp,Wang:2018wfj}.
In our PQCD evaluation we would take the conventional interval $\omega_B (1 \, {\rm GeV}) = 400 \pm 40 \, {\rm MeV}$.
The mean lifetimes of $B$ mesons entered in the observables are also taken from PDG as $\tau_{B^\pm} = 1.638 \times 10^{-12} \, {\rm s}$, $\tau_{B^0} = 1.520 \times 10^{-12} \, {\rm s}$.
In Table \ref{tab:input-paras} we present all the parameters of light meson LCDAs used in our evaluation.
The default scale is indicated at $1 \, {\rm GeV}$.

\begin{table}[tb]
\begin{center}\vspace{-4mm}
\caption{The inputs of parameters in the light meson LCDAs.} \label{tab:input-paras}
{\footnotesize
\begin{tabular}{|c|c c c c | }
\toprule
{\rm Meson} & \quad $\pi^\pm/\pi^0$ & \quad $K^\pm/K^0$ & \quad $\eta_q$ & \quad $\eta_s$ \quad \non
\hline
{\rm $m$ (GeV)} \cite{Workman:2022ynf} & \quad $0.140/0.135$  & \quad $0.494/0.498$  & \quad  $0.104$ \quad  & \quad $0.705$ \quad \non
{\rm $f$ (GeV)}  & \quad $0.130$ \cite{Workman:2022ynf} \quad & \quad $0.156$ \cite{Workman:2022ynf} \quad &
\quad $0.125$ \cite{Ottnad:2017bjt} \quad & \quad $0.177$ \cite{Ottnad:2017bjt} \quad \non
{\rm $m_0$ (GeV)} & \quad $1.400$ \quad & \quad $1.892$ \cite{Leutwyler:1996qg} \quad &
\quad $1.087$ \quad & \quad $1.990$ \quad \non
$a_1$ & \quad $0$ \quad & \quad $0.076 \pm 0.004$ \cite{Arthur:2010xf} \quad  & \quad $0$ \quad & \quad $0$ \quad \non
$a_2$ & \quad $0.270 \pm 0.047 $\cite{Cheng:2020vwr} \quad & \quad $0.221 \pm 0.082$ \cite{Arthur:2010xf} \quad &
\quad $0.250 \pm 0.150$ \cite{Offen:2013nma} \quad & \quad $0.250 \pm 0.150$ \cite{Offen:2013nma} \quad  \non
\toprule
{\rm Meson} & \quad $\rho^{\pm}/\rho^0$ & \quad $K^{\ast \pm}/K^{\ast 0}$ & \quad $\omega$ & \quad $\phi$ \non
\hline
{\rm $m$ (GeV)} \cite{Workman:2022ynf} & \quad $0.775$ & \quad $0.892$ & \quad $0.783$ &\quad $1.019$ \quad \non
{\rm $f^\parallel$ (GeV)} \cite{Straub:2015ica} & \quad $0.210/0.213$ & \quad $0.204$ & \quad $0.197$ & \quad $0.233$ \quad \non
{\rm $f^\perp$ (GeV)} & \quad $0.144/0.146$ \cite{Braun:2016wnx} & \quad $0.159$ \cite{Straub:2015ica} &
\quad $0.162$ \cite{Straub:2015ica} & \quad $0.191$ \cite{Straub:2015ica} \non
$a_1^\parallel$ & \quad $0$ & \quad  $0.060 \pm 0.040$ \cite{Dimou:2012un} & \quad $0$ & \quad $0$ \non
$a_1^\perp$ & \quad $0$ & \quad  $0.040 \pm 0.030$ \cite{Dimou:2012un} & \quad $0$ & \quad $0$ \non
$a_2^\parallel$ & \quad $0.180 \pm 0.037$ \cite{Braun:2016wnx} & \quad $0.160 \pm 0.090$ \cite{Dimou:2012un}
& \quad $0.150 \pm 0.120$ \cite{Dimou:2012un} & \quad $0.230 \pm 0.080$ \cite{Dimou:2012un} \non
$a_2^\perp$ & \quad $0.137 \pm 0.030$ \cite{Braun:2016wnx} & \quad $0.100 \pm 0.080$ \cite{Dimou:2012un}
& \quad $0.140 \pm 0.120$ \cite{Dimou:2012un} & \quad $0.140 \pm 0.070$ \cite{Dimou:2012un} \non
\toprule
\end{tabular} }
\end{center}
\end{table}%

\section{$B$ decay amplitudes at the leading order}\label{sec:amp-B}

The eight leading order Feynman diagrams in Fig.\ref{fig:PQCD} are classed into four groups: 
the naive factorizable diagrams (a) and (b), the hard scattering emission diagrams (c) and (d), 
the naive factorizable annihilation type diagrams (e) and (f) and the hard scattering annihilation type diagrams (g) and (h). 
The calculation is not trivial. We give the leading order formulas of the $B \to PP$, $PV$ and $VV$ decays amplitudes in the following.

\subsection{$B\to PP$ decay modes}\label{B2PP}

The decay amplitudes associated to Fig.\ref{fig:PQCD}(a,b) are detached in to the 
production of heavy-to-light form factors and the decay constant of the emission meson ($f_{M_2}$) based on the naive factorization hypothesis \cite{Keum:2000wi,Lu:2000em}. 
The complete expressions for the naive factorizable emission amplitudes for different type of operators are
\beq
\mathcal{E}_{M_3}^{\bf{LL}} &=& -\mathcal{E}_{M_3}^{\bf{LR}}=
8 \pi C_F m_B^4 f_{M_2} \int_0^1 dx_1 \, dx_3 \, \int_0^{1/\Lambda} b_1 db_1 \, b_3 db_3 \varphi_B(x_1,b_1)\, \non
&\cdot&  \Big\{ h_{e}(x_1,x_3,b_1,b_3) E_e(\mu_e) 
\Big[ (2\,r_b-\bar{x}_3)\varphi_{\pi}^a(x_3)-r_3\left(r_b-2\bar{x}_3 \right)\left(\varphi_{\pi}^p(x_3)+\varphi_{\pi}^t(x_3)\right)  \Big]   \non
&~& + h_{e}(x_3,x_1,b_3,b_1)E_{e^{\prime}}(\mu_{e^{\prime}}) \,2\,r_3\varphi_{\pi}^p(x_3) \Big\} \,, 
\label{eq:FE-B2PP-LL}\\
\mathcal{E}_{M_3}^{\bf{SP}} &=&
16 \pi r_2 C_F m_B^4 f_{M_2} \int_0^1 dx_1 \, dx_3 \, \int_0^{1/\Lambda} b_1 db_1 \, b_3 db_3 \varphi_B(x_1,b_1)\, \non
&\cdot&  \Big\{ h_{e}(x_1,x_3,b_1,b_3) E_e(\mu_e) \Big[ \left(2-r_b \right)\varphi_{\pi}^a(x_3) +r_3\left(4r_b+x_3-2\right)\varphi_{\pi}^p(x_3)-r_3x_3\varphi_{\pi}^t(x_3) \Big]  \non
&~& + h_{e}(x_3,x_1,b_3,b_1) E_{e^{\prime}}(\mu_{e^{\prime}})\, \left(x_1\varphi_{\pi}^a(x_3)+2r_3 \bar{x}_1 \varphi_{\pi}^p(x_3)\right) \Big\} \,. 
\label{eq:FE-B2PP-LL}
\eeq
We use $r_i \equiv m_0^{M_i}/m_B$ to denote the ratio between the chiral mass and the $B$ meson mass. 
The transversal-momenta integrated hard functions $h_{e^{(\prime)}}(x_i,b_j)$ and the Sudakov factor involved function $E_{e}(\mu)$ are collected in  
appendix \ref{app-hardfunction}. 
The subscript ${\bf LL, LR}$ and ${\bf SP}$ indicate the decay amplitudes generated by the corresponding 
${\rm (V-A) \otimes (V-A)}$, ${\rm (V-A) \otimes (V+A)}$ and ${\rm (S-P) \otimes (S+P)}$ types of the four-quark operators 
as shown in Eqs. (\ref{eq:operator-tree}-\ref{eq:operator-QED}). 
In contrast to previous PQCD calculations, we consider two more power corrections in proportional to $x_1$ and $r_b \equiv m_b/m_B$, 
which reflects the high order corrections of heavy quark effective theory in $B$ meson decays. 
We remark that the $r_b$ corrections are only considered in the numerators of invariant decaying amplitudes, 
while not considered in the denominators (hard functions) of them. 
The decay amplitudes of the hard scattering diagrams as shown in Fig.\ref{fig:PQCD}(c,d) read as 
\beq
\mathcal{E}_{NF,M_3}^{\bf{LL}} &=&
\frac{16 \sqrt {6}}{3} \pi C_F m_B^4 \int_0^1 dx_1 \,dx_2 \, dx_3 \, \int_0^{1/\Lambda} b_1 db_1 \, b_2 db_2 \varphi_B(x_1,b_1)\,\varphi_\pi(x_2) \non
&\cdot& \Big\{ h_{ne}(x_1,{\bar x}_2,x_3,b_1,b_2)E_{ne}(\mu_{ne}) 
\Big[ (\bar{x}_2-x_1)\varphi_{\pi}^a(x_3)-r_3x_3\left(\varphi_{\pi}^p(x_3)-\varphi_{\pi}^t(x_3)\right) \Big]  \non
&~& + h_{ne}(x_1,x_2,x_3,b_1,b_2) E_{ne}(\mu_{ne^{\prime}}) \Big[\left(x_1-x_2-x_3\right)\varphi_{\pi}^a(x_3) 
+ r_3x_3\left(\varphi_{\pi}^p(x_3)+\varphi_{\pi}^t(x_3)\right)\Big] \Big\} \,, 
\label{eq:NFE-B2PP-LL}
\eeq
\beq
\mathcal{E}_{NF,M_3}^{\bf{LR}} &=&
\frac{16 \sqrt {6}}{3} \pi C_F m_B^4 r_2 \int_0^1 dx_1 \,dx_2 \, dx_3 \, \int_0^{1/\Lambda} b_1 db_1 \, b_2 db_2 \varphi_B(x_1,b_1)\,\non
&\cdot&  \Big\{ h_{ne}(x_1,{\bar x}_2,x_3,b_1,b_2)E_{ne}(\mu_{ne}) 
\Big[ \left(\bar{x}_2-x_1\right)\varphi_{\pi}^a(x_3)\left(\varphi_{\pi}^p(x_2) +\varphi_{\pi}^t(x_2) \right) \non 
&~& + r_3x_3 \left(\varphi_{\pi}^p(x_2)-\varphi_{\pi}^t(x_2)  \right)\left(\varphi_{\pi}^p(x_3)+\varphi_{\pi}^t(x_3)\right) \non 
&~& + r_3(\bar{x}_2- x_1) \left(\varphi_{\pi}^p(x_2)+\varphi_{\pi}^t(x_2)  \right)\left(\varphi_{\pi}^p(x_3)-\varphi_{\pi}^t(x_3)\right)  \Big] \non 
&~& + h_{ne}(x_1,x_2,x_3,b_1,b_2) E_{ne}(\mu_{ne^{\prime}}) \Big[ -x_2\varphi_{\pi}^a(x_3)\left(\varphi_{\pi}^p(x_2)-\varphi_{\pi}^t(x_2)\right) \non 
&~& - r_3x_2\left(\varphi_{\pi}^p(x_2)-\varphi_{\pi}^t(x_2)\right)\left(\varphi_{\pi}^p(x_3)-\varphi_{\pi}^t(x_3)\right)  \non
&~& - r_2x_3\left(\varphi_{\pi}^p(x_2)+\varphi_{\pi}^t(x_2)\right)\left(\varphi_{\pi}^p(x_3)+\varphi_{\pi}^t(x_3)\right) \non 
&~& + x_1\left(\varphi_{\pi}^p(x_2)-\varphi_{\pi}^t(x_2)\right) \left(\varphi_{\pi}^a(x_3)+r_3(\varphi_{\pi}^p(x_3)-\varphi_{\pi}^t(x_3)) \right) \Big]\Big\} \,, 
\label{eq:NFE-B2PP-LR}\\
\mathcal{E}_{NF,M_3}^{\bf{SP}} &=&
\frac{16 \sqrt {6}}{3} \pi C_F m_B^4 \int_0^1 dx_1 \,dx_2 \, dx_3 \, \int_0^{1/\Lambda} b_1 db_1 \, b_2 db_2 \varphi_B(x_1,b_1)\,\varphi_\pi(x_2) \non
&\cdot&  \Big\{ h_{ne}(x_1,{\bar x}_2,x_3,b_1,b_2)E_{ne}(\mu_{ne}) 
\Big[\left(x_1-\bar{x}_2-x_3\right)\varphi_{\pi}^a(x_3) +r_3x_3\left(\varphi_{\pi}^p(x_3)+\varphi_{\pi}^t(x_3) \right) \Big]   \non
&~& + h_{ne}(x_1,x_2,x_3,b_1,b_2)E_{ne}(\mu_{ne^{\prime}}) \Big[\left( x_2 -x_1\right) \varphi_{\pi}^a(x_3)- r_3x_3\left(\varphi_{\pi}^p(x_3)-\varphi_{\pi}^t(x_3) \right) \Big]\Big\} \,. 
\label{eq:NFE-B2PP-SP}
\eeq

The Feynman diagrams in Fig.\ref{fig:PQCD}(e,f) can be naively factorized as the product of the $B$ meson decay constant and the light meson timelike form factor, 
since quark and anti-quark in the $B$ meson should form a color singlet state. 
The factorizable amplitudes of these diagrams for two-body $\bar{B}^0 \to P_2P_3$ decays are then collected as \cite{Mishima:2003wm,Li:2005kt,Xiao:2011tx,Bai:2013tsa,Fan:2012kn,Zhang:2014bsa,Zhang:2009zg}
\beq
\mathcal{A}^{\bf{LL}}_{M_3}&=&\mathcal{A}^{\bf{LR}}_{M_3}=
8 \pi C_F m_B^4 f_B \int_0^1 \,dx_2 \, dx_3 \, \int_0^{1/\Lambda} b_2 db_2 \, b_3 db_3 \non
&\cdot&  \Big\{ h_{a}(x_2,{\bar x}_3,b_2,b_3)E_a({\mu_a}) \Big[
-\bar{x}_3 \varphi_{\pi}^a(x_2)\varphi_{\pi}^a(x_3)+ 2r_2r_3 \varphi_{\pi}^p(x_2) \non 
&~& \cdot \big[ -  \left( \varphi_{\pi}^p(x_3) + \varphi_{\pi}^t(x_3) \right) 
+ {\bar x}_3  \left( \varphi_{\pi}^t(x_3) - \varphi_{\pi}^p(x_3) \right) \big] \Big] \non
&~& + h_{a}({\bar x}_3,x_2,b_3,b_2) E_{a^\prime}({\mu_{a^\prime}}) 
\Big[ x_2 \varphi_{\pi}^a(x_2)\varphi_{\pi}^a(x_3)  \non 
&~&+ 2r_2r_3 \varphi_{\pi}^p(x_3) \big[ \left( \varphi_{\pi}^p(x_2) - \varphi_{\pi}^t(x_2) \right) 
+ x_2 \left( \varphi_{\pi}^p(x_2) + \varphi_{\pi}^t(x_2) \right) \big] \Big]\Big\} \,, 
 \label{eq:FA-B2PP-LL}\\ 
\mathcal{A}^{\bf{SP}}_{M_3}&=&
16 \pi C_F m_B^4 f_B \int_0^1 \,dx_2 \, dx_3 \, \int_0^{1/\Lambda} b_2 db_2 \, b_3 db_3 \non
&\cdot&  \Big\{ h_{a}(x_2,{\bar x}_3,b_2,b_3)E_a({\mu_a}) \Big[ r_3\bar{x}_3\varphi_{\pi}^a(x_2) 
\left( \varphi_{\pi}^p(x_3)+\varphi_{\pi}^t(x_3)\right)+2r_2\varphi_{\pi}^p(x_2)\varphi_{\pi}^a(x_3) \Big] \non
&~& + h_{a}({\bar x}_3,x_2,b_3,b_2)E_{a^\prime}({\mu_{a^\prime}}) 
\Big[r_2x_2\varphi_{\pi}^a(x_3)\left(\varphi_{\pi}^p(x_2)-\varphi_{\pi}^t(x_2)\right) + 2r_3\varphi_{\pi}^a(x_2)\varphi_{\pi}^p(x_3) \Big]\Big\} \,. 
\label{eq:FA-B2PP-SP}
\eeq
It is easy to see that the decay amplitudes of these kinds of Feynman diagrams are independent of the LCDA of the $B$ meson. 
In the case with two identical particle final states,
only the scalar meson form factors contribute to the factorizable annihilation amplitudes in two-body $B$ meson decays,
while the contribution from ${\rm V-A}$ current is cancelled between electromagnetic form factors, due to the identical particle symmetry. 
The electromagnetic form factor are carried by the ${\rm (V-A)}$ and ${\rm (V+A)}$ currents of four fermion effective operators.
Besides it, the scalar density $\mathcal{J}_S = m_q \, \bar{q} q$ give contribution for the factorizable annihilation amplitudes too,
especially in the color suppressed channels, like $B \to \pi^0\pi^0, \rho^0\rho^0$, et.al.
This contribution is generated by the Fiertz transformation of weak decay operator from ${\rm (V-A) \otimes (V+A)}$ to ${\rm (S-P) \otimes (S+P)}$.
In fact, the ${\rm P}$ term is forbidden because two pion states can not been produced by the pseudo-scalar density operator
$\big \langle P_2P_2 \big\vert \mathcal{J}_{\rm P} \big\vert 0 \big\rangle = 0$.
But in other final states, such as one pseudo-scalar  and one vector meson final states, the ${\rm P}$ term gives the leading contribution.

The last piece of the decay amplitude is considered for the hard scattering annihilation diagrams shown in Fig.\ref{fig:PQCD}(g,h), 
\beq
\mathcal{A}^{\bf{LL}}_{NF,M_3}&=&
\frac{16 \sqrt {6}}{3} \pi C_F m_B^4 \int_0^1 dx_1 \,dx_2 \, dx_3 \, \int_0^{1/\Lambda} b_1 db_1 \, b_2 db_2 \varphi_B(x_1,b_1)\,\non
&\cdot& \Big\{ h_{na}(x_1,x_2,x_3,b_1,b_2)E_{na}(\mu_{na}) 
\Big[ \left(\bar x_1-r_b-x_2\right) \varphi_{\pi}^a(x_2)\varphi_{\pi}^a(x_3)-4r_br_2r_3\varphi_{\pi}^p(x_2)\varphi_{\pi}^p(x_3) \non 
&~& - r_2r_3\left(x_1-\bar{x}_2 \right)\left(\varphi_{\pi}^p(x_2)+\varphi_{\pi}^t(x_2)\right)\left(\varphi_{\pi}^p(x_3)-\varphi_{\pi}^t(x_3)\right) \non 
&~& + r_2r_3x_3 \left(\varphi_{\pi}^p(x_2)-\varphi_{\pi}^t(x_2)\right)\left(\varphi_{\pi}^p(x_3)+\varphi_{\pi}^t(x_3)\right)\Big]   \non
&~& + h_{na^\prime}(x_1,x_2,x_3,b_1,b_2)E_{na}(\mu_{na^{\prime}}) \Big[ \bar{x}_3\varphi_{\pi}^a(x_2)\varphi_{\pi}^a(x_3)\non
&~& + \bar{x}_3r_2r_3\left(\varphi_{\pi}^p(x_2)+\varphi_{\pi}^t(x_2)\right)\left(\varphi_{\pi}^p(x_3)-\varphi_{\pi}^t(x_3) \right) \non 
&~& + r_2r_3 (x_2-x_1) \left(\varphi_{\pi}^p(x_2)-\varphi_{\pi}^t(x_2)\right)\left(\varphi_{\pi}^p(x_3)+\varphi_{\pi}^t(x_3)\right) \Big]\Big\} \,, 
\label{eq:NFA-B2PP-LL}\\
\mathcal{A}^{\bf{LR}}_{NF,M_3}&=&
\frac{16 \sqrt {6}}{3} \pi C_F m_B^4 \int_0^1 dx_1 \,dx_2 \, dx_3 \, \int_0^{1/\Lambda} b_1 db_1 \, b_2 db_2 \varphi_B(x_1,b_1)\,\non
&\cdot&  \Big\{ h_{na}(x_1,x_2,x_3,b_1,b_2)E_{na}(\mu_{na}) \Big[ r_3\left(r_b+x_3\right) 
\varphi_{\pi}^a(x_2)\left(\varphi_{\pi}^t(x_3)-\varphi_{\pi}^p(x_3)\right) \non 
&~& + r_2\left(r_b-x_1+\bar{x}_2 \right)\varphi_{\pi}^a(x_3) \left(\varphi_{\pi}^p(x_2)+\varphi_{\pi}^t(x_2) \right)\Big]  \non
&~& + h_{na^\prime}(x_1,x_2,x_3,b_1,b_2)E_{na}(\mu_{na^{\prime}}) 
\Big[ r_3\bar{x}_3\varphi_{\pi}^a(x_2)\left(\varphi_{\pi}^t(x_3)-\varphi_{\pi}^p(x_3)\right) \non
&~& + r_2\left(x_2-x_1 \right)\varphi_{\pi}^a(x_3) \left(\varphi_{\pi}^p(x_2)+\varphi_{\pi}^t(x_2) \right) \Big]\Big\} \,, 
\label{eq:NFA-B2PP-LR}\\
\mathcal{A}^{\bf{SP}}_{NF,M_3}&=&
\frac{16 \sqrt {6}}{3} \pi C_F m_B^4 \int_0^1 dx_1 \,dx_2 \, dx_3 \, \int_0^{1/\Lambda} b_1 db_1 \, b_2 db_2 \varphi_B(x_1,b_1)\,\non
&\cdot&  \Big\{ h_{na}(x_1,x_2,x_3,b_1,b_2)E_{na}(\mu_{na}) 
\Big[ (x_3-r_b) \varphi_{\pi}^a(x_2)\varphi_{\pi}^a(x_3)-4r_2 r_3 \varphi_{\pi}^p(x_2)\varphi_{\pi}^p(x_3) \non
&~& + r_2r_3\left(\bar{x}_2-x_1 \right) \left(\varphi_{\pi}^p(x_2)-\varphi_{\pi}^t(x_2)\right)\left(\varphi_{\pi}^p(x_3)+\varphi_{\pi}^t(x_3)\right)\non
&~& + r_2r_3x_3 \left(\varphi_{\pi}^p(x_2)+\varphi_{\pi}^t(x_2)\right) \big( \varphi_{\pi}^p(x_3)-\varphi_{\pi}^t(x_2) \Big) \Big]  \non
&~& + h_{na^\prime}(x_1,x_2,x_3,b_1,b_2) E_{na}(\mu_{na^{\prime}}) \Big[ \left(x_2-x_1\right)\varphi_{\pi}^a(x_2)\varphi_{\pi}^a(x_3) \non 
&~& + r_2r_3\left(x_2- x_1\right)\left(\varphi_{\pi}^p(x_2)+\varphi_{\pi}^t(x_2)\right)\left(\varphi_{\pi}^p(x_3)-\varphi_{\pi}^t(x_3)\right)\non 
&~& + r_2r_3\bar{x}_3\left(\varphi_{\pi}^p(x_2)-\varphi_{\pi}^t(x_2)\right)\left(\varphi_{\pi}^p(x_3)+\varphi_{\pi}^t(x_3)\right)\Big]\Big\} \,. 
\label{eq:NFA-B2PP-SP}
\eeq

\subsection{$B\to PV$ decay modes}\label{B2M-ff}
 
Because of the angular momentum conservation, only the longitudinal polarization of vector meson contributes in the $B\to PV$ decay modes. 
With the similar spinor structures as in the case of $B \to PP$ transition matrix element,
one expect that the $B \to PV$ decaying amplitudes can be obtained by a certain substitution from that of $B \to PP$ decays \cite{Li:2006jv,Zhang:2008by,Rui:2011dr,Liu:2005mm,Zhang:2009zzn,Lu:2000hj,Hua:2020usv}.
For the channels with a spectator vector meson and an emission pseudoscalar meson, 
we introduce the substitutions $R1$ and $R2$, 
\beq
&&R1=\Big\{ \varphi_{\pi}^a(x_3) \to \varphi^{\parallel}_{\rho}(x_3)  \,, \;
\varphi_{\pi}^p(x_3) \to  \bar{\psi}_{\rho}^{s,\parallel}(x_3) \,,  \;
\varphi_{\pi}^t(x_3) \to   \varphi_{\rho}^{t,\parallel}(x_3) \Big\} \,, \\
&&R2=\Big\{ \varphi_{\pi}^a(x_3) \to \varphi^{\parallel}_{\rho}(x_3) \,, \;
\varphi_{\pi}^p(x_3) \to -  \bar{\psi}_{\rho}^{s,\parallel}(x_3) \,,  \;
\varphi_{\pi}^t(x_3) \to -  \varphi_{\rho}^{t,\parallel}(x_3)  \Big\} \,.
\label{eq:relations-rhopi-R2p}
\eeq
In this way, the decay amplitudes in $B \to P_2V_3$ channels are obtained by
\beq
&&\mathcal{E}^{{\bf LL}}_{{V_3}}=-\mathcal{E}^{{\bf LR}}_{{V_3}} \stackrel{R1}\Longleftarrow \mathcal{E}^{\bf{LL}}_{{M_3}} \,,  \quad\     
\mathcal{E}^{{\bf SP}}_{{V_3}} \stackrel{R1}\Longleftarrow -\mathcal{E}^{\bf{SP}}_{{M_3}}  \,,      \non
&&\mathcal{E}^{{\bf LL}}_{{NF,V_3}}  \stackrel{R1}\Longleftarrow \mathcal{E}^{\bf{LL}}_{{NF,M_3}}  \,, \quad
\mathcal{E}^{{\bf LR}}_{{NF,V_3}}  \stackrel{R1}\Longleftarrow -\mathcal{E}^{\bf{LR}}_{{NF,M_3}}  \,,  \quad 
\mathcal{E}^{{\bf SP}}_{{NF,V_3}} \stackrel{R1}\Longleftarrow \mathcal{E}^{\bf{SP}}_{{NF,M_3}}  \,,      \non
&&\mathcal{A}^{{\bf LL}}_{{V_3}}=-\mathcal{A}^{{\bf LR}}_{{V_3}} \stackrel{R2}\Longleftarrow \mathcal{A}^{\bf{LL}}_{{M_3}}  \,, \quad
\mathcal{A}^{{\bf SP}}_{{V_3}} \stackrel{R2}\Longleftarrow -\mathcal{A}^{\bf{SP}}_{{M_3}}  \,,      \non
&&\mathcal{A}^{{\bf LL}}_{{NF,V_3}}  \stackrel{R2}\Longleftarrow \mathcal{A}^{\bf{LL}}_{{NF,M_3}}  \,, \quad 
\mathcal{A}^{{\bf LR}}_{{NF,V_3}}  \stackrel{R2}\Longleftarrow -\mathcal{A}^{\bf{LR}}_{{NF,M_3}}  \,,  \quad 
\mathcal{A}^{{\bf SP}}_{{NF,V_3}} \stackrel{R2}\Longleftarrow -\mathcal{A}^{\bf{SP}}_{{NF,M_3}}  \,.      
\label{eq:NFA-B2PV}
\eeq
For the channels with a spectator pseudoscalar meson and an emission vector meson, 
we introduce the substitution $R$
\beq
R=\Big\{ \varphi_{\pi}^a(x_2) \to \varphi^{\parallel}_{\rho}(x_2) \,, \;
\varphi_{\pi}^p(x_2) \to   \bar{\psi}_{\rho}^{s,\parallel}(x_2) \,,  \; 
\varphi_{\pi}^t(x_2) \to   \varphi_{\rho}^{t,\parallel}(x_2) \Big\} \,.
\label{eq:relations-rhopi-R2}
\eeq    
The decay amplitudes of $B \to V_2P_3$ channels are then obtained by
\beq
&&\mathcal{E}^{{\bf LL}}_{{P_3}} =\mathcal{E}^{{\bf LR}}_{{P_3}} \stackrel{R}\Longleftarrow \mathcal{E}^{\bf{LL}}_{{M_3}}  \,, \quad
\mathcal{E}^{{\bf SP}}_{{P_3}} =0 \,,      \non
&&\mathcal{E}^{{\bf LL}}_{{NF,P_3}}  \stackrel{R}\Longleftarrow \mathcal{E}^{\bf{LL}}_{{NF,M_3}}  \,, \quad 
\mathcal{E}^{{\bf LR}}_{{NF,P_3}}  \stackrel{R}\Longleftarrow \mathcal{E}^{\bf{LR}}_{{NF,M_3}}  \,,  \quad 
\mathcal{E}^{{\bf SP}}_{{NF,P_3}} \stackrel{R}\Longleftarrow -\mathcal{E}^{\bf{SP}}_{{NF,M_3}}  \,,      \non
&&\mathcal{A}^{{\bf LL}}_{{P_3}}= -\mathcal{A}^{{\bf LR}}_{{P_3}} \stackrel{R}\Longleftarrow \mathcal{A}^{\bf{LL}}_{{M_3}}  \,, \quad 
\mathcal{A}^{{\bf SP}}_{{P_3}} \stackrel{R}\Longleftarrow \mathcal{A}^{\bf{SP}}_{{M_3}}  \,,      \non
&&\mathcal{A}^{{\bf LL}}_{{NF,P_3}}  \stackrel{R}\Longleftarrow \mathcal{A}^{\bf{LL}}_{{NF,P_3}}  \,, \quad 
\mathcal{A}^{{\bf LR}}_{{NF,P_3}}  \stackrel{R}\Longleftarrow \mathcal{A}^{\bf{LR}}_{{NF,P_3}}  \,,  \quad 
\mathcal{A}^{{\bf SP}}_{{NF,P_3}} \stackrel{R}\Longleftarrow -\mathcal{A}^{\bf{SP}}_{{NF,M_3}}  \,.      
\label{eq:NFA-B2VP}
\eeq

\subsection{$B\to VV$ decay modes}\label{B2M-vv}

We now consider the decays to two vector mesons final states \cite{Li:2004ti,Li:2005hg,Zou:2015iwa}. 
When both the final vector mesons are polarised in the longitudinal direction, 
the decay amplitudes can be obtained again by taken simple substitution from that of $B \to PP$ decays as
\beq
&&\mathcal{E}^{{\bf LL}}_{{V_3}}= \mathcal{E}^{{\bf LR}}_{{V_3}} \stackrel{R3} \Longleftarrow \mathcal{E}^{\bf{LL}}_{{M_3}}  \,, \quad 
\mathcal{E}^{{\bf SP}}_{{V_3}} =0 \,,      \non
&&\mathcal{E}^{{\bf LL}}_{{NF,V_3}}  \stackrel{R3}\Longleftarrow \mathcal{E}^{\bf{LL}}_{{NF,M_3}}  \,, \quad 
\mathcal{E}^{{\bf LR}}_{{NF,V_3}}  \stackrel{R3}\Longleftarrow -\mathcal{E}^{\bf{LR}}_{{NF,M_3}}  \,,  \quad 
\mathcal{E}^{{\bf SP}}_{{NF,V_3}} \stackrel{R3}\Longleftarrow -\mathcal{E}^{\bf{SP}}_{{NF,M_3}}  \,,      \non
&&\mathcal{A}^{{\bf LL}}_{{V_3}} = \mathcal{A}^{{\bf LR}}_{{V_3}} \stackrel{R4}\Longleftarrow \mathcal{A}^{\bf{LL}}_{{M_3}}  \,, \quad
\mathcal{A}^{{\bf SP}}_{{V_3}} \stackrel{R4}\Longleftarrow -\mathcal{A}^{\bf{SP}}_{{M_3}}  \,,      \non
&&\mathcal{A}^{{\bf LL}}_{{NF,V_3}}  \stackrel{R4}\Longleftarrow \mathcal{A}^{\bf{LL}}_{{NF,M_3}}  \,, \quad 
\mathcal{A}^{{\bf LR}}_{{NF,V_3}}  \stackrel{R4}\Longleftarrow -\mathcal{A}^{\bf{LR}}_{{NF,M_3}}  \,,  \quad 
\mathcal{A}^{{\bf SP}}_{{NF,V_3}} \stackrel{R4}\Longleftarrow \mathcal{A}^{\bf{SP}}_{{NF,M_3}}  \,.      
\label{eq:NFA-B2VV}
\eeq
Here the new substitutions read as
\beq
&&R3=\Big\{ \varphi^a_{\pi}(x_{2(3)}) \to \varphi^\parallel_{\rho}(x_{2(3)}) \,, \;
 \varphi_{\pi}^p(x_{2(3)}) \to  \bar{\psi}_{\rho}^{s,\parallel}(x_{2(3)}) \,,  \;
 \varphi_{\pi}^t(x_{2(3)}) \to   \varphi_{\rho}^{t,\parallel}(x_{2(3)}) \Big\} \,,\\
&&R4=\Big\{ \varphi^a_{\pi}(x_{2(3)}) \to \varphi^\parallel_{\rho}(x_{2(3)}) \,, \;
 \varphi_{\pi}^p(x_{2(3)}) \to (-)  \bar{\psi}_{\rho}^{s,\parallel}(x_{2(3)}) \,,   \;
 \varphi_{\pi}^t(x_{2(3)}) \to (-)  \varphi_{\rho}^{t,\parallel}(x_{2(3)})  \Big\} \,.
\label{eq:relations-rhopi-R4p}
\eeq

For the $B \to VV$ decays with both final two vector mesons being transversal polarised, 
each of the decay amplitude can be decomposed into two independent ingredients, polarisation along the parallel and perpendicular directions, respectively. 
Up to the power $\mathcal{O}(r^2)$ with $r_2 \equiv m_{V_2}/m_B$ and $r_3 \equiv m_{V_3}/m_B$, 
the results of factorizable emission diagrams are 
\beq
\mathcal{E}^{\bf{LL,N}}_{V_3}&=&
8 \pi C_F m_B^4 f_{V_2}^\parallel r_2 \int_0^1 dx_1 \, dx_3 \, \int_0^{1/\Lambda} b_1 db_1 \, b_3 db_3 \varphi_B(x_1,b_1)\, \non
&\cdot&  \Big\{ h_{e}(x_1,x_3,b_1,b_3) E_e(\mu_e) \Big[ \left(2-r_b \right)\varphi_{\rho}^{\perp}(x_3) 
- r_3x_3\bar{\psi}_{\rho}^{s,\perp}(x_3) \non 
&~& \hspace{4.2cm} +r_3\left(4r_b+x_3-2\right)\varphi_{\rho}^{t,\perp}(x_3) \Big]   \non
&~& + h_{e}(x_3,x_1,b_3,b_1)E_{e^{\prime}}(\mu_{e^{\prime}}) 
\Big[r_3 (x_1+1)\bar{\psi}_{\rho}^{s,\perp}(x_3)+r_3 \bar x_1 \varphi_{\rho}^{t,\perp}(x_3)\Big]\Big\} \,,  
\label{eq:FE-B2VV_N-LL}\\
\mathcal{E}^{\bf{LL,T}}_{V_3}&=& 
8 \pi C_F m_B^4 f_{V_2}^\parallel r_2 \int_0^1 dx_1 \, dx_3 \, \int_0^{1/\Lambda} b_1 db_1 \, b_3 db_3 \varphi_B(x_1,b_1)\, \non
&\cdot& \Big\{ h_{e}(x_1,x_3,b_1,b_3) E_e(\mu_e) \Big[ \left(2-r_b \right)\varphi_{\rho}^{\perp}(x_3) 
+ r_3\left(4r_b+x_3-2\right)\bar{\psi}_{\rho}^{s,\perp}(x_3) \non 
&~& \hspace{4.2cm} -r_3x_3\varphi_{\rho}^{t,\perp}(x_3) \Big] \non
&~& + h_{e}(x_3,x_1,b_3,b_1)E_{e^{\prime}}(\mu_{e^{\prime}})  
\Big[r_3 \bar{x}_1\bar{\psi}_{\rho}^{s,\perp}(x_3)+r_3(1+x_1)\varphi_{\rho}^{t,\perp}(x_3) \Big]\Big\} \,, 
\label{eq:FE-B2VV_T-LL}\\
\mathcal{E}^{\bf{LR,N}}_{V_3}&=&\mathcal{E}^{\bf{LL,N}}_{V_3}\,,  \quad 
\mathcal{E}^{\bf{LR,T}}_{V_3}=\mathcal{E}^{\bf{LL,T}}_{V_3}\,, \quad
\mathcal{E}^{\bf{SP,N}}_{V_3}=\mathcal{E}^{\bf{SP,T}}_{V_3}=0\,. 
\label{eq:FE-B2VV_NT-LR-SP}
\eeq
The decay amplitudes of the hard scattering emission diagrams shown in Fig.\ref{fig:PQCD}(c) and (d) are 
\beq
\mathcal{E}^{\bf{LL,N}}_{NF,V_3}&=&
\frac{16 \sqrt {6}}{3} \pi C_F m_B^4 r_2 \int_0^1 dx_1 \,dx_2 \, dx_3 \, \int_0^{1/\Lambda} b_1 db_1 \, b_2 db_2 \varphi_B(x_1,b_1)\,  \non
&\cdot&  \Big\{ h_{ne}(x_1,{\bar x}_2,x_3,b_1,b_2)E_{ne}(\mu_{ne}) 
\Big[ (\bar x_2-x_1)\left(\bar{\psi}_{\rho}^{s,\perp}(x_2)+\varphi_{\rho}^{t,\perp}(x_2)\right) \varphi_{\rho}^{\perp}(x_3) \Big]  \non
&~& + h_{ne}(x_1,x_2,x_3,b_1,b_2) E_{ne}(\mu_{ne^{\prime}}) 
\Big[(x_2-x_1)\left(\bar{\psi}_{\rho}^{s,\perp}(x_2)+\varphi_{\rho}^{t,\perp}(x_2)\right)\varphi_{\rho}^{\perp}(x_3)\non 
&~& + 2r_3(x_1-x_2-x_3)\left(\bar{\psi}_{\rho}^{s,\perp}(x_2)\bar{\psi}_{\rho}^{s,\perp}(x_3)
+\varphi_{\rho}^{t,\perp}(x_2)\varphi_{\rho}^{t,\perp}(x_3)\right)\Big] \Big\} \,, 
\label{eq:NFE-B2VV_N-LL} \\
\mathcal{E}^{\bf{LL,T}}_{NF,V_3}&=&  
\frac{16 \sqrt {6}}{3} \pi C_F m_B^4 r_2 \int_0^1 dx_1 \,dx_2 \, dx_3 \, \int_0^{1/\Lambda} b_1 db_1 \, b_2 db_2 \varphi_B(x_1,b_1)\,  \non
&\cdot&  \Big\{ h_{ne}(x_1,{\bar x}_2,x_3,b_1,b_2)E_{ne}(\mu_{ne}) 
\Big[ (\bar x_2-x_1 )\left(\bar{\psi}_{\rho}^{s,\perp}(x_2)+\varphi_{\rho}^{t,\perp}(x_2)\right) \varphi_{\rho}^{\perp}(x_3)  \Big]  \non
&~& + h_{ne}(x_1,x_2,x_3,b_1,b_2) E_{ne}(\mu_{ne^{\prime}}) 
\Big[(x_2-x_1)\left(\bar{\psi}_{\rho}^{s,\perp}(x_2)+\varphi_{\rho}^{t,\perp}(x_2)\right)\varphi_{\rho}^{\perp}(x_3)\non 
&~& + 2r_3(x_1-x_2-x_3)\left(\bar{\psi}_{\rho}^{s,\perp}(x_2)\varphi_{\rho}^{t,\perp}(x_3)
+\varphi_{\rho}^{t,\perp}(x_2)\bar{\psi}_{\rho}^{s,\perp}(x_3)\right)\Big] \Big\} \,, 
\label{eq:NFE-B2VV_T-LL} \\
\mathcal{E}^{\bf{LR,N}}_{NF,V_3}&=&\mathcal{E}^{\bf{LR,T}}_{NF,V_3}=
\frac{16 \sqrt {6}}{3}  \pi C_F m_B^4 r_3 x_3 \int_0^1 dx_1 \,dx_2 \, dx_3 \, \int_0^{1/\Lambda} b_1 db_1 \, b_2 db_2 \non
&\cdot& \varphi_B(x_1,b_1)\, \varphi_{\rho}^{\perp}(x_2)  \Big[\bar{\psi}_{\rho}^{s,\perp}(x_3) - \varphi_{\rho}^{t,\perp}(x_3) \Big]  \non
&\cdot& \Big\{ h_{ne}(x_1,{\bar x}_2,x_3,b_1,b_2)E_{ne}(\mu_{ne}) + h_{ne}(x_1,x_2,x_3,b_1,b_2) E_{ne}(\mu_{ne^{\prime}}) \Big\} \,,
\label{eq:NFE-B2VV_N-LR} 
\eeq
\beq
\mathcal{E}^{\bf{SP,N}}_{NF,V_3}&=&\frac{16 \sqrt {6}}{3} \pi C_F m_B^4 r_2 \int_0^1 dx_1 \,dx_2 \, dx_3 \, \int_0^{1/\Lambda} b_1 db_1 \, b_2 db_2 \varphi_B(x_1,b_1)  \non
&\cdot& \Big\{ h_{ne}(x_1,{\bar x}_2,x_3,b_1,b_2)E_{ne}(\mu_{ne}) 
\Big[ ( x_1 -\bar{x}_2 )\left(\varphi_{\rho}^{t,\perp}(x_2)-\bar{\psi}_{\rho}^{s,\perp}(x_2)\right)\varphi_{\rho}^{\perp}(x_3) \non 
&~& + 2r_3(x_1+\bar{x}_2+x_3)\left(\varphi_{\rho}^{t,\perp}(x_2)\varphi_{\rho}^{t,\perp}(x_3)-\bar{\psi}_{\rho}^{s,\perp}(x_2)\bar{\psi}_{\rho}^{s,\perp}(x_3)\right)\Big]  \non
&~& + h_{ne}(x_1,x_2,x_3,b_1,b_2) E_{ne}(\mu_{ne^{\prime}}) 
\Big[(x_2-x_1)\left( \bar{\psi}_{\rho}^{s,\perp}(x_2)-\varphi_{\rho}^{t,\perp}(x_2)\right)  \varphi_\rho ^\perp (x_3) \Big] \Big\} \,, 
\label{eq:NFE-B2VV_N-SP}\\
\mathcal{E}^{\bf{SP,T}}_{NF,V_3}&=&
\frac{16 \sqrt {6}}{3} \pi C_F m_B^4 r_2  \int_0^1 dx_1 \,dx_2 \, dx_3 \, \int_0^{1/\Lambda} b_1 db_1 \, b_2 db_2 \varphi_B(x_1,b_1)\,  \non
&\cdot&  \Big\{ h_{ne}(x_1,{\bar x}_2,x_3,b_1,b_2) E_{ne}(\mu_{ne}) 
\Big[ (x_1-\bar{x}_2)\left(\varphi_{\rho}^{t,\perp}(x_2)-\bar{\psi}_{\rho}^{s,\perp}(x_2)\right)\varphi_{\rho}^{\perp}(x_3)\non 
&~& + 2r_3(x_1-\bar{x}_2-x_3)\left(\bar{\psi}_{\rho}^{s,\perp}(x_2)\varphi_{\rho}^{t,\perp}(x_3)-\varphi_{\rho}^{t,\perp}(x_2)\bar{\psi}_{\rho}^{s,\perp}(x_3)\right) \Big]  \non
&~& + h_{ne}(x_1,x_2,x_3,b_1,b_2) E_{ne}(\mu_{ne^{\prime}}) 
\Big[(x_2-x_1)\left(\bar{\psi}_{\rho}^{s,\perp}(x_2)-\varphi_{\rho}^{t,\perp}(x_2) \right) \varphi_\rho ^\perp (x_3)\Big] \Big\} \,. 
\label{eq:NFE-B2VV_T-SP} 
\eeq
The decay amplitudes for the naive factorizable annihilation diagrams shown in Fig.\ref{fig:PQCD}(e,f) read as 
\beq
\mathcal{A}^{\bf{LL,N}}_{V_3}&=&\mathcal{A}^{\bf{LR,N}}_{V_3}=
8 \pi C_F m_B^4 f_B r_2 r_3 \int_0^1 \,dx_2 \, dx_3 \, \int_0^{1/\Lambda} b_2 db_2 \, b_3 db_3 \non
&\cdot&  \Big\{ h_{a}(x_2,{\bar x}_3,b_2,b_3)E_a({\mu_a}) 
\Big[(x_3-2)(\bar{\psi}_{\rho}^{s,\perp}(x_2)\bar{\psi}_{\rho}^{s,\perp}(x_3)+\varphi_{\rho}^{t,\perp}(x_2)\varphi_{\rho}^{t,\perp}(x_3)) \non
&~& - x_3(\bar{\psi}_{\rho}^{s,\perp}(x_2)\varphi_{\rho}^{t,\perp}(x_3)+\varphi_{\rho}^{t,\perp}(x_2)\bar{\psi}_{\rho}^{s,\perp}(x_3) )\Big] \non
&~&+ h_{a}({\bar x}_3,x_2,b_3,b_2) E_{a^\prime}({\mu_{a^\prime}}) 
\Big[(x_2+1)\left (\varphi_{\rho}^{t,\perp}(x_2)  \varphi_{\rho}^{t,\perp}(x_3)+\bar{\psi}_{\rho}^{s,\perp}(x_2)\bar{\psi}_{\rho}^{s,\perp}(x_3)  \right) \non &~& - \bar x_2 \left(\varphi_{\rho}^{t,\perp}(x_2)\bar{\psi}_{\rho}^{s,\perp}(x_3) + \bar{\psi}_{\rho}^{s,\perp}(x_2) \varphi_{\rho}^{t,\perp}(x_3) \right) 
 \Big]\Big\} \,, 
 \label{eq:FA-B2VV_N-LL}\\
\mathcal{A}^{\bf{LL,T}}_{V_3}&=& -\mathcal{A}^{\bf{LR,T}}_{V_3}=
-8 \pi C_F m_B^4 f_B r_2 r_3 \int_0^1 \,dx_2 \, dx_3 \, \int_0^{1/\Lambda} b_2 db_2 \, b_3 db_3 \non    
&\cdot&  \Big\{h_{a}(x_2, {\bar x}_3,b_2,b_3)E_a({\mu_a}) 
\Big[2\left(\varphi_{\rho}^{t,\perp}(x_2)\bar{\psi}_{\rho}^{s,\perp}(x_3)+\bar{\psi}_{\rho}^{s,\perp}(x_2)\varphi_{\rho}^{t,\perp}(x_3)\right) \non 
&~& + x_3\left(\varphi_{\rho}^{t,\perp}(x_2)-\bar{\psi}_{\rho}^{s,\perp}(x_2)\right)\left(\varphi_{\rho}^{t,\perp}(x_3)-\bar{\psi}_{\rho}^{s,\perp}(x_3)\right)  \Big]  \non
&~& + h_{a}({\bar x}_3,x_2,b_3,b_2) E_{a^\prime}({\mu_{a^\prime}}) 
\Big[\bar x_2 \left (\varphi_{\rho}^{t,\perp}(x_2)  \varphi_{\rho}^{t,\perp}(x_3)+\bar{\psi}_{\rho}^{s,\perp}(x_2)\bar{\psi}_{\rho}^{s,\perp}(x_3)  \right) \non &~& - (1+x_2) \left(\varphi_{\rho}^{t,\perp}(x_2)\bar{\psi}_{\rho}^{s,\perp}(x_3) + \bar{\psi}_{\rho}^{s,\perp}(x_2) \varphi_{\rho}^{t,\perp}(x_3) \right) 
\Big]\Big\} \,,  
\label{eq:FA-B2VV_T-LL}\\
\mathcal{A}^{\bf{SP,N}}_{V_3} &=& \mathcal{A}^{\bf{SP,T}}_{V_3}=
16 \pi C_F m_B^4 f_B \int_0^1 \,dx_2 \, dx_3 \, \int_0^{1/\Lambda} b_2 db_2 \, b_3 db_3 \non
&\cdot&  \Big\{ h_{a}(x_2,{\bar x}_3,b_2,b_3) E_a({\mu_a}) \, r_2 \varphi_\rho ^\perp (x_3) 
\Big[\bar{\psi}_{\rho}^{s,\perp}(x_2)+\varphi_{\rho}^{t,\perp}(x_2)\Big]  \non 
&~& + h_{a}({\bar x}_3,x_2,b_3,b_2) E_{a^\prime}({\mu_{a^\prime}}) \, r_3 \varphi_\rho ^\perp (x_2) 
\Big[\varphi_{\rho}^{t,\perp}(x_3)-\bar{\psi}_{\rho}^{s,\perp}(x_3) \Big]\Big\} \, 
\label{eq:FA-B2VV_N-SP}  \,.
\eeq
For the decay amplitudes corresponding to the hard scattering annihilation diagrams shown in Fig.\ref{fig:PQCD}(g,h),  the PQCD formulas are
\beq
\mathcal{A}^{\bf{LL,N}}_{NF,V_3} &=& 
-\frac{16 \sqrt {6}}{3}  \pi C_F m_B^4 \int_0^1 dx_1 \,dx_2 \, dx_3 \, \int_0^{1/\Lambda} b_1 db_1 \, b_2 db_2 \varphi_B(x_1,b_1)\,\non
&\cdot&  \Big\{ h_{na}(x_1,x_2,x_3,b_1,b_2)E_{na}(\mu_{na}) 2r_2r_3r_b
\Big[ \varphi_{\rho}^{t,\perp}(x_2)\varphi_{\rho}^{t,\perp}+\bar{\psi}_{\rho}^{s,\perp}(x_2)\bar{\psi}_{\rho}^{s,\perp}(x_3)\Big] \non
&~& + h_{na^\prime}(x_1,x_2,x_3,b_1,b_2)E_{na}(\mu_{na^{\prime}}) 
\Big[ \varphi_\rho^\perp (x_2) \varphi_\rho^\perp (x_3) [r_3^2(x_3-1)-r_2^2x_2] \big] \Big\}  \,, 
\label{eq:NFA-B2VV_N-LL} 
\eeq
\beq
\mathcal{A}^{\bf{LL,T}}_{NF,V_3}&=& 
-\frac{16 \sqrt {6}}{3} \pi C_F m_B^4 \int_0^1 dx_1 \,dx_2 \, dx_3 \, \int_0^{1/\Lambda} b_1 db_1 \, b_2 db_2 \varphi_B(x_1,b_1)\,\non
&\cdot& \Big\{h_{na}(x_1,x_2,x_3,b_1,b_2)E_{na}(\mu_{na}) 
2r_2r_3r_b \Big[ \varphi_{\rho}^{t,\perp}(x_2)\bar{\psi}_{\rho}^{s,\perp}(x_3)+\bar{\psi}_{\rho}^{s,\perp}(x_2)\varphi_{\rho}^{t,\perp}(x_3)\Big] \non
&~& + h_{na^\prime}(x_1,x_2,x_3,b_1,b_2) E_{na}(\mu_{na^{\prime}}) 
\Big[ \varphi_\rho^\perp (x_2) \varphi_\rho^\perp (x_3) [r_3^2(x_3-1) + r_2^2x_2] \big] \Big\} \,, 
\label{eq:NFA-B2VV_T-LL}\\
\mathcal{A}^{\bf{LR,N}}_{NF,V_3}&=& 
\frac{16 \sqrt {6}}{3} \pi C_F m_B^4 \int_0^1 dx_1 \,dx_2 \, dx_3 \, \int_0^{1/\Lambda} b_1 db_1 \, b_2 db_2 \varphi_B(x_1,b_1)\,\non
&\cdot& \Big\{ h_{na}(x_1,x_2,x_3,b_1,b_2)E_{na}(\mu_{na}) 
\Big[ r_2(r_b-x_1+\bar{x}_2)\varphi_{\rho}^{\perp}(x_3)\left(\varphi_{\rho}^{t,\perp}(x_2)+\bar{\psi}_{\rho}^{s,\perp}(x_2)\right)\non 
&~& + r_3(r_b+x_3)\varphi_{\rho}^{\perp}(x_2)\left(\bar{\psi}_{\rho}^{s,\perp}(x_3)-\varphi_{\rho}^{t,\perp}(x_3)\right)\Big]   \non
&~& + h_{na^\prime}(x_1,x_2,x_3,b_1,b_2)E_{na}(\mu_{na^{\prime}}) 
\Big[ r_3 \bar x_3 \varphi_{\rho}^{\perp}(x_2)\left(\bar{\psi}_{\rho}^{s,\perp}(x_3)-\varphi_{\rho}^{t,\perp}(x_3)\right)\non 
&~& - r_2(x_1-x_2)\varphi_{\rho}^{\perp}(x_3)\left(\varphi_{\rho}^{t,\perp}(x_2)+\bar{\psi}_{\rho}^{s,\perp}(x_2)\right)\Big]\Big\} \,, 
\label{eq:NFA-B2VV_N-LR} \\
\mathcal{A}^{\bf{LR,T}}_{NF,V_3}&=&\mathcal{A}^{\bf{LR,N}}_{NF,V_3} \,, \quad
\mathcal{A}^{\bf{SP,N}}_{NF,V_3}=\mathcal{A}^{\bf{LL,N}}_{NF,V_3} \,, \quad
\mathcal{A}^{\bf{SP,T}}_{NF,V_3}=-\mathcal{A}^{\bf{LL,T}}_{NF,V_3} \,. 
\label{eq:NFA-B2VV_NT-SP} 
\eeq

\section{The next-to-leading order corrections}\label{PQCD-NLO}

\begin{figure}[t]
\vspace{-0.5cm}
\includegraphics[scale=0.9]{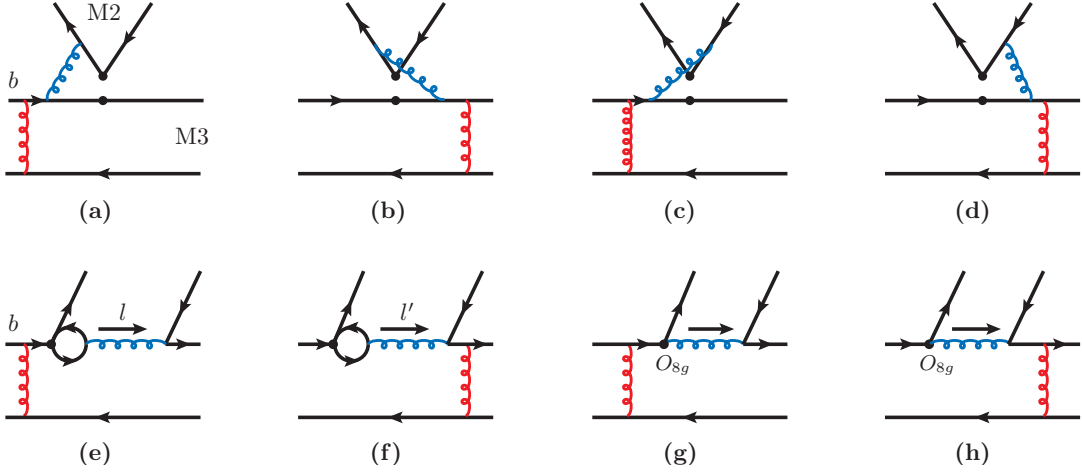}
\caption{Typical Feynman diagrams for the NLO corrections to the emission amplitudes, via the vertex (a-d), 
the quark-loop (e-f) and the chromomagnetic dipole operator (g-h).}
\label{fig:PQCD-NLO1}
\end{figure}

\begin{figure}[t]
\includegraphics[scale=0.9]{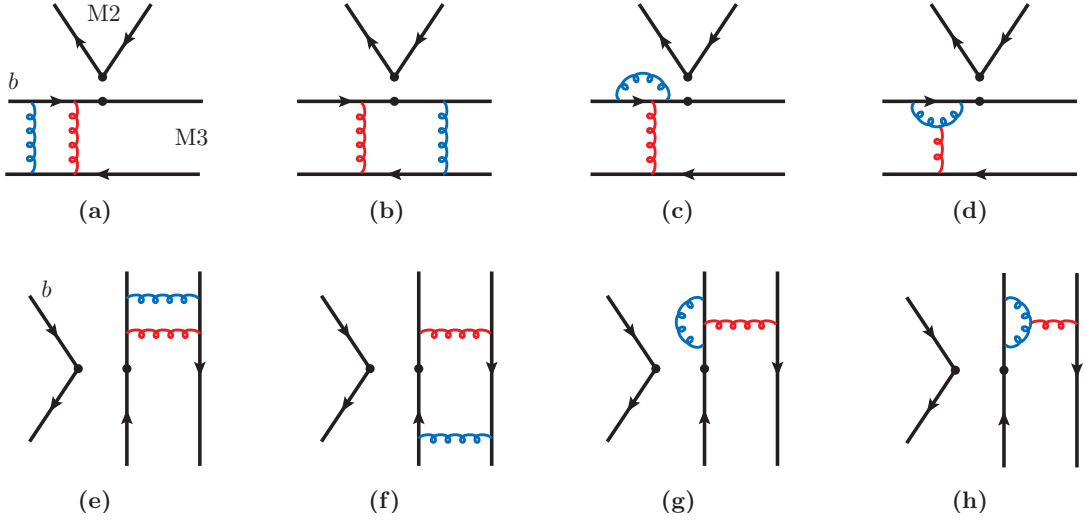}
\caption{Typical Feynman diagrams for the NLO corrections to the naive factorizable amplitudes:
  the $B \to M_2$ transition (a-d) and the timelike $M_2 \to M_3$ (e-h) form factor corrections.}
\label{fig:PQCD-NLO2}
\end{figure}

\begin{figure}[t]
\vspace{-0.5cm}
\includegraphics[scale=0.9]{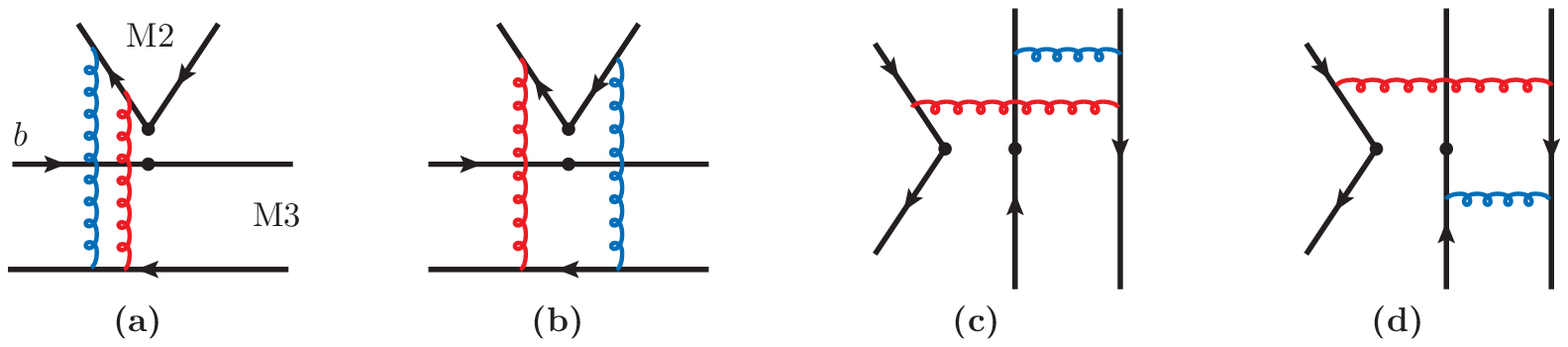}
\caption{Typical Feynman diagrams for the NLO corrections to the hard scattering amplitudes.}
\label{fig:PQCD-NLO3}
\end{figure}

In the last twenty years, a lot of efforts have been made to improve the accuracy of the PQCD calculation in non-leptonic B decays. 
The NLO QCD corrections are supplemented in the PQCD approach to calculate the factorizable emission amplitudes
as depicted in Fig.~\ref{fig:PQCD-NLO1}. 
These corrections include the vertex corrections (a-d), the quark-loop contributions (e-f),
the chromo-magnetic penguin ${\cal O}_{8g}$ contributions (g-h) \cite{Li:2005kt,Mishima:2003wm}.
The Feynman diagrams of NLO QCD corrections to $B \to M_3$ transition form factors are shown in  Fig.~\ref{fig:PQCD-NLO2} (a-d), 
while QCD corrections to the timelike $M_2\to M_3$ form factors are shown in Fig.~\ref{fig:PQCD-NLO2} (e-h) \cite{Li:2012nk,Cheng:2014fwa}.
The NLO QCD corrections to the hard scattering amplitudes as depicted in Fig.~\ref{fig:PQCD-NLO3} (a,b)
are the Glauber gluon (red curves) contributions to the spectator amplitudes \cite{Li:2014haa,Liu:2015sra,Liu:2015upa}, 
which are essential to explain the $\pi\pi$ and $\pi K$ puzzle. 
The diagrams in Fig.\ref{fig:PQCD-NLO3}(c,d) show the NLO contributions to the hard scattering annihilation amplitudes.
These types of NLO corrections are still under study now. Therefore, the current NLO QCD correction to charmless $B$ decays is not a complete one.


\subsection{The vertex corrections}\label{PQCD-NLO-vertex}

The NLO vertex corrections play an important role to reduce the dependence of the Wilson coefficient on the renormalization/factorization scale.
Since this type corrections do not involve the end-point singularity in the collinear factorization theorem,
the PQCD approach results without $k_T$ dependence are  the same as the  QCDF result of the vertex corrections as given in Ref. \cite{Beneke:2001ev}.
The hard gluon of vertex corrections attaches two different fermion lines among the four quark operators,
so the corrections can be absorbed into the effective Wilson coefficients according to the effective operators
\beq
a_{1,2}(\mu) &\to& a_{1,2}(\mu) + \frac{\alpha_s(\mu)C_F}{4\pi} \frac{C_{1,2}(\mu)}{N_C} \, V_{1,2}(M_2) \,, \non
a_{i,i+1}(\mu) &\to&  a_{i,i+1}(\mu) + \frac{\alpha_s(\mu)C_F}{4\pi} \frac{C_{i+1,i}}{N_C} \, V_{i,i+1}(M_2) \, \quad {\rm with} \, i = 3,5,7,9 \,. 
\label{eq:VC-WC}
\eeq
The infrared divergence in the NLO QCD correction is factorized into the LCDAs of the emission meson. So the above NLO results are dependent of the meson type. 
The function $V_{1,2}$ in the naive dimensional regularization scheme for the pseudoscalar meson are
\beq
V_i(P) &=& 12 \ln\frac{m_b}{\mu} - 18 +\int_0^1 dx \, \varphi_{P}^a(x) \, g(x) \, \quad {\rm with} \, i = 1-4,9,10 \,, \non  
V_i(P) &=& - 12 \ln\frac{m_b}{\mu} + 6 - \int_0^1 dx \, \varphi_{P}^a(x) \, g(1-x) \, \quad {\rm with} \, i = 5,7 \,, \non
V_i(P) &=& - 6 + \int_0^1 dx \, \varphi^{{\rm p}}_{P}(x) \, h(x) \,
\quad {\rm with} \, i = 6,8 \,, \label{eq:VC-functions}
\eeq
where $\varphi_{\rm p}^{(a)}$ is the LCDAs of the pseudoscalar meson. The hard kernel functions are
\beq
g(x) &=& 3 \left( \frac{1-2x}{1-x} \ln x - i \pi \right) + \left[ 2 \, {\rm Li_2}(x) - \ln^2 x + \frac{2 \, \ln x}{1-x} + (3+2i\pi) \, \ln x  - \{ x \leftrightarrow 1-x\} \right] \,, \non
h(x) &=&  2 \, {\rm Li_2}(x) - \ln^2 x - (1+2i \pi) \, \ln x - \{ x \leftrightarrow 1-x \} \,.
\label{eq:VC-hard-functions}
\eeq
When the emission particle is a vector meson, the correction functions in Eq. (\ref{eq:VC-functions}) are modified slightly by the simple substitutions of the LCDAs
$\varphi_{P}^a(x) \to \varphi_{V}^\parallel(x), \, \varphi_{P}^{\rm p}(x) \to \bar{\psi}_{V}^{s,\parallel}(x)$.

The vertex correction function contributes an imaginary part, whose effect is mainly embodied on $a_2, a_3$ and $a_{10}$.
Wilson coefficients $a_3$ and $a_{10}$ are suppressed compared with $a_4$ and $a_9$ for the QCD penguin and electroweak penguins, respectively,
so this correction leads to significant change to the color-suppressed amplitudes, rather than the penguin amplitudes.
For example, the vertex corrections enhance the branching fraction of $B^0 \to \pi^0\pi^0$ by a factor of $\sim 1.5$
and change the sign of its direct ${\bf CP}$ violation from minus to plus, which are shown in the beginning of next section.

Besides the naive factorizable emission amplitudes as shown already in figure \ref{fig:PQCD-NLO1}(a-d), 
the vertex correction to the hard scattering emission amplitudes are also an important part of the complete NLO results. 
This contribution is argued to be small in contrast to the naive factorizable ones for most charged channels. 
But it is important in the color-suppressed decay channels with neutral meson final states, 
since it may change the relative sign between the naive factorizable and hard scattering emission amplitudes, especially the imaginary parts.
A further implementation of the vertex correction in the annihilation topological amplitudes,
might provide significant changes for the ${\bf CP}$ violation with the potential correction effects to the strong phase.
Although the difference may be as small as estimated in Ref. \cite{Li:2005kt}, 
the direct calculation of the NLO vertex corrections in the framework of the PQCD approach is still an important work to be done in the near future.

\subsection{The quark-loop corrections}\label{PQCD-NLO-QL}

Because the quark-loop correction as depicted in Fig. \ref{fig:PQCD-NLO1} (e-f) does not involve the end-point singularity
and does not cause the momenta redistribution in the hard kernel, the result of this NLO correction has the same form as the QCDF calculation\cite{Beneke:2001ev}.
\beq
&&{\cal C}^{(u,c)}(\mu, l^2) = \left[ {\cal G}^{(u,c)}(\mu, l^2) - \frac{2}{3} \right] C_2(\mu) \,,
\label{eq:ql-uc-function}\\
&&{\cal C}^{(t)} (\mu, l^2)= \left[ {\cal G}^{(s)}(\mu, l^2) - \frac{2}{3} \right] C_3(\mu)
+ \sum_{q^{\prime\prime}}^{u,d,s,c} {\cal G}^{(q^{\prime\prime})}(\mu, l^2) \left[ C_4(\mu) + C_6(\mu) \right] \,,
\label{eq:ql-t-function-C}
\eeq
where $l^2$ is the invariant mass of the gluon attached to the quark loop.
We here   only include  the quark-loop QCD corrections from the tree and QCD penguin operators. The   quark-loop corrections from electroweak penguin operators are neglected due to their smallness.
The correction to operator $O_5$ is special with a pure ultraviolet effect,
which is absorbed into the effective Wilson coefficient by the redefinition $C_{8g}^{{\rm eff}} = C_{8g} + C_5$ of the chromomagnetic dipole operator shown later.
For the case of massive charm quark\footnote{For the cases with light quarks in the loop,
the expressions are similar but with the replacements $m_c \to m_{u,d,s}$.}, the function ${\cal G}$ reads as
\beq
{\cal G}^{(c)}(\mu, l^2) = - 4 \int_0^1 \, dx \, x(1-x) \ln \left[ \frac{m_c^2 - x(1-x) l^2}{\mu^2} \right] \,,
\label{eq:ql-t-function-G}
\eeq
with the real and imaginary parts
\beq
{\rm Re} \left[ {\cal G}^{(c)}(\mu, l^2) \right] &=& \frac{2}{3} \left( \frac{5}{3} + \frac{4 m_c^2}{l^2} - \ln \frac{m_c^2}{\mu^2} \right) \non
&+& \frac{2}{3} \left( 1 + \frac{2 m_c^2}{l^2} \right) \left\{
\begin{array}{cc}
\sqrt{\beta} \ln \frac{\sqrt{\beta} - 1}{\sqrt{\beta} + 1}\,,  & l^2\in (- \infty, 0) \,, \\
-2 \sqrt{- \beta} \cot^{-1} \sqrt{- \beta} \,, & l^2\in [0, 4m_c^2) \,, \\
-2 \sqrt{\beta} \,,   &   l^2 = 4 m_c^2 \,,\\
\sqrt{\beta} \ln \frac{1 - \sqrt{\beta}}{1 + \sqrt{\beta}} \,,  & l^2\in (4 m_c^2, \infty) \,, \end{array} \right. \, \label{eq:ql-t-function-G-real} \\
{\rm Im} \left[ {\cal G}^{(c)}(\mu, l^2) \right] &=& \frac{2 \pi}{3} \left( 1 + \frac{2 m_c^2}{l^2} \right)  \sqrt{\beta} \, \Theta[\beta] \,,
\label{eq:ql-t-function-G-imag}
\eeq 
where $\Theta[\beta] $ is the Heaviside function.
The phase space factor of quark loop invariant mass is defined by $\beta \equiv \beta(l^2) = 1 - 4 m_c^2/l^2$.
We mark that the above expressions with the number of active quark flavour $n_f =4$ is a little different from the QCDF result with $n_f = 5$, 
since the characteristic hard scale is ${\cal O}(\sqrt{\bar{\Lambda} m_b} \sim 1.5 {\rm GeV})$ in the PQCD.

The gluon momentum $l^2$ of the quark-loop corrections is the sum of  two quark momenta from two final state mesons as shown in Fig.\ref{fig:PQCD-NLO1} (e-f), 
which makes the decay amplitude of nonleptonic two-body $ {B} $ decays are dependent of three meson wave functions
\beq     
{\cal M}^{({\rm ql})}_{B \to P_2 P_3}    
&=& - \frac{8}{\sqrt{6}} C_F^2  m_B^4 \int_0^1 dx_1dx_2 \, dx_3 \, \int_0^{1/\Lambda} b_1 db_1 \, b_3 db_3 \varphi_B(x_1,b_1) \, 
 \Big\{h_{e}(x_1,x_3,b_1,b_3) E_q(\mu_q)\non
&~& \Big[ \varphi_{\pi}^a(x_2)\big[ \left(2r_b-\bar{x}_3\right)\varphi_{\pi}^a(x_3) 
- r_3\left(r_b-2\bar{x}_3\right)\left(\varphi_{\pi}^p(x_3)+\varphi_{\pi}^t(x_3)\right)\big] \non 
&~& - 2r_2\varphi_{\pi}^p(x_2)\big[\left(r_b-2\right)\varphi_{\pi}^a(x_3) - r_3\left(4r_b+x_3-2\right)\varphi_{\pi}^p(x_3) 
+r_3x_3\varphi_{\pi}^t(x_3)\big] \Big]   \non
&~& + h_{e}(x_3,x_1,b_3,b_1)E_{q^{\prime}}(\mu_{q^{\prime}}) \, 2 \big[ r_3\varphi_{\pi}^a(x_2)\varphi_{\pi}^p(x_3) \non 
&~& + r_2\varphi_{\pi}^p(x_2) (x_1\varphi_{\pi}^a(x_3)+2r_3\bar{x}_1\varphi_{\pi}^p(x_3) )\big] \Big\} \,, 
\label{eq:QL-amplitudes-B2PP}\\
{\cal M}^{({\rm ql})}_{ B\to P_2V_3}    
&=& \frac{8}{\sqrt{6}} C_F^2  m_B^4 \int_0^1 dx_1dx_2 \, dx_3 \, \int_0^{1/\Lambda} b_1 db_1 \, b_3 db_3 \varphi_B(x_1,b_1) \, 
\Big\{ \, h_{e}(x_1,x_3,b_1,b_3) E_q(\mu_q) \non
&~& \Big[ \varphi_{\pi}^a(x_2) \big[ \left(\bar{x}_3- 2 r_b\right)\varphi_\rho^\parallel(x_3)+r_3\left(r_b-2\bar{x}_3\right) \left(\bar{\psi}_{\rho}^{s,\parallel}(x_3)+\varphi_{\rho}^{t,\parallel}(x_3)\right) \big] \non 
&~& - 2r_2\varphi_{\pi}^p(x_2)\big[\left(r_b-2\right)\varphi_\rho^\parallel(x_3)-r_3\left(4r_b+x_3-2\right)\bar{\psi}_{\rho}^{s,\parallel}(x_3)+r_3x_3\varphi_{\rho}^{t,\parallel}(x_3) \big] \Big]   \non
&~& + h_{e}(x_3,x_1,b_3,b_1)E_{q^{\prime}}(\mu_{q^{\prime}}) \, 2 \big[ -r_3 \varphi_{\pi}^a(x_2) \bar{\psi}_{\rho}^{s,\parallel}(x_3) \non 
&~& + r_2 \varphi_{\pi}^p(x_2)  (x_1 \varphi_{\rho}^{\parallel}(x_3) +2 r_3 \bar x_1 \bar{\psi}_{\rho}^{s,\parallel}(x_3) ) \big] \Big\} \,, 
\label{eq:QL-amplitudes-B2PV}\\
{\cal M}^{({\rm ql})}_{B\to V_2P_3} &\stackrel{R}\Longleftarrow&  {\cal M}^{({\rm ql})}_{B \to P_2 P_3}  , 
\label{eq:QL-amplitudes-B2VP}   \quad\quad\quad\quad
{\cal M}^{({\rm ql})}_{B \to V_2V_3}  \stackrel{R}\Longleftarrow   {\cal M}^{({\rm ql})}_{ B\to P_2V_3}  \,, 
\label{eq:QL-amplitudes-B2VV_L}\\
{\cal M}^{({\rm ql}), {\bf N}}_{B\to V_2V_3} &=& 
-\frac{8}{\sqrt{6}} C_F^2  m_B^4 r_2 \int_0^1 dx_1dx_2 \, dx_3 \, \int_0^{1/\Lambda} b_1 db_1 \, b_3 db_3 \varphi_B(x_1,b_1)\, 
\Big\{ h_{e}(x_1,x_3,b_1,b_3) E_q(\mu_q) \non
&~& \Big[ r_3\big[(4r_b+x_3-2)\left(\bar{\psi}_{\rho}^{s,\perp}(x_2)\bar{\psi}_{\rho}^{s,\perp}(x_3) 
+\varphi_{\rho}^{t,\perp}(x_2)\varphi_{\rho}^{t,\perp}(x_3)\right)\non 
&~& - x_3\left(\bar{\psi}_{\rho}^{s,\perp}(x_2)\varphi_{\rho}^{t,\perp}(x_3)+\varphi_{\rho}^{t,\perp}(x_2)\bar{\psi}_{\rho}^{s,\perp}(x_3)\right) \non
&~& - (r_b-2)\varphi_{\rho}^{\perp}(x_3)\left(\bar{\psi}_{\rho}^{s,\perp}(x_2)+\varphi_{\rho}^{t,\perp}(x_2)\right)\big]\Big]   \non
&~& + h_{e}(x_3,x_1,b_3,b_1)E_{q^{\prime}}(\mu_{q^{\prime}}) \, r_3\Big[\bar{\psi}_{\rho}^{s,\perp}(x_2)\left(\bar{x}_1\bar{\psi}_{\rho}^{s,\perp}(x_3)+(1+x_1)\varphi_{\rho}^{t,\perp}(x_3)\right)\non 
&~& + \varphi_{\rho}^{t,\perp}(x_2)\left((1+x_1)\bar{\psi}_{\rho}^{s,\perp}(x_3)+\bar{x}_1\varphi_{\rho}^{t,\perp}(x_3)\right)\Big]\Big\} \,, 
\label{eq:QL-amplitudes-B2VV_N}\\
{\cal M}^{({\rm ql}), {\bf T}}_{B\to V_2V_3} &=&
-\frac{8}{\sqrt{6}} C_F^2  m_B^4 r_2 \int_0^1 dx_1dx_2 \, dx_3 \, \int_0^{1/\Lambda} b_1 db_1 \, b_3 db_3 \varphi_B(x_1,b_1)\, 
\Big\{ \, h_{e}(x_1,x_3,b_1,b_3) E_q(\mu_q) \non
&~& \Big[r_3 \big[ x_3\left(\varphi_{\rho}^{t,\perp}(x_2)-\bar{\psi}_{\rho}^{s,\perp}(x_2)\right)\left(\bar{\psi}_{\rho}^{s,\perp}(x_3)-\varphi_{\rho}^{t,\perp}(x_3)\right)\non 
&~& + 2(2 r_b-1)\left(\bar{\psi}_{\rho}^{s,\perp}(x_2)\varphi_{\rho}^{t,\perp}(x_3)+\varphi_{\rho}^{t,\perp}(x_2)\bar{\psi}_{\rho}^{s,\perp}(x_3)\right) \big] \non 
&~& - (r_b-2)\varphi_{\rho}^{\perp}(x_3)\left(\bar{\psi}_{\rho}^{s,\perp}(x_2)+\varphi_{\rho}^{t,\perp}(x_2)\right) \Big]   \non
&~& + h_{e}(x_3,x_1,b_3,b_1)E_{q^{\prime}}(\mu_{q^{\prime}}) \,r_3 
\Big[ \bar{\psi}_{\rho}^{s,\perp}(x_2) \big[ (1+x_1)\bar{\psi}_{\rho}^{s,\perp}(x_3)+\bar{x}_1\varphi_{\rho}^{t,\perp}(x_3) \big] \non 
&~& + \varphi_{\rho}^{t,\perp}(x_2) \big[ \bar{x}_1\bar{\psi}_{\rho}^{s,\perp}(x_3)+(1+x_1)\varphi_{\rho}^{t,\perp}(x_3) \big] \Big] \Big\} \,.
\label{eq:QL-amplitudes-B2VV_T}   
\eeq
The Sudakov factors functions $E_{q^{(\prime)}}$ in these amplitudes are also collected in the appendix \ref{app-hardfunction}.

\subsection{The chromomagnetic-penguin corrections}\label{PQCD-NLO-MP}

As depicted in Fig.~\ref{fig:PQCD-NLO1} (g,h), the chromomagnetic dipole operator $O_{\rm 8g}$ will contribute another NLO correction.
Similar to the quark-loop corrections, it introduces another independent amplitude in the two body $B$ decays
\beq
{\cal M}^{({\rm mp})}_{B\to P_2P_3} &=& 
-\frac{8}{\sqrt{6}} C_F^2  m_B^6 \int_0^1 dx_1 dx_2\, dx_3 \, \int_0^{1/\Lambda} b_1 db_1 b_2 db_2 \, b_3 db_3 \varphi_B(x_1,b_1) \non 
&\cdot&\Big\{ h_g(\alpha_g,\beta_g,\gamma_g,b_1,b_2,b_3) E_{8g}(\mu_g) 
\Big[-\bar{x}_3\varphi_{\pi}^a(x_2)\Big(2(r_b-2)\varphi_{\pi}^a(x_3) \non
&~& +r_3 \big[ \varphi_{\pi}^t(x_3)(-2r_b+x_3+1)- \varphi_{\pi}^p(x_3)\left(6r_b-\bar{x}_3\right) \big] \Big)  \non 
&~& +r_2 \big[ x_2\left(3\varphi_{\pi}^p(x_2)-\varphi_{\pi}^t(x_2)\right) 
\big(\varphi_{\pi}^a(x_3)(2r_b-\bar{x}_3)-r_3(r_b-2\bar{x}_3) \left(\varphi_{\pi}^p(x_3)+\varphi_{\pi}^t(x_3)\right)\big) \non 
&~& - r_3(r_b-2)\bar{x}_3\left(3\varphi_{\pi}^p(x_2)+\varphi_{\pi}^t(x_2)\right)\left(\varphi_{\pi}^p(x_3)-\varphi_{\pi}^t(x_3)\right) \big] \Big]   \non
&~& + h_g(\alpha_{g^\prime},\beta_{g^\prime},\gamma_{g^\prime},b_1,b_2,b_3) E_{8g^{\prime}}(\mu_{g^{\prime}}) 
\Big[x_1\varphi_{\pi}^a(x_3)\big[ \varphi_{\pi}^a(x_2)+r_2\left(3\varphi_{\pi}^p(x_2)+\varphi_{\pi}^t(x_2)\right)\big] \non 
&~& - 2 r_3\varphi_{\pi}^p(x_3) \big[(x_1-2)\varphi_{\pi}^a(x_2)+r_2\left(6x_1\varphi_{\pi}^p(x_2)+x_2\left(\varphi_{\pi}^t(x_2)-3\varphi_{\pi}^p(x_2)\right)\right)\big] \Big]\Big\} \,,
\label{eq:MP-amplitudes-B2PP} \\
{\cal M}^{({\rm mp})}_{B\to P_2V_3} &=& 
\frac{8}{\sqrt{6}} C_F^2  m_B^6 \int_0^1 dx_1 dx_2\, dx_3 \, \int_0^{1/\Lambda} b_1 db_1 b_2 db_2 \, b_3 db_3 \varphi_B(x_1,b_1)\, \non
&\cdot&  \Big\{ h_g(\alpha_g,\beta_g,\gamma_g,b_1,b_2,b_3) E_{8g}(\mu_g)
\Big[ \bar{x}_3\varphi_{\pi}^a(x_2)\Big(2(r_b-2)\varphi_\rho^\parallel(x_3) \non 
&~& + r_3\big[ \varphi_{\rho}^{t,\parallel}(x_3)(-2r_b+x_3+1) - \bar{\psi}_\rho^{s,\parallel}(x_2)\left(6r_b-\bar x_3 \right) \Big) \big] \non 
&~& + r_2 \big[ x_2\left(3\varphi_{\pi}^p(x_2)-\varphi_{\pi}^t(x_2)\right)\big(\varphi_\rho^\parallel(x_3)(2r_b-\bar{x}_3)
- r_3(r_b-2\bar{x}_3) (\bar{\psi}_\rho^{s,\parallel}(x_2)+\varphi_{\rho}^{t,\parallel}(x_3) )\big) \non 
&~& + r_3(r_b-2)\bar{x}_3\left(3\varphi_{\pi}^p(x_2)+\varphi_{\pi}^t(x_2)\right)\left( \varphi_{\rho}^{t,\parallel}(x_3)-\bar{\psi}_\rho^{s,\parallel}(x_3)\right) \big] \Big]   \non
&~& + h_g(\alpha_{g^\prime},\beta_{g^\prime},\gamma_{g^\prime},b_1,b_2,b_3)E_{8g^{\prime}}(\mu_{g^{\prime}}) \Big[x_1\varphi_\rho^\parallel(x_3)\big[ r_2\left(3\varphi_{\pi}^p(x_2)+\varphi_{\pi}^t(x_2)\right)-\varphi_{\pi}^a(x_2) \big] \non 
&~& + 2 r_3\bar{\psi}_\rho^{s,\parallel}(x_3)\big[ (x_1-2)\varphi_{\pi}^a(x_2)-r_2\left(6x_1\varphi_{\pi}^p(x_2)+x_2\left(\varphi_{\pi}^t(x_2)-3\varphi_{\pi}^p(x_2)\right)\right)\big] \Big]\Big\} \,,
\label{eq:MP-amplitudes-B2PV} \\
{\cal M}^{({\rm mp}), {\bf N}}_{B\to V_2V_3} &=&
-\frac{8}{\sqrt{6}} C_F^2  m_B^6 \int_0^1 dx_1dx_2 \, dx_3 \, \int_0^{1/\Lambda} b_1 db_1 b_2 db_2 \, b_3 db_3 \varphi_B(x_1,b_1)\, \non
&\cdot& \Big\{ h_g(\alpha_g,\beta_g,\gamma_g,b_1,b_2,b_3) E_{8g}(\mu_g) \non
&~& \Big[r_2 r_3\Big( \big[ 2\left(1+x_2\right)\bar{x}_3-r_b\left(x_2+\bar{x}_3\right) \big]
\big[ \bar{\psi}_{\rho}^{s,\perp}(x_2)\bar{\psi}_{\rho}^{s,\perp}(x_3) + \varphi_{\rho}^{t,\perp}(x_2)\varphi_{\rho}^{t,\perp}(x_3)\big] \non 
&~&- \big[ 2\bar{x}_2\bar{x}_3+r_b\left(x_2-\bar{x}_3\right)\big] 
\big[\bar{\psi}_{\rho}^{s,\perp}(x_2)\varphi_{\rho}^{t,\perp}(x_3)+\varphi_{\rho}^{t,\perp}(x_2)\bar{\psi}_{\rho}^{s,\perp}(x_3)\big]\Big)\non 
&~& + \left( \bar{x}_3-2r_b \right) \big[ r_3\bar{x}_3\varphi_{\rho}^{\perp}(x_2)\left(\bar{\psi}_{\rho}^{s,\perp}(x_3)-\varphi_{\rho}^{t,\perp}(x_3)\right) \non
&~& - r_2x_2\varphi_{\rho}^{\perp}(x_3)\left(\bar{\psi}_{\rho}^{s,\perp}(x_2)+\varphi_{\rho}^{t,\perp}(x_2)\right) \big] \Big]   \non
&~& + h_g(\alpha_{g^\prime},\beta_{g^\prime},\gamma_{g^\prime},b_1,b_2,b_3) E_{8g^{\prime}}(\mu_{g^{\prime}}) \, r_3 
\Big[ x_1\varphi_{\rho}^{\perp}(x_2)\left(\bar{\psi}_{\rho}^{s,\perp}(x_3)-\varphi_{\rho}^{t,\perp}(x_3)\right) \non
&~& + r_2\big[ x_2\left(\bar{\psi}_{\rho}^{s,\perp}(x_2)+\varphi_{\rho}^{t,\perp}(x_2)\right) 
\left(\bar{\psi}_{\rho}^{s,\perp}(x_3)+\varphi_{\rho}^{t,\perp}(x_3)\right) \non
&~& - 2x_1\left(\bar{\psi}_{\rho}^{s,\perp}(x_2)\bar{\psi}_{\rho}^{s,\perp}(x_3)+\varphi_{\rho}^{t,\perp}(x_2)\varphi_{\rho}^{t,\perp}(x_3)\right)\big] \Big] \Big\} \,,
\label{eq:MP-amplitudes-B2VV_N} 
\eeq
\beq
{\cal M}^{({\rm mp}), {\bf T}}_{B\to V_2V_3} &=&
-\frac{8}{\sqrt{6}} C_F^2  m_B^6 \int_0^1 dx_1dx_2 \, dx_3 \, \int_0^{1/\Lambda} b_1 db_1 b_2 db_2 \, b_3 db_3 \varphi_B(x_1,b_1)\, \non
&\cdot& \Big\{h_g(\alpha_g,\beta_g,\gamma_g,b_1,b_2,b_3) E_{8g}(\mu_g) \non
&~& \Big[r_2 r_3\Big(-[ 2\bar{x}_2\bar{x}_3+r_b\left(x_2-\bar{x}_3\right) \big]  
\big[ \bar{\psi}_{\rho}^{s,\perp}(x_2)\bar{\psi}_{\rho}^{s,\perp}(x_3) + \varphi_{\rho}^{t,\perp}(x_2)\varphi_{\rho}^{t,\perp}(x_3) \big] \non 
&~& + \big[ 2\left(1+x_2 \right)\bar{x}_3-r_b\left(x_2+\bar{x}_3\right)\big] 
\big[ \bar{\psi}_{\rho}^{s,\perp}(x_2)\varphi_{\rho}^{t,\perp}(x_3)+\varphi_{\rho}^{t,\perp}(x_2)\bar{\psi}_{\rho}^{s,\perp}(x_3)\big] \Big) \non 
&~& + \left(\bar{x}_3-2r_b\right) \big[ r_3\bar{x}_3\varphi_{\rho}^{\perp}(x_2)\left(\bar{\psi}_{\rho}^{s,\perp}(x_3)-\varphi_{\rho}^{t,\perp}(x_3)\right) \non
&~& - r_2x_2\varphi_{\rho}^{\perp}(x_3)\left(\bar{\psi}_{\rho}^{s,\perp}(x_2)+\varphi_{\rho}^{t,\perp}(x_2)\right) \big] \Big] \non
&~& + h_g(\alpha_{g^\prime},\beta_{g^\prime},\gamma_{g^\prime},b_1,b_2,b_3) E_{8g^{\prime}}(\mu_{g^{\prime}})  \, r_3 
\Big[ x_1\varphi_{\rho}^{\perp}(x_2)\left(\bar{\psi}_{\rho}^{s,\perp}(x_3)-\varphi_{\rho}^{t,\perp}(x_3)\right) \non 
&~& + r_2\big[ x_2\left(\bar{\psi}_{\rho}^{s,\perp}(x_2)+\varphi_{\rho}^{t,\perp}(x_2)\right) 
\left(\bar{\psi}_{\rho}^{s,\perp}(x_3)+\varphi_{\rho}^{t,\perp}(x_3)\right) \non
&~& - 2x_1\left(\bar{\psi}_{\rho}^{s,\perp}(x_2)\varphi_{\rho}^{t,\perp}(x_3)+\varphi_{\rho}^{t,\perp}(x_2)\bar{\psi}_{\rho}^{s,\perp}(x_3)\right)\big] \Big] \Big\} \,,
\label{eq:MP-amplitudes-B2VV_T} \\
{\cal M}^{({\rm mp})}_{B\to V_2P_3} &\stackrel{R}\Longleftarrow&  {\cal M}^{({\rm mp})}_{B\to P_2P_3} \,,
\label{eq:MP-amplitudes-B2VP}    \quad 
{\cal M}^{({\rm mp})}_{B\to V_2V_3}  \stackrel{R}\Longleftarrow  {\cal M}^{({\rm mp})}_{B\to P_2V_3} \,.
\label{eq:MP-amplitudes-B2VV_L} 
\eeq
The hard functions in these decay amplitudes are also collected in the appendix \ref{app-hardfunction}.

\subsection{The \texorpdfstring{$B \to \pi$}{} transition form factor at NLO}\label{PQCD-NLO-B2pi-ff}

The NLO QCD corrections to the naive factorizable diagrams of hadronic B decays are the same as the $B \to \pi$ transition form factors. 
The typical Feynman diagrams shown in the first row of Fig. \ref{fig:PQCD-NLO2} are calculated in the framework of $k_T$ factorization \cite{Li:2012nk,Cheng:2014fwa}.
The resulting NLO correction kernels are of the following form 
\beq
{\rm F_{T2}^{(1)}}(x_i, \zeta_3) &=& \frac{\alpha_s(\mu_f)C_F}{4\pi} \Big[ \frac{21}{4}\ln \frac{\mu^2}{m_B^2} -
\left( \ln \frac{m_B^2}{\zeta_1^2} + \frac{13}{2} \right) \ln \frac{\mu_f^2}{m_B^2} + \frac{7}{16} \ln^2(x_1x_3) \non
&~& + \frac{1}{8} \ln^2 x_1 + \frac{1}{4} \ln x_1 \ln x_3 + \left( 2 \ln \frac{m_B^2}{\zeta_1^2} - \frac{1}{4} \right) \ln x_1 - \frac{3}{2} \ln x_3 \non
&~&  - \left( \frac{3}{2} \ln \frac{m_B^2}{\zeta_1^2} + 1 \right) \ln \frac{m_B^2}{\zeta_1^2} 
+ \frac{101}{48} \pi^2 + \frac{219}{16} \Big] \,,
\label{eq:B2pi-ff2-NLO}\\
{\rm F_{T3}^{(1)}}(x_i, \zeta_3 )&=& \frac{\alpha_s(\mu_f)C_F}{4\pi} \Big[ \frac{21}{4}\ln \frac{\mu^{\prime 2}}{m_B^2} -
\left( \frac{1}{2} \ln \frac{m_B^2}{\zeta_1^2} + 3 \right) \ln \frac{\mu_f^2}{m_B^2} + \frac{7}{16} \ln^2 x_1 \non 
&~&  - \frac{3}{8} \ln^2 x_3 + \frac{9}{8} \ln x_1 \ln x_3 + \left( \ln \frac{m_B^2}{\zeta_1^2} - \frac{29}{8} \right) \ln x_1 +  \ln \frac{m_B^2}{\zeta_3^2} \non
&~& + \left( \ln \frac{m_B^2}{\zeta_3^2} - \frac{25}{16} \right) \ln x_3 - \left( \frac{1}{4} \ln \frac{m_B^2}{\zeta_1^2} 
- \frac{1}{2} \right) \ln \frac{m_B^2}{\zeta_1^2} + \frac{37}{32} \pi^2 + \frac{91}{32} \Big] \, .
\label{eq:B2pi-ff1-NLO}
\eeq

To evade the light-cone singularities in the above calculations, we introduce the dimensional parameters
$\zeta_1^2 \equiv 4 (n_1 \cdot p_1)^2/\vert n_1^2 \vert$ and $\zeta_3^2 \equiv 4 (n_3 \cdot p_3)^2/\vert n_3^2 \vert$
by varying the Wilson links in the transversal momentum dependent definition of mesons away from the light-cone ($n_1^2, n_3^2  \neq 0$).
These additional scales bring another scheme dependence knowing as the rapidity singularities,
which in principle should be resummed to all orders by resolving the evolution equation of $B$ and $\pi$ meson wave functions
on the dimensionless scales $\zeta_1^2/m_B^2$ and $\zeta_3^2/m_B^2$, respectively.
The solutions demonstrate that the resummation effect suppresses the shape of B meson LCDA $\varphi_+(k^+, b, \mu)$ near the end point $k^+ \to 0$ \cite{Li:2012md},
and the dependence on the choice of $\zeta_3$ is not significant in pion meson wave function since
the joint resummation-improved pion wave function does not show a sizable corrections \cite{Li:2013xna}.
This is not accidental because the $k_T$ and threshold resummations have reshaped the end-point behaviours of meson wave functions already \cite{Cheng:2020fcx}.
For the sake of simplicity, we can take $\zeta_3^2 = m_B^2$ to eliminate the logarithm term $\ln (\zeta_3^2/m_B^2)$, 
and take a large value of $\zeta_1^2$ at $25^2 m_B^2$ to reduce the uncertainty of the constant terms from the loop calculation by different momentum regions.

The factorization scale in the NLO kernels is chosen at the hard scale in the corresponding decay amplitudes,
and the choice of renormalization scale is purposed to eliminate all the single logarithm,
with the same argument to reduce the uncertainties as much as possible.
\beq
\mu \equiv \mu(\mu_f) &=& \Big\{ {\rm Exp} \Big[ c_1 + \left( \ln \frac{m_B^2}{\zeta_1^2} + \frac{5}{4} \right) \ln \frac{\mu_f^2}{m_B^2} \Big]
x_1^{c_2} x_2^{c_3}\Big\}^{2/21} \mu_f \,, \label{eq:B2pi-ff1-NLO-scale} \\
\mu^\prime \equiv \mu^\prime(\mu_f) &=& \Big\{ {\rm Exp} \Big[ c^\prime_1 + \left( \frac{1}{2}\ln \frac{m_B^2}{\zeta_1^2} - \frac{9}{4} \right)
\ln \frac{\mu_f^2}{m_B^2} \Big] x_1^{c_2^\prime} x_2^{c_3^\prime}\Big\}^{2/21} \mu_f \,.
\label{eq:B2pi-ff2-NLO-scale}
\eeq
The parameters entered in Eqs.(\ref{eq:B2pi-ff1-NLO-scale},\ref{eq:B2pi-ff2-NLO-scale}) are
\beq
c_1 &=& \left( \frac{3}{2} \ln \frac{m_B^2}{\zeta_1^2} + 1 \right) \ln \frac{m_B^2}{\zeta_1^2}
- \frac{101}{48} \pi^2 - \frac{219}{16} \,, \quad
c_2 = - \left( 2 \ln \frac{m_B^2}{\zeta_1^2} - \frac{1}{4} \right) \,, \quad c_3 = + \frac{3}{2} \,, \non
c_1^\prime &=& \left( \frac{1}{4} \ln \frac{m_B^2}{\zeta_1^2} - \frac{1}{2} \right) \ln \frac{m_B^2}{\zeta_1^2} - \frac{37}{32} \pi^2 - \frac{91}{32} \,, \quad 
c_2^\prime = - \left( \ln \frac{m_B^2}{\zeta_1^2} - \frac{29}{8} \right) \,, \quad c_3^\prime = + \frac{25}{16} \,.
\label{eq:B2pi-ff2-NLO-scale-para}
\eeq
branching ratios
We mark that the kernels ${\rm F_{T2}^{(1)}}$ and ${\rm F_{T3}^{(1)}}$ are the corrections to the 
decay amplitude of naive factorizable emission diagrams in $B \to PP$ decays as shown in Eq. (\ref{eq:FE-B2PP-LL}), 
which attached to the terms proportional to leading twist $\pi$ meson LCDA $\varphi_\pi$ and the twist-3 LCDA $\varphi_\pi^p$, respectively. 
The NLO expression is modified to 
\beq
\mathcal{E}_{M_3}^{\bf{LL}, {\rm NLO}} &=&  -\mathcal{E}_{M_3}^{\bf{LR}, {\rm NLO}}=    
8 \pi C_F m_B^4 f_{M_2} \int_0^1 dx_1 \, dx_3 \, \int_0^{1/\Lambda} b_1 db_1 \, b_3 db_3 \varphi_B(x_1,b_1)\, \non
&\cdot&  \Big\{ h_{e}(x_1,x_3,b_1,b_3) E_e(\mu_e) 
\Big[ (2\,r_b-\bar{x}_3) \left[ 1 + {\rm F_{T2}^{(1)}}(x_i, \zeta_3) \right] \varphi_{\pi}^a(x_3) \non 
&~& - r_3\left(r_b-2\bar{x}_3 \right)\left(\varphi_{\pi}^p(x_3)+\varphi_{\pi}^t(x_3)\right)  \Big]   \non
&~& + h_{e}(x_3,x_1,b_3,b_1) E_{e^{\prime}}(\mu_{e^{\prime}}) \,2\,r_3 \left[ 1 + {\rm F_{T3}^{(1)}}(x_i, \zeta_3) \right] \varphi_{\pi}^p(x_3) \Big\} \,.
\label{eq:FE-B2PP-LL-NLO}
\eeq
Several comments are given in order as following,
\begin{enumerate}
\item[(1)]
The ultraviolet divergence at NLO is treated by the traditional renormalization group technique in the $\overline{\rm{MS}}$ scheme.
It is regulated by the counter terms of quark fields and strong coupling constant.
\item[(2)]
The infrared divergences are regulated by the transversal momenta of light quarks and also the soft gluon mass,
where the later one is introduced to protect $b$ quark on shell required by the standard effective heavy field in the $k_T$ factorization theorem.
The cancellation of infrared divergences happens between the QCD quark diagrams and the effective diagrams of meson wave function, 
up to sub-leading twist LCDAs of pion.
The retaining infrared finite NLO hard kernel still have large logarithms, 
in which the double logarithms in terms of $\ln^2 (x_i)$ is resummed to the threshold Sudakov exponent and the single logarithm $\ln(x_i)$
can be diminished by the choice of renormalization scale to the physical mass in the process.
\item[(3)]
The infrared finite NLO hard kernel is $k_T$ dependent because it is obtained by taking the difference of NLO quark amplitude
and the convolution of NLO wave functions with LO hard kernel.
Meanwhile, the small $k_T$ contribution is highly suppressed by the $k_T$ resummed wave functions.
\item[(4)]
From the phenomenological side, the $\sim 30\%$ enhancement from NLO correction with leading twist distribution amplitude of pion meson
is partly cancelled by the $\sim 20\%$ decrease of NLO correction with sub-leading twist distribution amplitude.
Therefore it results in about $8\%$ NLO correction to the leading order contributions of $B \to \pi$ transition form factors.
\end{enumerate}
In fact, only the NLO correction for the $B \to \pi$ transition form factors induced by vector current is finished. 
The NLO calculations for the $B \to \pi$ transition induced by scalar quark current and those for $B \to \rho$ transition are still absent.  
A lot of efforts are clearly required.

\subsection{The timelike pion form factor at NLO}\label{PQCD-NLO-timelike-pi-ff}

Concerning the color suppressed channels with the $W$-exchange diagram contribution, like $B^0 \to \pi^0\pi^0, \pi^0 \eta_q, \eta_q\eta_q$,
the annihilation type amplitudes provide not only the main source of large ${\bf CP}$ asymmetry, but also an indispensable contribution to the total decay amplitude.
The annihilation amplitudes are more important than the emission ones when the strong destructive interaction
between the color suppressed tree amplitude ${\rm C}$ and the penguin amplitude ${\rm P}$  become true.
The NLO QCD corrections to the naive factorizable annihilation amplitudes, 
have been calculated to improve the PQCD precision for those relevant B meson decays \cite{Hu:2012cp,Cheng:2015qra,Cheng:2014rka,Zhang:2015mxa,Hua:2018kho}.
The most important contribution of this type NLO correction is the timelike form factors as depicted in the second row of figure \ref{fig:PQCD-NLO2}, 
and the NLO hard kernels are quoted as \cite{Hu:2012cp,Cheng:2015qra}
\beq
{\rm G}_{\pi, {\rm T2}}^{(1)} &=&  \frac{\alpha_s(\mu) C_F}{4 \pi} \Big[ -\frac{3}{4} \ln \frac{\mu^2}{m_B^2}
- \frac{1}{4} \ln^2 \left(\frac{4x_2x_3}{m_B^2b_2^2}\right) - \frac{17}{4} \ln^2 x_3  + \frac{27}{8} \ln x_2 \ln x_3  \non
&~& + \left( \frac{17}{8} \ln x_3 + \frac{23}{16} + \gamma_E + i \frac{\pi}{2} \right) \ln \left( \frac{4x_2x_3}{m_B^2b_2^2} \right)
- \left( \frac{13}{8} + \frac{17}{4} \gamma_E - i \frac{17}{8} \pi \right) \ln x_3  \non
&~& + \frac{31}{16} \ln x_2 - \frac{\pi^2}{2} + (1 - 2 \gamma_E) \pi + \frac{\ln 2}{2} + \frac{53}{4} - \frac{23 \gamma_E}{8}
- \gamma_E^2 + i \pi \left( \frac{171}{16} + \gamma_E \right) \Big] \,,
\label{eq:pi-ff-tl-em-nlo-t2}\\
{\rm G}_{\pi, {\rm T3, 1}}^{(1)} &=& \frac{\alpha_s(\mu) C_F}{4 \pi} \Big[ \frac{9}{4} \ln \frac{\mu^2}{m_B^2}
- \frac{53}{32} \ln \left(\frac{4x_2x_3}{m_B^2b_2^2} \right) - \frac{23}{32} \ln \left( \frac{4x_3}{m_B^2b_3^2}\right)
- \frac{1}{8} \ln^2 x_2 - \frac{9}{8} \ln x_2 \non
&~& - \frac{137}{96} \pi^2 + \frac{19}{4} \gamma_E + \frac{337}{64} + i \pi \frac{39}{8}  \Big] \,,
\label{eq:pi-ff-tl-em-nlo-t3-1}\\
{\rm G}_{\pi, {\rm T3, 2}}^{(1)} &=&  \frac{\alpha_s(\mu) C_F}{4 \pi} \Big[ \frac{9}{4} \ln \frac{\mu^2}{m_B^2}
- 2 \ln \left(\frac{4x_2x_3}{m_B^2b_2^2} \right) - \frac{1}{8} \ln^2 \left( \frac{4x_3}{m_B^2b_3^2}\right)
+ \left( \frac{\gamma_E}{2} + \frac{3}{4} i \pi \right) \ln \left( \frac{4x_3}{m_B^2b_3^2}\right) \non
&~& + 2\ln x_2 - \frac{\pi^2}{4} - \frac{\gamma_E^2}{2} + 4 \gamma_E + \frac{\ln 2}{4} + \frac{11}{2}
+ i \pi \left( \frac{15}{4} - \frac{3}{2} \gamma_E \right)\Big]\,.
\label{eq:pi-ff-tl-em-nlo-t3-2}
\eeq
The $B \to PP$ decay amplitudes, with ${\rm (V-A) \otimes (V \pm A)}$ current in naive factorizable annihilation diagrams at NLO accuracy, is then modified to 
\beq
\mathcal{A}^{\bf{LL}, {\rm NLO}}_{M_3}&=&\mathcal{A}^{\bf{LR}, {\rm NLO}}_{M_3}=
8 \pi C_F m_B^4 f_B \int_0^1 \,dx_2 \, dx_3 \, \int_0^{1/ \Lambda} b_2 db_2 \, b_3 db_3 \non
&\cdot&  \Big\{ h_{a}(x_2,{\bar x}_3,b_2,b_3)E_a({\mu_a}) \Big[
-\bar{x}_3 \left( 1 + {\rm G}_{\pi, {\rm T2}}^{(1)} \right)  \varphi_{\pi}^a(x_2)\varphi_{\pi}^a(x_3) + 2r_2r_3 \varphi_{\pi}^p(x_2) \non 
&~& \cdot \big[ - \left( 1 + {\rm G}_{\pi, {\rm T3, 2}}^{(1)}  \right) \left( \varphi_{\pi}^p(x_3) + \varphi_{\pi}^t(x_3) \right) 
+ {\bar x}_3 \left( 1 + {\rm G}_{\pi, {\rm T3, 1}}^{(1)}  \right) \left( \varphi_{\pi}^t(x_3) - \varphi_{\pi}^p(x_3) \right) \big] \Big] \non
&~& + h_{a}({\bar x}_3,x_2,b_3,b_2) E_{a^\prime}({\mu_{a^\prime}}) 
\Big[ x_2 \left( 1 + {\rm G}_{\pi, {\rm T2}}^{(1)} \right) \varphi_{\pi}^a(x_2)\varphi_{\pi}^a(x_3) + 2r_2r_3 \varphi_{\pi}^p(x_3)  \non 
&~& \cdot \big[\left( 1 + {\rm G}_{\pi, {\rm T3, 1}}^{(1)}  \right) \left( \varphi_{\pi}^p(x_2) - \varphi_{\pi}^t(x_2) \right) 
+ x_2 \left( 1 + {\rm G}_{\pi, {\rm T3, 2}}^{(1)}  \right) \left( \varphi_{\pi}^p(x_2) + \varphi_{\pi}^t(x_2) \right) \big] \Big]\Big\} \,. 
\label{eq:FA-B2PP-LL_NLO}
\eeq
This correction does not bring contribution to the decaying channels with two identical mesons in the final states.   
While its contribution to the channels $B^0 \to \pi^0 \eta, \pi^0 \eta^\prime$ could be expected around $30 \%$ with the generally recognition of the $SU(3)$ flavour breaking.

For the channel $B^0 \to \pi^0 \pi^0$, ${\cal A}^{{\bf LL}} = {\cal A}^{{\bf LR}} = 0$, 
and hence the naive factorizable annihilation is only stemmed from the ${\rm (S+P) \otimes (S-P)}$ current 
with the amplitude ${\cal A}^{\bf SP}$ as given in Eq. (\ref{eq:FA-B2PP-SP}). 
With considering the NLO correction, this piece of decay amplitude is modified to 
\beq
\mathcal{A}^{\bf{SP}}_{M_3}&=&
16 \pi C_F m_B^4 f_B \int_0^1 \,dx_2 \, dx_3 \, \int_0^{1/\Lambda} b_2 db_2 \, b_3 db_3 \non
&\cdot&  \Big\{ h_{a}(x_2,{\bar x}_3,b_2,b_3)E_a({\mu_a}) \Big[ r_3\bar{x}_3\varphi_{\pi}^a(x_2) \left( \varphi_{\pi}^p(x_3)+\varphi_{\pi}^t(x_3)\right) \non
&~& + 2 r_2 \left( 1 + \Gamma^{(1)}(Q^2)\right) \varphi_{\pi}^p(x_2)\varphi_{\pi}^a(x_3) \Big] \non
&~& + h_{a}({\bar x}_3,x_2,b_3,b_2)E_{a^\prime}({\mu_{a^\prime}}) \Big[ r_2 x_2\varphi_{\pi}^a(x_3)\left(\varphi_{\pi}^p(x_2)-\varphi_{\pi}^t(x_2)\right) \non
&~& + 2 r_3 \left( 1 + \Gamma^{(1)}(Q^2)\right) \varphi_{\pi}^a(x_2)\varphi_{\pi}^p(x_2) \Big]\Big\} \,
\label{eq:FA-B2PP-SP}
\eeq
with the correction kernel function \cite{Cheng:2014rka}
\beq
\Gamma^{(1)} &=& \frac{\alpha_s C_F}{8\pi} \Big\{ \frac{5}{2} \ln \frac{\mu^2}{m_B^2} - \frac{1}{16} \ln^2 \left( \frac{4 x_2 (1-x_3)}{m_B^2 b_3^2} \right)
+ \left( \frac{33}{8} - \frac{\gamma_E}{4} \right) \ln \left( \frac{4 x_2  (1-x_3)}{m_B^2 b_3^2} \right) \non
&~& - \ln^2  (1-x_3)  - \frac{3}{8} \ln^2 x_2  + \frac{1}{2} \ln x_2 \ln  (1-x_3) - \frac{71}{4} \ln  (1-x_3) - \frac{7}{4} \ln x_2 \non
&~& - \frac{105}{48} \pi^2 + \frac{41}{2}
- \frac{\gamma_E^2}{2} - \frac{33 \gamma_E}{4} + i \pi \frac{5}{2} \Big\} \,.
\label{eq:tl-ff-pi-S-nlo}
\eeq
For the timelike scalar pion form factor, 
the positive real part and the negative imaginary part at LO play the dominant role in producing the large ${\bf CP}$ asymmetry in $B^0 \to \pi^0\pi^0$ decay.
The numerical results show that the NLO contribution is tiny in both of the magnitude and the phase, 
which only slightly modify the LO PQCD predictions for the $B \to \pi\pi$ decays.

The timelike $\rho\pi$ electromagnetic form factor is also studied at NLO in the framework of $k_T$ factorization \cite{Hua:2018kho}. 
In the $B \to \rho\pi$ decays, it is found that this contribution proportional to timelike electromagnetic form factor is power suppressed even at LO, 
which can be neglected in the numerical calculations.
As for the leading power contribution to the annihilation amplitude comes from the timelike $\rho\pi$ transition form factor
induced by the axial-vector current and pseudoscalar density operators, whose NLO correction is still unknown at present.

\subsection{Decay amplitudes for different channels at NLO}\label{subsec:decayamp}

We have presented the various NLO contributions for the two body $B$ meson decays. 
These corrections together with the leading order perturbation calculations are channel dependent, 
which are shown as the different effective four quark operators involved in different $B$ meson decay channels.
With the inclusion of various kinds of the NLO corrections presented above,
the total decay amplitudes for different $B \to PP$ channels are presented as following.
\beq
{\cal M}(B^+ & \to & \pi^+ K^0) = 
\frac{G_F}{\sqrt{2}} V^{\ast}_{ub} V_{us} \Big[ a_1 {\cal A}_{\pi}^{{\bf LL}} + C_1 {\cal A}_{NF, \pi}^{{\bf LL}}
+{\cal M}^{({\rm ql,u})}_{B \to K^0 \pi^+}  \Big]+\frac{G_F}{\sqrt{2}} V^{\ast}_{cb} V_{cs}{\cal M}^{({\rm ql,c})}_{B \to K^0 \pi^+}  \non   
&~& - \frac{G_F}{\sqrt{2}} \, V^{\ast}_{tb} V_{ts} \Big[ \left( a_4-\frac{a_{10}}{2} \right) {\cal E}_{\pi}^{{\bf LL}} 
+ \left( a_6 - \frac{a_{8}}{2} \right) {\cal E}_{\pi}^{{\bf SP}}
+ \left( C_3 - \frac{C _{9}}{2} \right) {\cal E}_{NF, \pi}^{{\bf LL}}  \non
&~& + \left( C_5 - \frac{C_{7}}{2} \right) {\cal E}_{NF, \pi}^{{\bf LR}} + \left( a_4 + a_{10} \right) {\cal A}_{\pi}^{{\bf LL}} 
+ \left( a_6 + a_{8} \right) {\cal A}_{\pi}^{{\bf SP}} + \left( C_3 + C_{9} \right) {\cal A}_{NF, \pi}^{{\bf LL}}  \non
&~& + \left( C_5 + C_{7} \right) {\cal A}_{NF, \pi}^{{\bf LR}} + {\cal M}^{({\rm ql,t })}_{B \to K^0 \pi^+}+ {\cal M}^{({\rm mp})}_{B \to K^0 \pi^+}   \Big] \,, 
\label{eq:pi+K0}\\
\sqrt{2}{\cal M}(B^+ &\to& \pi^0 K^+)  =  \frac{G_F}{\sqrt{2}} V^{\ast}_{ub }V_{us}
\Big[ a_1 \left( {\cal E}_{\pi}^{{\bf LL}} + {\cal A}_{\pi}^{{\bf LL}} \right) + a_2 {\cal E}_{K}^{{\bf LL}}
+ C_1 \left( {\cal E}_{NF \pi}^{{\bf LL}} + {\cal A}_{NF, \pi}^{{\bf LL}} \right) + C_2 {\cal E}_{NF, K}^{{\bf LL}}  \non
&~& +{\cal M}^{({\rm ql,u})}_{B \to K^+ \pi^0} \Big] + \frac{G_F}{\sqrt{2}} V^{\ast}_{cb }V_{cs} {\cal M}^{({\rm ql,c})}_{B \to K^+ \pi^0}  \non
&~& - \frac{G_F}{\sqrt{2}} V^{\ast}_{tb}V_{ts} \Big[ \left( a_4+a_{10} \right) {\cal E}_{\pi}^{{\bf LL}} + \left( a_6 + a_{8} \right) {\cal E}_{\pi}^{{\bf SP}}
+\frac{3a_{9}}{2} {\cal E}_{K}^{{\bf LL}} + \frac{3a_{7}}{2} {\cal E}_{K}^{{\bf LR}} \, \non
&~& + \frac{3C_{10}}{2} {\cal E}_{NF, K}^{{\bf LL}} + \frac{3C_{8}}{2} {\cal E}_{NF, K}^{{\bf SP}}
+ \left( C_3 + C_{9} \right) {\cal E}_{NF, \pi}^{{\bf LL}} + \left( C_5 + C_{7} \right) {\cal E}_{NF, \pi}^{{\bf LR}} \non
&~& + \left( a_4 + a_{10} \right) {\cal A}_{\pi}^{{\bf LL}} + \left( a_6 + a_{8} \right) {\cal A}_{\pi}^{{\bf SP}}
+ \left( C_3 + C_{9} \right) {\cal A}_{NF, \pi}^{{\bf LL}} + \left( C_5 + C_{7} \right) {\cal A}_{NF, \pi}^{{\bf LR}} \non &~&+ {\cal M}^{({\rm ql,t})}_{B \to K^+ \pi^0} + {\cal M}^{({\rm mp})}_{B \to K^+ \pi^0}  \Big] \,,
\label{eq:pi0K+}  \\
{\cal M}(B^+ &\to& \eta_s K^+)  =  \frac{G_F}{\sqrt{2}} \, V^{\ast}_{ub} V_{us} \Big[ a_1 {\cal A}_{K}^{\bf{LL}} + C_1 {\cal A}_{NF, K}^{\bf{LL}} + {\cal M}^{({\rm ql,u})}_{B \to \eta_s K^+} \Big] + \frac{G_F}{\sqrt{2}} \, V^{\ast}_{cb} V_{cs}  {\cal M}^{({\rm ql,c})}_{B \to \eta_s K^+}             \non
&~& - \frac{G_F}{\sqrt{2}} \, V^{\ast}_{tb} V_{ts} \Big[  \left( a_3 + a_4-\frac{a_{9}+a_{10}}{2} \right) {\cal E}_{K}^{\bf{LL}}
+ \left( a_{5}-\frac{a_{7}}{2} \right) {\cal E}_{K}^{\bf{LR}} + \left( a_6 - \frac{a_{8}}{2} \right) {\cal E}_{K}^{\bf{SP}} \non
&~& + \left( C_3 + C_4 - \frac{C_{9}+C_{10}}{2} \right) {\cal E}_{NF, K}^{\bf{LL}}
+ \left( C_5 - \frac{C_{7}}{2} \right) {\cal E}_{NF, K}^{\bf{LR}} + \left( C_6 - \frac{C_{8}}{2} \right) {\cal E}_{NF, K}^{\bf{SP}} \non
&~& + \left( a_4 + a_{10} \right) {\cal A}_{K}^{\bf{LL}} + \left( a_6 + a_{8} \right) {\cal A}_{K}^{\bf{SP}}
+ \left( C_3 + C_{9} \right) {\cal A}_{NF, K}^{\bf{LL}} + \left( C_5 + C_{7} \right) {\cal A}_{NF, K}^{\bf{LR}} \non  &~& +{\cal M}^{({\rm ql,t})}_{B \to \eta_s K^+} +{\cal M}^{({\rm mp})}_{B \to \eta_s K^+} \Big]    \,,
\label{eq:etasK+}
\eeq
\beq
\sqrt{2}{\cal M}(B^+ &\to& \eta_q K^+) =  \frac{G_F}{\sqrt{2}} V^{\ast}_{ub} V_{us}
\Big[ a_1 \left( {\cal E}_{\eta_q}^{\bf{LL}} + {\cal A}_{\eta_q}^{\bf{LL}} \right) + a_2 {\cal E}_{K}^{\bf{LL}}
+ C_1 \left( {\cal E}_{NF, \eta_q}^{\bf{LL}} + {\cal A}_{NF, \eta_q}^{\bf{LL}} \right) + C_2 {\cal E}_{NF, K}^{\bf{LL}}  \non &~&+ {\cal M}^{({\rm ql,u})}_{B \to K^+ \eta_q }  \Big]  + \frac{G_F}{\sqrt{2}} V^{\ast}_{cb} V_{cs} {\cal M}^{({\rm ql,c})}_{B \to K^+ \eta_q } \non
&~& - \frac{G_F}{\sqrt{2}} \, V^{\ast}_{tb}\, V_{ts} \Big[ \left( a_4 + a_{10} \right) {\cal E}_{\eta_q}^{\bf{LL}} + \left( a_6 + a_{8} \right) {\cal E}_{\eta_q}^{\bf{SP}}
+ \left( C_3 + C_{9} \right) {\cal E}_{NF, \eta_q}^{\bf{LL}} + \left( C_5 + C_{7} \right) {\cal E}_{NF, \eta_q}^{\bf{LR}} \non
&~& + \left( 2 a_{3} + \frac{a_{9}}{2} \right) {\cal E}_{K}^{\bf{LL}} + \left( 2a_{5} +\frac{a_{7}}{2} \right) {\cal E}_{K}^{\bf{LR}}
+ \left( 2C_4 + \frac{C_{10}}{2} \right) {\cal E}_{NF, K}^{\bf{LL}} + \left( 2C_6 + \frac{C_{8}}{2} \right) {\cal E}_{NF, K}^{\bf{SP}} \non
&~& + \left( a_4 + a_{10} \right) {\cal A}_{\eta_q}^{\bf{LL}} + \left( a_6 + a_{8} \right) {\cal A}_{\eta_q}^{\bf{SP}}
+ \left( C_3 + C_{9} \right) {\cal A}_{NF, \eta_q}^{\bf{LL}} + \left( C_5 + C_{7} \right) {\cal A}_{NF,\eta_q}^{\bf{LR}} \non
&~& + {\cal M}^{({\rm ql,t})}_{B \to K^+ \eta_q } + {\cal M}^{({\rm mp})}_{B \to K^+ \eta_q }  \Big]  \,,
\label{eq:etaqK+}   \\
{\cal M}(B^+ &\to& K^+ \bar K^0)  =  \frac{G_F}{\sqrt{2}} V^{\ast}_{ub} V_{ud} \Big[ a_1 {\cal A}_{K^+}^{\bf{LL}} + C_1 {\cal A}_{NF, K^+}^{\bf{LL}} + {\cal M}^{({\rm ql,u})}_{B \to \bar K^0 K^+ }  \Big] + \frac{G_F}{\sqrt{2}} V^{\ast}_{cb} V_{cd} {\cal M}^{({\rm ql,c})}_{B \to \bar K^0 K^+ } \non
&~& - \frac{G_F}{\sqrt{2}} \, V^{\ast}_{tb} V_{td} \Big[ \left( a_4 - \frac{a_{10}}{2} \right) {\cal E}_{K^+}^{\bf{LL}}
+ \left( a_6 - \frac{a_{8}}{2} \right) {\cal E}_{K^+}^{\bf{SP}} + \left( C_3 - \frac{C_{9}}{2} \right) {\cal E}_{NF, K^+}^{\bf{LL}} \non
&~& + \left( C_5 - \frac{C_{7}}{2} \right) {\cal E}_{NF, K^+}^{\bf{LR}} + \left( a_4 + a_{10} \right) {\cal A}_{K^+}^{\bf{LL}} 
+ \left( a_6 + a_{8} \right) {\cal A}_{K^+}^{\bf{SP}} + \left( C_3 + C_{9} \right) {\cal A}_{NF, K^+}^{\bf{LL}}  \non
&~& + \left( C_5 + C_{7} \right) {\cal A}_{NF, K^+}^{\bf{LR}} + {\cal M}^{({\rm ql,t})}_{B \to \bar K^0 K^+ } +{\cal M}^{({\rm mp})}_{B \to \bar K^0 K^+ } \Big]  \,,
\label{eq:K+K0}\\
\sqrt{2}{\cal M}(B^+ &\to& \pi^0 \pi^+ )  =  \frac{G_F}{\sqrt{2}} \, V^{\ast}_{ub} V_{ud}
\Big[ a_2   {\cal E}_{\pi^+}^{\bf{LL}} +  C_2   {\cal E}_{NF, \pi^+}^{\bf{LL}}
+ a_1   {\cal E}_{\pi^0}^{\bf{LL}} +  C_1   {\cal E}_{NF, \pi^0}^{\bf{LL}} \non 
&~& +  {\cal M}^{({\rm ql,u})}_{B \to  \pi^+  \pi^0}  - {\cal M}^{({\rm ql,u})}_{B \to \pi^0\pi^+ }  \Big] 
 +  \frac{G_F}{\sqrt{2}} \, V^{\ast}_{cb} V_{cd} ( {\cal M}^{({\rm ql,c})}_{B \to  \pi^+  \pi^0}  - {\cal M}^{({\rm ql,c})}_{B \to \pi^0\pi^+ }  ) \non
&~& - \frac{G_F}{\sqrt{2}} \, V^{\ast}_{tb} V_{td} \Big[   \left( \frac{3a_{9}+a_{10}}{2} -a_4 \right) {\cal E}_{\pi^+}^{\bf{LL}}
+ \frac{3a_7}{2} {\cal E}_{\pi^+}^{\bf{LR}} +(\frac{a_8}{2}-a_6)  {\cal E}_{\pi^+}^{\bf{SP}} \non
&~& + \left(\frac{ (3 C_{10} + C_9 )}{2}-C_3\right) {\cal E}_{NF, \pi^+}^{\bf{LL}} + \left(\frac{C_7}{2}-C_5\right) {\cal E}_{NF, \pi^+}^{\bf{LR}}
+\frac{3C_8}{2} {\cal E}_{NF, \pi^+}^{\bf{SP}} \non
&~& + (a_4+a_{10}){\cal E}_{\pi^0}^{\bf{LL}} + (a_6+a_8){\cal E}_{\pi^0}^{\bf{SP}} +(C_3+C_9) {\cal E}_{NF, \pi^0}^{\bf{LL}}
+ (C_5+C_7) {\cal E}_{NF, \pi^0}^{\bf{LR}} \non
&~& + {\cal M}^{({\rm ql,t})}_{B \to  \pi^+  \pi^0}  - {\cal M}^{({\rm ql,t})}_{B \to \pi^0\pi^+ }+ {\cal M}^{({\rm mp})}_{B \to  \pi^+  \pi^0} 
- {\cal M}^{({\rm mp})}_{B \to \pi^0\pi^+ } \Big]   \,, 
\label{eq:pi+pi0}\\   
{\cal M}(B^+ &\to&  \pi^+ \eta_s) = -\frac{G_F}{\sqrt{2}} \, V^{\ast}_{tb} V_{td}
\Big[ \left( a_3-\frac{a_{9}}{2} \right) {\cal E}_{\pi}^{\bf{LL}} + \left(a_5-\frac{a_{7}}{2} \right) {\cal E}_{\pi}^{\bf{LR}} \non
&~& + \left( C_4-\frac{C_{10}}{2} \right) {\cal E}_{NF, \pi}^{\bf{LL}} +
\left( C_6-\frac{C_{8}}{2} \right) {\cal E}_{NF, \pi}^{\bf{SP}} \Big] \,,
\label{eq:pi+etas}
\eeq
\beq
\sqrt{2}{\cal M}(B^+ &\to& \pi^+ \eta_q) =  \frac{G_F}{\sqrt{2}} \, V^{\ast}_{ub} V_{ud}
\Big[ a_1\left( {\cal E}_{\eta_q}^{\bf{LL}} + {\cal A}_{\eta_q}^{\bf{LL}} + {\cal A}_{\pi}^{\bf{LL}} \right)
+ a_2 {\cal E}_{\pi}^{\bf{LL}} + C_2 {\cal E}_{NF, \pi}^{\bf{LL}} \non
&~& + C_1 \left( {\cal E}_{NF, \eta_q}^{\bf{LL}} + {\cal A}_{NF, \eta_q}^{\bf{LL}} + {\cal A}_{NF, \pi}^{\bf{LL}} \right) + {\cal M}^{({\rm ql,u})}_{B \to  \pi^+\eta_q } +{\cal M}^{({\rm ql,u})}_{B \to  \eta_q \pi^+ } \Big] \non
&~& +\frac{G_F}{\sqrt{2}} \, V^{\ast}_{cb} V_{cd} ({\cal M}^{({\rm ql,c})}_{B \to  \pi^+\eta_q } +{\cal M}^{({\rm ql,c})}_{B \to  \eta_q \pi^+ })   \non
&~& - \frac{G_F}{\sqrt{2}} \, V^{\ast}_{tb} V_{td} \Big[ \left( 2a_3 + a_4 + \frac{a_{9}-a_{10} }{2} \right) {\cal E}_{\pi}^{\bf{LL}}
+ \left( 2a_{5}+\frac{a_{7}}{2} \right) {\cal E}_{\pi}^{\bf{LR}} + \left(a_6-\frac{a_{8}}{2} \right) {\cal E}_{\pi}^{\bf{SP}} \non
&~& + \left( C_3+2C_4 - \frac{C_{9}-C_{10}}{2} \right) {\cal E}_{NF, \pi}^{\bf{LL}}
+ \left( C_5-\frac{C_{7}}{2} \right) {\cal E}_{NF, \pi}^{\bf{LR}}
+ \left(2 C_6 + \frac{C_{8}}{2} \right) {\cal E}_{NF, \pi}^{\bf{SP}} \non
&~& + \left( a_4+a_{10} \right) {\cal E}_{\eta_q}^{\bf{LL}} + \left(a_6+a_{8}\right) {\cal E}_{\eta_q}^{\bf{SP}}
+ \left( C_3+C_{9} \right) {\cal E}_{NF, \eta_q}^{\bf{LL}} + \left( C_5+C_{7} \right) {\cal E}_{NF, \eta_q}^{\bf{LR}} \non
&~& + \left( a_4+a_{10} \right) \left({\cal A}_{\pi}^{\bf{LL}} + {\cal A}_{\eta_q}^{\bf{LL}} \right) + \left(a_6+a_{8}\right) \left({\cal A}_{\pi}^{\bf{SP}} + {\cal A}_{\eta_q}^{\bf{SP}} \right)\non
&~& + \left(C_3+C_{9} \right) \left({\cal A}_{NF, \pi}^{\bf{LL}} + {\cal A}_{NF, \eta_q}^{\bf{LL}} \right)
+ \left( C_5+C_{7} \right) \left( {\cal A}_{NF, \pi}^{\bf{LR}} + {\cal A}_{NF, \eta_q}^{\bf{LR}} \right) \non
&~&+{\cal M}^{({\rm ql,t})}_{B \to  \pi^+\eta_q } +{\cal M}^{({\rm ql,t})}_{B \to  \eta_q \pi^+ } 
+  {\cal M}^{({\rm mp})}_{B \to \pi^+\eta_q} +{\cal M}^{({\rm mp})}_{B \to \eta_q \pi^+ } \Big] \,,
\label{eq:pi+eatq}\\
{\cal M}(B^0 &\to& \pi^- K^+) = \frac{G_F}{\sqrt{2}} \, V^{\ast}_{ub} V_{us} \Big[ a_1 {\cal E}_{\pi}^{\bf{LL}} + C_1 {\cal E}_{NF, \pi}^{\bf{LL}} + {\cal M}^{({\rm ql,u})}_{B \to  K^+ \pi^- } \Big]+ \frac{G_F}{\sqrt{2}} \, V^{\ast}_{cb} V_{cs} {\cal M}^{({\rm ql,c})}_{B \to  K^+ \pi^- }  \non
&~& - \frac{G_F}{\sqrt{2}} \, V^{\ast}_{tb} V_{ts} \Big[ \left(a_4+a_{10}\right) {\cal E}_{\pi}^{\bf{LL}} + \left(a_6+a_{8}\right) {\cal E}_{\pi}^{\bf{SP}}
+ \left( C_3+C_{9} \right) {\cal E}_{NF, \pi}^{\bf{LL}} \non
&~& + \left( C_5+C_{7} \right) {\cal E}_{NF, \pi}^{\bf{LR}} + \left(a_4-\frac{a_{10}}{2} \right) {\cal A}_{\pi}^{\bf{LL}} 
+ \left(a_6-\frac{a_{8}}{2}\right) {\cal A}_{\pi}^{\bf{SP}} + \left(C_3-\frac{C_{9}}{2}\right) {\cal A}_{NF, \pi}^{\bf{LL}}  \non
&~& + \left(C_5-\frac{C_{7}}{2} \right) {\cal A}_{NF, \pi}^{\bf{LR}}+ {\cal M}^{({\rm ql,t})}_{B \to  K^+ \pi^- } 
+{\cal M}^{({\rm mp})}_{B \to  K^+ \pi^-} \Big] \,,
\label{eq:pi-K+}\\
\sqrt{2}{\cal M}(B^0 &\to& \pi^0 K^0) = \frac{G_F}{\sqrt{2}} \, V^{\ast}_{ub} V_{us} \Big[ a_2 {\cal E}_{K}^{\bf{LL}} + C_2 {\cal E}_{NF, K}^{\bf{LL}} -{\cal M}^{({\rm ql,u})}_{B \to  K^0 \pi^0 }  \Big] -\frac{G_F}{\sqrt{2}} \, V^{\ast}_{cb} V_{cs} {\cal M}^{({\rm ql,c})}_{B \to  K^0 \pi^0 } \non
&~& - \frac{G_F}{\sqrt{2}} \, V^{\ast}_{tb} V_{ts} \Big[ \left(-a_4+\frac{a_{10}}{2} \right) {\cal E}_{\pi}^{\bf{LL}}
+ \left(-a_6+\frac{a_{8}}{2} \right) {\cal E}_{\pi}^{\bf{SP}} +\frac{3a_{9}}{2} {\cal E}_{K}^{\bf{LL}} + \frac{3a_{7}}{2} {\cal E}_{K}^{\bf{LR}} \non
&~& + \frac{3C_{10}}{2} {\cal E}_{NF, K}^{\bf{LL}} + \frac{3C_{8}}{2} {\cal E}_{NF, K}^{\bf{SP}}
+ \left(-C_3+\frac{C_{9}}{2} \right) {\cal E}_{NF, \pi}^{\bf{LL}} + \left(-C_5+\frac{C_{7}}{2} \right) {\cal E}_{NF, \pi}^{\bf{LR}} \non
&~& + \left(-a_4+\frac{a_{10}}{2} \right) {\cal A}_{\pi}^{\bf{LL}} + \left(-a_6+\frac{a_{8}}{2} \right) {\cal A}_{\pi}^{\bf{SP}}
+ \left(-C_3+\frac{C_{9}}{2} \right) {\cal A}_{NF, \pi}^{\bf{LL}}  \non
&~& + \left(-C_5+\frac{C_{7}}{2} \right) {\cal A}_{NF, \pi}^{\bf{LR}}-{\cal M}^{({\rm ql,t})}_{B \to  K^0 \pi^0 } 
- {\cal M}^{({\rm mp})}_{B \to  K^0 \pi^0}\Big]   \,,
\label{eq:pi0K0}
\eeq
\beq
{\cal M}(B^0 &\to& \eta_s K^0) = \frac{G_F}{\sqrt{2}} \, V^{\ast}_{ub} V_{us} {\cal M}^{({\rm ql,u})}_{B \to  \eta_s K^0  } +\frac{G_F}{\sqrt{2}} \, V^{\ast}_{cb} V_{cs} {\cal M}^{({\rm ql,c})}_{B \to  \eta_s K^0  }        \non
&~& -\frac{G_F}{\sqrt{2}} \, V^{\ast}_{tb} V_{ts} \Big[ \left(a_3+a_4-\frac{a_{9}+a_{10}}{2} \right) {\cal E}_{K}^{\bf{LL}}
+ \left( a_{5}-\frac{a_{7}}{2} \right) {\cal E}_{K}^{\bf{LR}}  \non
&~& + \left(a_6-\frac{a_{8}}{2} \right) {\cal E}_{K}^{\bf{SP}} + \left(C_3+C_4-\frac{C_{9}+C_{10}}{2} \right) {\cal E}_{NF, K}^{\bf{LL}}
+ \left(C_5-\frac{C_{7}}{2} \right) {\cal E}_{NF, K}^{\bf{LR}}   \non
&~& + \left(C_6-\frac{C_{8}}{2} \right) {\cal E}_{NF, K}^{\bf{SP}} + \left(a_4-\frac{a_{10}}{2} \right) {\cal A}_{K}^{\bf{LL}} 
+ \left(a_6-\frac{a_{8}}{2} \right) {\cal A}_{K}^{\bf{SP}} + \left(C_3-\frac{C_{9}}{2} \right) {\cal A}_{NF, K}^{\bf{LL}} \non
&~& + \left(C_5-\frac{C_{7}}{2} \right) {\cal A}_{NF, K}^{\bf{LR}} + {\cal M}^{({\rm ql,t})}_{B \to  \eta_s K^0 }+ {\cal M}^{({\rm mp})}_{B \to \eta_s K^0 } \Big]   \,,
\label{eq:etasK0}\\   
\sqrt{2}{\cal M}(B^0 &\to& \eta_q K^0) = \frac{G_F}{\sqrt{2}} \, V^{\ast}_{ub} V_{us} \Big[a_2 {\cal E}_{K}^{\bf{LL}} + C_2 {\cal E}_{NF, K}^{\bf{LL}} + {\cal M}^{({\rm ql,u})}_{B \to  K^0 \eta_q  }  \Big]   + \frac{G_F}{\sqrt{2}} \, V^{\ast}_{cb} V_{cs} {\cal M}^{({\rm ql,c})}_{B \to  K^0 \eta_q  }   
\non
&-& \frac{G_F}{\sqrt{2}} \, V^{\ast}_{tb} V_{ts} \Big[ \left(a_4-\frac{a_{10}}{2} \right) {\cal E}_{\eta_q}^{\bf{LL}}
+ \left( a_6-\frac{a_{8}}{2} \right) {\cal E}_{\eta_q}^{\bf{SP}} + \left( 2a_{3}+\frac{a_{9}}{2} \right) {\cal E}_{K}^{\bf{LL}} \non
&~& +\left(2a_{5}+\frac{a_{7}}{2} \right) {\cal E}_{K}^{\bf{LR}} + \left(2C_4+\frac{C_{10}}{2} \right) {\cal E}_{NF, K}^{\bf{LL}}
+ \left( 2C_6+\frac{C_{8}}{2} \right) {\cal E}_{NF, K}^{\bf{SP}}  \non
&~& + \left( C_3-\frac{C_{9}}{2} \right) {\cal E}_{NF, \eta_q}^{\bf{LL}} + \left(C_5-\frac{C_{7}}{2} \right) {\cal E}_{NF, \eta_q}^{\bf{LR}} 
+ \left(a_4-\frac{a_{10}}{2} \right) {\cal A}_{\eta_q}^{\bf{LL}} + \left(a_6-\frac{a_{8}}{2} \right) {\cal A}_{\eta_q}^{\bf{SP}}  \non
&~& + \left(C_3-\frac{C_{9}}{2} \right) {\cal A}_{NF, \eta_q}^{\bf{LL}} + \left(C_5-\frac{C_{7}}{2} \right) {\cal A}_{NF, \eta_q}^{\bf{LR}} + {\cal M}^{({\rm ql,t})}_{B \to  K^0 \eta_q  }+ {\cal M}^{({\rm mp})}_{B \to K^0 \eta_q }\Big]  \,,
\label{eq:etaqK0}\\
{\cal M}(B^0 &\to& K^0 \bar K^0) =\frac{G_F}{\sqrt{2}} \, V^{\ast}_{ub} V_{ud} {\cal M}^{({\rm ql,u})}_{B \to  \bar K^0  K^0 } +\frac{G_F}{\sqrt{2}} \, V^{\ast}_{cb} V_{cd} {\cal M}^{({\rm ql,c})}_{B \to  \bar K^0  K^0 } \non
&~&  -\frac{G_F}{\sqrt{2}} \, V^{\ast}_{tb} V_{td} \Big[ \left(a_4-\frac{a_{10}}{2}\right) {\cal E}_{K}^{\bf{LL}}
+ \left(a_6-\frac{a_{8}}{2} \right) {\cal E}_{K}^{\bf{SP}} + \left(C_3-\frac{C_{9}}{2} \right) {\cal E}_{NF, K}^{\bf{LL}}\non
&~& \left(C_5-\frac{C_{7}}{2} \right) {\cal E}_{NF, K}^{\bf{LR}} + \left(a_3+a_4-\frac{a_{9}+a_{10}}{2} \right) {\cal A}_{K}^{\bf{LL}} 
+ \left(a_{5}-\frac{a_{7}}{2} \right) {\cal A}_{K}^{\bf{LR}}\non
&~& + \left(a_6-\frac{a_{8}}{2} \right) {\cal A}_{K}^{\bf{SP}} + \left(C_3+C_4-\frac{C_{9}+C_{10}}{2} \right) {\cal A}_{NF, K}^{\bf{LL}}
+ \left(C_5-\frac{C_{7}}{2} \right) {\cal A}_{NF, K}^{\bf{LR}}  \non
&~& + \left(C_6-\frac{C_{8}}{2} \right) {\cal A}_{NF, K}^{\bf{SP}} + \left( a_3 - \frac{a_9}{2} \right) {\cal A}_{\bar{K}}^{\bf{LL}} 
+ \left( a_5 - \frac{a_7}{2} \right) {\cal A}_{\bar{K}}^{\bf{LR}} + \left( C_4^\prime - \frac{C_{10}^\prime}{2} \right) {\cal A}_{NF, \bar{K}}^{\bf{LL}}  \non
&~& + \left( C_6^\prime - \frac{C_8^\prime}{2} \right) {\cal A}_{NF, \bar{K}}^{\bf{SP}} 
+{\cal M}^{({\rm ql,t})}_{B \to  \bar K^0  K^0 } + {\cal M}^{({\rm mp})}_{B \to \bar K^0  K^0 } \Big]  \,,
\label{eq:K0K0}\\
{\cal M}(B^0 &\to& K^+ K^-)  =  \frac{G_F}{\sqrt{2}} \, V^{\ast}_{ub} V_{ud} \Big[ a_2 {\cal A}_{K^+}^{\bf{LL}} + C_2 {\cal A}_{NF, K^+}^{\bf{LL}} \Big] \non
&~& - \frac{G_F}{\sqrt{2}} \, V^{\ast}_{tb} V_{td} \Big[ \left(a_3-\frac{a_{9}}{2} \right) {\cal A}_{K^+}^{\bf{LL}} 
+ \left(a_5-\frac{a_{7}}{2} \right) {\cal A}_{K^+}^{\bf{LR}} + \left(C_4-\frac{C_{10}}{2} \right) {\cal A}_{NF, K^+}^{\bf{LL}} \non
&~&  + \left(C_6-\frac{C_8}{2} \right) {\cal A}_{NF, K^+}^{\bf{SP}} + \left(a_3+a_{9} \right) {\cal A}_{K^-}^{\bf{LL}} 
+ \left(a_5+a_7 \right) {\cal A}_{K^-}^{\bf{LR}}\non
&~&  + \left(C_4+C_{10} \right) {\cal A}_{NF, K^-}^{\bf{LL}} + \left(C_6+C_{8}\right) {\cal A}_{NF, K^-}^{\bf{SP}} \Big] \,,
\label{eq:K+K-}
\eeq
\beq
{\cal M}(B^0 &\to& \pi^+\pi^-) = \frac{G_F}{\sqrt{2}} \, V^{\ast}_{ub} V_{ud} \Big[ a_1 {\cal E}_{\pi^-}^{\bf{LL}} + a_2 {\cal A}_{\pi^+}^{\bf{LL}}
+ C_1 {\cal E}_{NF, \pi^-}^{\bf{LL}} + C_2 {\cal A}_{NF, \pi^+}^{\bf{LL}} + {\cal M}^{({\rm ql,u})}_{B \to \pi^+ \pi^-}  \Big] \non
&~& +\frac{G_F}{\sqrt{2}} \, V^{\ast}_{cb} V_{cd} {\cal M}^{({\rm ql,c})}_{B \to \pi^+ \pi^-}  - \frac{G_F}{\sqrt{2}} \, V^{\ast}_{tb} V_{td} \Big[ \left(a_4+a_{10} \right) {\cal E}_{\pi^-}^{\bf{LL}} + \left(a_6+a_{8}\right) {\cal E}_{\pi^-}^{\bf{SP}} \non  
&~& + \left(C_3+C_{9} \right) {\cal E}_{NF, \pi^-}^{\bf{LL}}  + \left(C_5+C_{7} \right) {\cal E}_{NF, \pi^-}^{\bf{LR}} + \left(a_3+a_4-\frac{a_{9}+a_{10}}{2} \right) {\cal A}_{\pi^-}^{\bf{LL}} \non
&~& + \left(a_{5}-\frac{a_{7}}{2} \right) {\cal A}_{\pi^-}^{\bf{LR}}
+ \left(a_6-\frac{a_{8}}{2} \right) {\cal A}_{\pi^-}^{\bf{SP}} + \left(C_3+C_4-\frac{C_{9}+C_{10}}{2} \right) {\cal A}_{NF, \pi^-}^{\bf{LL}}  \non
&~&  + \left(C_5-\frac{C_{7}}{2} \right) {\cal A}_{NF, \pi^-}^{\bf{LR}} + \left(C_6-\frac{C_{8}}{2} \right) {\cal A}_{NF, \pi^-}^{\bf{SP}} + \left(a_3+a_9 \right) {\cal A}_{\pi^+}^{\bf{LL}} + \left(a_{5}+ a_{7} \right) {\cal A}_{\pi^+}^{\bf{LR}} \non
&~& + \left(C_4+C_{10} \right) {\cal A}_{NF, \pi^+}^{\bf{LL}}+\left(C_6+C_{8} \right) {\cal A}_{NF, \pi^+}^{\bf{SP}} + {\cal M}^{({\rm ql,t})}_{B \to \pi^+ \pi^-} +{\cal M}^{({\rm mp})}_{B \to \pi^+ \pi^- } \Big]  \,,
\label{eq:pi+pi-}\\
\sqrt{2}{\cal M}(B^0 &\to& \pi^0\eta_s) = -\frac{G_F}{\sqrt{2}} \, V^{\ast}_{tb}V_{td} \Big[ \left(-a_3+\frac{a_{9}}{2} \right) {\cal E}_{\pi}^{\bf{LL}}
+ \left(-a_5+\frac{a_{7}}{2} \right) {\cal E}_{\pi}^{\bf{LR}} \non
&~& + \left(-C_4+\frac{C_{10}}{2} \right) {\cal E}_{NF, \pi}^{\bf{LL}} + \left(-C_6+\frac{C_{8}}{2} \right) {\cal E}_{NF, \pi}^{\bf{SP}} \Big] \,,
\label{eq:pi0etas}\\
{\cal M}(B^0 &\to& \eta_s\eta_s)  = - \frac{G_F} {\sqrt{2}} \, V^{\ast}_{tb}V_{td} \, 2 \Big[ \left(a_3-\frac{a_{9}}{2}\right) {\cal A}_{\eta_s}^{\bf{LL}}
+ \left(a_5-\frac{a_{7}}{2} \right) {\cal A}_{\eta_s}^{\bf{LR}} \non
&~& + \left(C_4-\frac{C_{10}}{2} \right) {\cal A}_{NF, \eta_s}^{\bf{LL}} + \left(C_6-\frac{C_{8}}{2} \right) {\cal A}_{NF, \eta_s}^{\bf{SP}} \Big]  \,,
\label{eq:etasetas}\\
\sqrt{2} {\cal M}(B^0 &\to& \pi^0\pi^0) = \frac{G_F}{\sqrt{2}} \, V^{\ast}_{ub}V_{ud} \Big[a_2 \left( {\cal A}_{\pi}^{\bf{LL}} - {\cal E}_{\pi}^{\bf{LL}}  \right)
+ C_2 \left( {\cal A}_{NF, \pi}^{\bf{LL}} - {\cal E}_{NF, \pi}^{\bf{LL}} \right) + {\cal M}^{({\rm ql,u})}_{B \to \pi^0 \pi^0}  \Big]  \non
&~& +\frac{G_F}{\sqrt{2}} \, V^{\ast}_{cb}V_{cd}  {\cal M}^{({\rm ql,c})}_{B \to \pi^0 \pi^0}  - \frac{G_F}{\sqrt{2}} \, V^{\ast}_{tb} V_{td} \Big[ \left(a_4-\frac{3a_{9}}{2}-\frac{a_{10}}{2} \right) {\cal E}_{\pi}^{\bf{LL}}
+ \left(-\frac{3a_{7}}{2} \right) {\cal E}_{\pi}^{\bf{LR}} \non
&~& + \left(a_6-\frac{a_{8}}{2} \right) {\cal E}_{\pi}^{\bf{SP}} + \left(C_3-\frac{C_{9}}{2}-\frac{3C_{10}}{2} \right) {\cal E}_{NF, \pi}^{\bf{LL}}
+ \left(C_5-\frac{C_{7}}{2} \right) {\cal E}_{NF, \pi}^{\bf{LR}} \non
&~& - \left(\frac{3C_{8}}{2} \right) {\cal E}_{NF, \pi}^{\bf{SP}}  + \left(2a_3+a_4+\frac{a_{9}-a_{10}}{2} \right) {\cal A}_{\pi}^{\bf{LL}} + \left(2a_{5}+\frac{a_{7}}{2} \right) {\cal A}_{\pi}^{\bf{LR}}   \non
&~&+ \left(a_6-\frac{a_{8}}{2} \right) {\cal A}_{\pi}^{\bf{SP}} + \left(C_3+2C_4+\frac{-C_{9}+C_{10}}{2} \right) {\cal A}_{NF, \pi}^{\bf{LL}}
+ \left(C_5-\frac{C_{7}}{2} \right) {\cal A}_{NF, \pi}^{\bf{LR}} \non
&~&  + \left(2C_6+\frac{C_{8}}{2} \right) {\cal A}_{NF, \pi}^{\bf{SP}} + {\cal M}^{({\rm ql,t})}_{B \to \pi^0 \pi^0} +  {\cal M}^{({\rm mp})}_{B \to \pi^0 \pi^0 }  \Big]  \,,
\label{eq:pi0pi0}
\eeq
\beq
2{\cal M}(B^0 &\to& \pi^0\eta_q) =  \frac{G_F}{\sqrt{2}} \, V^{\ast}_{ub} V_{ud} \Big[a_2 \left({\cal E}_{\eta_q}^{\bf{LL}} - {\cal E}_{\pi}^{\bf{LL}}
+ {\cal A}_{\eta_q}^{\bf{LL}} + {\cal A}_{\pi}^{\bf{LL}} \right) \non
&~& + C_2 \left( {\cal E}_{NF, \eta_q}^{\bf{LL}} - {\cal E}_{NF, \pi}^{\bf{LL}} + {\cal A}_{NF, \pi}^{\bf{LL}} + {\cal A}_{NF, \eta_q}^{\bf{LL}} \right)  - {\cal M}^{({\rm ql,u})}_{B \to \eta_q \pi^0 }-{\cal M}^{({\rm ql,u})}_{B \to \pi^0  \eta_q }  \Big] \non
&~& - \frac{G_F}{\sqrt{2}} \, V^{\ast}_{cb} V_{cd} ( {\cal M}^{({\rm ql,c})}_{B \to \eta_q \pi^0 }+{\cal M}^{({\rm ql,c})}_{B \to \pi^0  \eta_q } )  \non
&~& - \frac{G_F}{\sqrt{2}} \, V^{\ast}_{tb}V_{td}
\Big[ \left(-a_4+\frac{3a_{9}}{2}+\frac{a_{10}}{2} \right) \left( {\cal E}_{\eta_q}^{\bf{LL}} + {\cal A}_{\eta_q}^{\bf{LL}} + {\cal A}_{\pi}^{\bf{LL}} \right) \non
&~& + \left(\frac{3a_{7}}{2} \right) \left( {\cal E}_{\eta_q}^{\bf{LR}} + {\cal A}_{\eta_q}^{\bf{LR}} + {\cal A}_{\pi}^{\bf{LR}} \right)
+ \left(-a_6+ \frac{a_{8}}{2} \right) \left( {\cal E}_{\pi}^{\bf{SP}} + {\cal E}_{\eta_q}^{\bf{SP}} + {\cal A}_{\pi}^{\bf{SP}} + {\cal A}_{\eta_q}^{\bf{SP}} \right) \non
&~& + \left(-2a_3-a_4-\frac{a_{9}}{2}+\frac{a_{10}}{2} \right) {\cal E}_{\pi}^{\bf{LL}} + \left(-2a_5-\frac{a_{7}}{2} \right) {\cal E}_{\pi}^{\bf{LR}} \non
&~& + \left(-C_3+\frac{C_{9}}{2}+\frac{3C_{10}}{2} \right) \left({\cal E}_{NF, \eta_q}^{\bf{LL}} + {\cal A}_{NF, \eta_q}^{\bf{LL}}
+ {\cal A}_{NF, \pi}^{\bf{LL}} \right) \non
&~& + \left(-C_5+\frac{C_{7}}{2} \right) \left( {\cal E}_{NF, \pi}^{\bf{LR}} + {\cal E}_{NF, \eta_q}^{\bf{LR}}
+ {\cal A}_{NF, \pi}^{\bf{LR}} + {\cal A}_{NF, \eta_q}^{\bf{LR}} \right) \non
&~& + \left(\frac{3C_{8}}{2} \right) \left( {\cal E}_{NF, \eta_q}^{\bf{SP}} + {\cal A}_{NF, \pi}^{\bf{SP}} + {\cal A}_{NF, \eta_q}^{\bf{SP}} \right) 
+ \left(-C_3-2C_4+\frac{C_{9}-C_{10}}{2} \right) {\cal E}_{NF, \pi}^{\bf{LL}} \non
&~& + \left(-2C_6-\frac{C_{8}}{2} \right) {\cal E}_{NF, \pi}^{\bf{SP}}-{\cal M}^{({\rm ql,t})}_{B \to \eta_q \pi^0 }- {\cal M}^{({\rm ql,t})}_{B \to \eta_q \pi^0 }-{\cal M}^{({\rm mp})}_{B \to \eta_q \pi^0 }- {\cal M}^{({\rm mp})}_{B \to \eta_q \pi^0 }\Big]   \,,
\label{:pi0etaq}\\
{\cal M}(B^0 &\to& \eta_q\eta_q)  =  \frac{G_F}{\sqrt{2}} \, V^{\ast}_{ub}V_{ud}
\Big[a_2 \left( {\cal A}_{\eta_q}^{\bf{LL}} + {\cal E}_{\eta_q}^{\bf{LL}} \right) + C_2 \left({\cal A}_{NF, \eta_q}^{\bf{LL}} + {\cal E}_{NF, \eta_q}^{\bf{LL}} \right) +{\cal M}^{({\rm ql,u})}_{B \to \eta_q \eta_q} \Big] \non
&~& + \frac{G_F}{\sqrt{2}} \, V^{\ast}_{cb}V_{cd} {\cal M}^{({\rm ql,c})}_{B \to \eta_q \eta_q} 
- \frac{G_F}{\sqrt{2}} \, V^{\ast}_{tb} V_{td} \Big[ \left(2a_3+a_4+\frac{a_{9}-a_{10}}{2} \right)
\left( {\cal A}_{\eta_q}^{\bf{LL}} + {\cal E}_{\eta_q}^{\bf{LL}} \right) \non
&~&  + \left(2a_{5}+\frac{a_{7}}{2} \right)\left({\cal A}_{\eta_q}^{\bf{LR}}+{\cal E}_{\eta_q}^{\bf{LR}} \right) + \left(a_6-\frac{a_{8}}{2} \right) \left( {\cal A}_{\eta_q}^{\bf{SP}} + {\cal E}_{\eta_q}^{\bf{SP}} \right) \non
&~& 
+ \left(C_3+2C_4+\frac{-C_{9}+C_{10}}{2}\right) \left( {\cal A}_{NF, \eta_q}^{\bf{LL}} + {\cal E}_{NF, \eta_q}^{\bf{LL}} \right) \non
&~&  + \left(C_5-\frac{C_{7}}{2} \right) \left({\cal A}_{NF, \eta_q}^{\bf{LR}}  + {\cal E}_{NF, \eta_q}^{\bf{LR}} \right)  
+ \left(2C_6+\frac{C_{8}}{2} \right) \left({\cal A}_{NF, \eta_q}^{\bf{SP}} + {\cal E}_{NF, \eta_q}^{\bf{SP}} \right)  \non &~& + {\cal M}^{({\rm ql,t})}_{B \to \eta_q \eta_q} +{\cal M}^{({\rm mp})}_{B \to \eta_q \eta_q } \Big] \,,
\label{eq:etaqetaq}\\
\sqrt{2}{\cal M}(B^0 &\to& \eta_q\eta_s) = -\frac{G_F}{\sqrt{2}} \, V^{\ast}_{tb} V_{td}
\Big[ \left(a_3-\frac{a_{9}}{2} \right) {\cal E}_{\eta_q}^{\bf{LL}} + \left(a_5-\frac{a_{7}}{2} \right) {\cal E}_{\eta_q}^{\bf{LR}} \non
&~& + \left(C_4-\frac{C_{10}}{2} \right) {\cal E}_{NF, \eta_q}^{\bf{LL}}
+ \left(C_6-\frac{C_{8}}{2} \right) {\cal E}_{NF, \eta_q}^{\bf{SP}} \Big] \,.
\label{eq:etaqetas}
\eeq  
The meson notated in the subscript of typical decaying amplitudes is the spectator meson, 
and the Wilson coefficients $a_i$ and $C_i^\prime$ appeared in the above decay amplitudes are defined in the form of 
\beq
a_1&=& \frac{C_1}{3} + C_2\,, \quad a_2 =\frac{C_2}{3}+C_1 \,, \non
a_i &=& \frac{C_{i+1}}{3} + C_i \quad {\rm with~i=3,5,7,9} \,, \quad
a_i = \frac{C_{i-1}}{3} + C_i \quad {\rm with~i=4,6,8,10} \,.
\eeq
It is noted that the above Wilson coefficients are scale dependent, which should be set at the same as the scale in the integration of leading order or NLO decay amplitudes. 
In the above formulas, we show only the flavour singlet final states $\eta_q$ and $\eta_s$. To compare with the experimental data, 
one can derive the $B\to \eta^{(\prime)} \eta^{(\prime)}$ decay amplitudes from the above formulas easily using the mixing angle in Eq. (\ref{eq:eta-etap-mix}).

For the decay channels with at least one vector meson in the final states, 
the total decay amplitudes have the same configuration of Wilson coefficients since 
the vector mesons have the same quark-antiquark components as their partner pseudoscalar mesons. 
We summary in Table \ref{tab:Wilson-decomp} the possible contributions as a general Wilson coefficient decomposition to each of the four typical decay amplitudes.
The mode $[q_1,q_2,q_3]$ in the first column indicates the light quarks of four fermion operator at the $b$ quark weak decay vertex, 
with the configuration that $q_1$ and $q_2$ are the component quark flavours in the emission meson $M_2$,
and $q_3$ being the quark flavour in the spectator meson $M_3$.

\begin{table}
\begin{center}
\caption{The general decomposition of Wilson coefficients for each certain effective weak vertex.   \label{tab:Wilson-decomp}     }
\vspace{4mm}
\begin{tabular}{| c | c c | c c |}
\toprule
Weak vertex & Typical amplitudes &  \quad\quad Wilson coefficients   \\
\hline
$[s,s,s], \quad [d,d,d]$ \quad & \quad ${\cal E}^{\bf{LL}}/{\cal A}^{\bf{LL}}, \quad {\cal E}_{NF}^{\bf{LL}}/{\cal A}_{NF}^{\bf{LL}}$ \quad &
\quad \quad \quad $a_3+a_4-\frac{a_9+a_{10}}{2}, \quad C_3+C_4-\frac{C_9+C_{10}}{2}$  \quad \\
~  & \quad ${\cal E}^{\bf{LR}}/{\cal A}^{\bf{LR}}, \quad {\cal E}_{NF}^{\bf{LR}}/{\cal A}_{NF}^{\bf{LR}}$ \quad &
\quad \quad \quad $a_5-\frac{a_7}{2}, \quad C_5-\frac{C_7}{2}$  \quad \\
~  & \quad ${\cal E}^{\bf{SP}}/{\cal A}^{\bf{SP}}, \quad {\cal E}_{NF}^{\bf{SP}}/{\cal A}_{NF}^{\bf{SP}}$ \quad &
\quad \quad \quad $a_6-\frac{a_8}{2}, \quad C_6-\frac{C_8}{2}$  \quad \\
\sglinespb
$[d,s,s], \quad [s,d,d]$ \quad & \quad ${\cal E}^{\bf{LL}}/{\cal A}^{\bf{LL}}, \quad {\cal E}_{NF}^{\bf{LL}}/{\cal A}_{NF}^{\bf{LL}}$ \quad &
\quad \quad \quad $a_4-\frac{a_{10}}{2}, \quad C_3-\frac{C_9}{2}$  \quad \\
~  & \quad ${\cal E}^{\bf{LR}}/{\cal A}^{\bf{LR}}, \quad {\cal E}_{NF}^{\bf{LR}}/{\cal A}_{NF}^{\bf{LR}}$ \quad &
\quad \quad \quad $a_6-\frac{a_8}{2}, \quad C_5-\frac{C_7}{2}$  \quad \\
\sglinespb
$[s,s,d], \quad [d,d,s]$ \quad & \quad ${\cal E}^{\bf{LL}}/{\cal A}^{\bf{LL}}, \quad {\cal E}_{NF}^{\bf{LL}}/{\cal A}_{NF}^{\bf{LL}}$ \quad &
\quad \quad \quad $a_3-\frac{a_{9}}{2}, \quad C_4-\frac{C_{10}}{2}$  \quad \\
~  & \quad ${\cal E}^{\bf{LR}}/{\cal A}^{\bf{LR}}, \quad {\cal E}_{NF}^{\bf{LR}}/{\cal A}_{NF}^{\bf{LR}}$ \quad &
\quad \quad \quad $a_5-\frac{a_7}{2}, \quad C_6-\frac{C_8}{2}$  \quad \\
\sglinespb
$[u,u,s], \quad [u,u,d]$ \quad & \quad ${\cal E}^{\bf{LL}}/{\cal A}^{\bf{LL}}, \quad {\cal E}_{NF}^{\bf{LL}}/{\cal A}_{NF}^{\bf{LL}}$ \quad &
\quad \quad \quad $a_2, \quad C_2$  \quad \\
~  & \quad ${\cal E}^{\bf{LR}}/{\cal A}^{\bf{LR}}, \quad {\cal E}_{NF}^{\bf{LR}}/{\cal A}_{NF}^{\bf{LR}}$ \quad &
\quad \quad \quad $a_3+a_9, \quad C_4+C_{10}$  \quad \\
~  & \quad ${\cal E}^{\bf{SP}}/{\cal A}^{\bf{SP}}, \quad {\cal E}_{NF}^{\bf{SP}}/{\cal A}_{NF}^{\bf{SP}}$ \quad &
\quad \quad \quad $a_5+a_7, \quad C_6+C_{8}$  \quad \\
\sglinespb
$[s,u,u], \quad [d,u,u]$ \quad & \quad ${\cal E}^{\bf{LL}}/{\cal A}^{\bf{LL}}, \quad {\cal E}_{NF}^{\bf{LL}}/{\cal A}_{NF}^{\bf{LL}}$ \quad &
\quad \quad \quad $a_1, \quad C_1$  \quad \\
~  & \quad ${\cal E}^{\bf{LR}}/{\cal A}^{\bf{LR}}, \quad {\cal E}_{NF}^{\bf{LR}}/{\cal A}_{NF}^{\bf{LR}}$ \quad &
\quad \quad \quad $a_4+a_{10}, \quad C_3+C_{9}$  \quad \\
~  & \quad ${\cal E}^{\bf{SP}}/{\cal A}^{\bf{SP}}, \quad {\cal E}_{NF}^{\bf{SP}}/{\cal A}_{NF}^{\bf{SP}}$ \quad &
\quad \quad \quad $a_6+a_8, \quad C_5+C_{7}$  \quad \\
\toprule
\end{tabular} 
\end{center} 
\end{table}

\section{ Numerical Results and Discussions}\label{sec:b2pp}

\begin{table}[t]
\caption{Anatomy of the NLO corrections to the branching ratios (${\cal B}$, in unit of $10^{-6}$) 
as well as the ${\bf CP}$ asymmetry parameters ($\acp, C_f, S_f$, in unit of $10^{-2}$) of $B \to \pi\pi, K\pi$ decays in the PQCD approach. }
\label{tab:B2pipi-Kpi-NLO}
\begin{center}
\begin{tabular}{l | r | r | r | r | r | r }
\toprule
Mode \quad\quad & \quad\quad ${\rm LO}$ &\quad\quad  $+ {\rm VC}$ &\quad\quad  $+ {\rm QL}$  &
\quad\quad $+ {\rm MP}$ & \quad\quad $+ {\cal F}^{\rm NLO}$ & \quad \quad PDG \cite{Workman:2022ynf} \non
\hline
${\cal B}(B^+ \to \pi^+\pi^0)$ &$3.58$\quad&$3.89$&$\cdots$&$\cdots$&$4.18^{+1.32}_{-0.97}$& $5.5 \pm 0.4$ \non
$\acp$ &$-0.05$&$0.09$&$\cdots$&$\cdots$&$0.08^{+0.09}_{-0.09}$& $3 \pm 4$ \non
${\cal B}(B^0 \to \pi^+\pi^-)$ &$6.97$&$6.82$&$6.92$&$6.76$&$7.31^{+2.38}_{-1.72}$& $5.12 \pm 0.19$ \non
$C_{\pi^+\pi^-}$ &$-23.4$&$-27.6$&$-13.8$&$-13.3$&$-12.8^{+3.5}_{-3.3}$& $-32 \pm 4$\non
$S_{\pi^+\pi^-}$ &$-31.1$&$-35.5$&$-46.4$&$-37.0$&$-36.4^{+1.5}_{-1.5}$& $-65 \pm 4$\non
${\cal B}(B^0 \to \pi^0\pi^0)$ &$0.14$&$0.29$&$0.30$&$0.22$&$0.23^{+0.07}_{-0.05}$& $1.59 \pm 0.26$ \non
$C_{\pi^0\pi^0}$ &$-3.1$&$60.1$&$73.6$&$77.6$&$80.2^{+5.2}_{-6.7}$& $33 \pm 22$ \non
\hline
${\cal B}(B^+ \to \pi^+K^0)$ &$17.0$&$20.8$&$28.0$&$19.4$&$20.3^{+6.3}_{-4.4}$& $23.7 \pm 0.8$ \non
$\acp$ &$-1.19$&$-0.95$&$-0.06$&$-0.08$&$-0.08^{+0.08}_{-0.09}$& $-1.7 \pm 1.6$ \non
${\cal B}(B^+ \to \pi^0K^+)$ &$10.0$&$12.75$&$16.76$&$11.92$&$12.3^{+3.8}_{-2.7}$& $12.9 \pm 0.5$ \non
$\acp$ &$-10.9$&$-5.20$&$2.26$&$2.48$&$2.28^{+1.61}_{-1.74}$& $3.7 \pm 2.1$ \non
${\cal B}(B^0 \to \pi^-K^+)$ &$14.3$&$18.0$&$23.9$&$16.4$&$17.1^{+5.2}_{-3.7}$& $19.6 \pm 0.5$ \non
$\acp$ &$-15.2$&$-14.2$&$-4.16$&$-5.42$&$-5.43^{+2.24}_{-2.34}$& $-8.3 \pm 0.4$ \non
${\cal B}(B^0 \to \pi^0K^0)$ &$5.90$&$8.12$&$10.4$&$6.99$&$7.38^{+2.11}_{-1.50}$& $9.9 \pm 0.5$ \non
$C_{\pi^0K^0}$ &$-2.62$&$-7.31$&$-6.57$&$-7.97$&$-7.70^{+0.21}_{-0.13}$& $0 \pm 13$ \non
$S_{\pi^0K^0}$ &$70.1$&$73.5$&$71.6$&$71.9$&$71.9^{+0.6}_{-0.6}$& $58 \pm 17$ \non
\toprule
\end{tabular} 
\end{center}  
\end{table}

All the charmless two-body $B$ meson decays have been calculated in the leading order of PQCD approach by different authors \cite{ Li:2005hg, Zou:2015iwa,Liu:2005mm, Lu:2005be, Zhu:2005rt, Wang:2005bk, Liu:2005mm, Huang:2005if, Lu:2000hj, Lu:2000em}. 
They are updated by various authors at the next-to-leading order recently \cite{ Liu:2015upa, Xiao:2011tx, Bai:2013tsa, Fan:2012kn, Zhang:2014bsa, Zhang:2009zg, Li:2006jv, Zhang:2008by, Rui:2011dr, Zhang:2009zzn, Li:2006cva}.
In order to show the effects of the NLO corrections from different sources explicitly,
we present the PQCD results for the branching ratios\footnote{As the convention of experiments, 
the branching ratios shown in this review are averages of the $B$ and $\bar B$ decays.} 
and ${\bf CP}$ asymmetries for $B \to \pi\pi$ and $B \to K \pi$ decay modes at the LO (second column) 
and with the inclusion of each of the four kinds of the NLO corrections iteratively:
i.e.  the vertex corrections, the quark loop, the magnetic penguin  and the heavy-to-light form factor (${\cal F}^{{\rm NLO}}$) corrections 
(3rd-6rd columns) in Table \ref{tab:B2pipi-Kpi-NLO}. 
For the definitions of the CP asymmetry parameters $\acp, C_f$ and $S_f$, we take the same convention as adopted by Particle Data Group in Ref. \cite{Workman:2022ynf}.  
As shown in Eq. (\ref{eq:pi+pi0}) that the quark loop and magnetic penguin diagrams do not contribute in $B^+ \to \pi^+ \pi^0$ decay.  
We see that 
\begin{itemize}
\item[(1)]
For the seven $B \to \pi\pi, K\pi$ decays, the corrections from quark loop and magnetic penguin associated with higher dimension operators
cancel with each other in the prediction of branching ratios.
The four-fermion vertex corrections and $B \to \pi$ form factors NLO corrections do not bring significant effect to branching ratios.
\item[(2)]
The four-fermion vertex correction flaps the sign of the prediction of direct ${\bf CP}$ asymmetries in $B^+ \to \pi^+ \pi^0$ and $B^0 \to \pi^0 \pi^0$ modes \cite{Zhang:2014bsa}. 
For the ${\bf CP}$ asymmetries of seven considered decay modes, 
the agreement between the PQCD predictions with the measured ones do become much better with the inclusion of the NLO corrections.
\item[(3)]
There are more sizable NLO corrections to the color suppressed decay modes, 
such as the $B^0\to \pi^0\pi^0$ and $B^0 \to \pi^0 K^0$ decays than the color favored decay modes. 
\item[(4)] 
The NLO corrections change the ${\bf CP}$ asymmetry parameters more significant than the branching ratios.
For example, the direct ${\bf CP}$ asymmetry of $B^+ \to \pi^0K^+$ decay is similar as that of  $B^0 \to \pi^-K^+$ decay at leading order. 
But the former one change sign with the inclusion of NLO corrections, which explicitly explains the so-called $\pi K$ puzzle \cite{Li:2005kt}.
\item[(5)] 
For the largest direct ${\bf CP}$ asymmetries measured in $B^0\to \pi^+\pi^-$ and $B^0\to K^+\pi^-$ decays, 
the NLO corrections reduce the size significantly comparing with the LO results. 
This indicates that the strong phase from charm quark loop correction gives destructive contribution to the annihilation type diagram at LO.
\end{itemize}

\subsection{\texorpdfstring{$B \to PP$}{} decay modes}

\begin{table}[t]
\vspace{-2mm}
\caption{The updated PQCD results for the branching ratios of $B\to PP$ decays (in unit of $10^{-6}$).}
\label{tab:BR-B2PP}
\begin{center}
\begin{tabular}{l | r | r  r | r | r}
\toprule
Mode     &   \quad PQCD \quad\quad & \quad SCET1 \cite{Williamson:2006hb} \quad & \quad SCET2 \cite{Williamson:2006hb} \quad&
\quad QCDF \cite{Cheng:2009cn} \quad &    PDG \cite{Workman:2022ynf} \non
\hline
$B^+ \to \pi^+ K^0$  &$20.3^{+6.3+0.1}_{-4.4-0.1}$&$\cdots$   & $\cdots$ & $21.7^{+13.4}_{-9.1}$ & $23.7 \pm 0.8$ \non
$B^+ \to \pi^0 K^+$  &$12.3^{+3.8+0.1}_{-2.7-0.1}$&$\cdots$     & $\cdots$   &   $12.5^{+6.8}_{-4.8}$ & $12.9 \pm 0.5$ \non
$B^+ \to \eta^\prime K^+$  
&$52.0^{+15.0+2.1}_{-10.8-0.7}$& $69.5\pm 28.4$     & $69.3\pm 27.7$ & $74.5^{+63.6}_{-31.6}$ & $70.4 \pm 2.5$ \non
$B^+ \to \eta K^+$  &$6.68^{+2.26+1.85}_{-1.60-0.96}$& $2.7\pm4.8$   & $2.3\pm 4.5$  & $2.2^{+2.0}_{-1.3}$ & $2.4 \pm 0.4$ \non
$B^+ \to K^+\bar{K}^0$  &$1.56^{+0.48+0.02}_{-0.34-0.02}$&$\cdots$    & $\cdots$   & $1.8^{+1.1}_{-0.7}$ & $1.31\pm 0.17$ \non
$B^+ \to \pi^0 \pi^+$  &$4.18^{+1.30+0.22}_{-0.94-0.22}$& $\cdots$     & $\cdots$ & $5.9^{+2.6}_{-1.6}$ & $5.5 \pm 0.4$ \non
$B^+ \to \pi^+\eta^\prime$  &$2.00^{+0.57+0.36}_{-0.42-0.31}$&$2.4\pm 1.3$      & $2.8\pm 1.3$ & $3.8^{+1.6}_{-0.8}$ & $2.7 \pm 0.9$ \non
$B^+ \to \pi^+\eta$  &$2.62^{+0.78+0.45}_{-0.57-0.40}$&$4.9\pm 2.0$    & $5.0\pm2.1$ & $5.0^{+1.5}_{-0.9}$ & $4.02 \pm 0.27$ \non
\hline
$B^0 \to \pi^- K^+$  &$17.1^{+5.2+0.1}_{-3.7-0.1}$&$\cdots$    & $\cdots$ & $19.3^{+11.4}_{-7.8}$ & $19.6 \pm 0.5$ \non
$B^0 \to \pi^0 K^0$  &$7.38^{+2.11+0.03}_{-1.50-0.04}$& $\cdots$    & $\cdots$ & $8.6^{+5.4}_{-3.6}$ & $9.9 \pm 0.5$ \non
$B^0 \to \eta^\prime K^0$  &$52.3^{+14.9+2.1}_{-10.8-0.3}$& $63.2\pm 26.3$    & $62.2\pm 25.4$   & $70.9^{+59.1}_{-29.8}$ & $66 \pm 4$  \non
$B^0 \to \eta K^0$  &$4.63^{+1.57+1.51}_{-1.09-0.79}$& $2.4\pm 4.4$      & $2.3\pm 4.4$  & $1.5^{+1.7}_{-1.1}$ & $1.23^{+0.27}_{-0.24}$ \non
$B^0 \to K^0 \bar{K}^0$  &$1.48^{+0.47+0.01}_{-0.33-0.00}$&$\cdots$      & $\cdots$  & $2.1^{+1.3}_{-0.8}$ & $1.21 \pm 0.16$ \non
$B^0 \to K^+K^-$  &$0.046^{+0.058+0.009}_{-0.039-0.008}$& $\cdots$       & $\cdots$ & $0.1 \pm 0.04$ & $0.078 \pm 0.015$ \non
$B^0 \to \pi^+ \pi^-$  &$7.31^{+2.35+0.38}_{-1.68-0.36}$&$\cdots$     & $\cdots$ & $7.0^{+0.8}_{-1.0}$ & $5.12 \pm 0.19$ \non
$B^0 \to \pi^0 \pi^0$  &$0.23^{+0.07+0.01}_{-0.05-0.01}$&$\cdots$     & $\cdots$   & $1.1^{+1.2}_{-0.5}$ & $1.59 \pm 0.26$ \non
$B^0 \to \pi^0 \eta^\prime $  &$0.20^{+0.05+0.02}_{-0.03-0.01}$&$2.3\pm 2.8$    & $1.3\pm 0.6$   & $0.42^{+0.28}_{-0.15}$ & $1.2 \pm 0.6$ \non
$B^0 \to \pi^0 \eta$  &$0.20^{+0.06+0.02}_{-0.04-0.01}$&$0.88\pm 0.68$     & $0.68\pm 0.62$  & $0.36^{+0.13}_{-0.11}$ & $0.41 \pm 0.17$ \non
$B^0 \to \eta\eta$  &$0.37^{+0.09+0.08}_{-0.07-0.07}$&$0.69\pm 0.71$    & $1.0\pm 1.5$  & $0.32^{+0.15}_{-0.08}$ & $ < 1$ \non
$B^0 \to \eta\eta^\prime$  &$0.29^{+0.07+0.06}_{-0.05-0.06}$&$1.0\pm 1.6$    & $2.2\pm 5.5 $  & $0.36^{+0.27}_{-0.13}$ & $< 1.2 $ \non
$B^0 \to \eta^\prime\eta^\prime$  &$0.42^{+0.09+0.13}_{-0.07-0.11}$&$0.57\pm 0.73$    & $1.2\pm 3.7$   & $0.22^{+0.16}_{-0.08}$ & $< 1.7 $ \\
\toprule
\end{tabular} \end{center}
\end{table}%

\begin{table}[t]
\vspace{-2mm}
\caption{The updated PQCD results for the ${\bf CP}$ asymmetries of $B\to PP$ decays (in unit of $10^{-2}$).}
\label{tab:CPV-B2PP} 
\begin{center}
\begin{tabular}{l | r | r  r | r | r}
\toprule
Mode     &   \quad PQCD \quad\quad   & \quad SCET1 \cite{Williamson:2006hb} & \quad SCET2 \cite{Williamson:2006hb} &
\quad QCDF \cite{Cheng:2009cn} \quad &    PDG \cite{Workman:2022ynf}  \non
\hline
$B^+ \to \pi^+ K^0_S$  &$-0.08^{+0.08+0.02}_{-0.09-0.02}$&$\cdots $    &$\cdots$   & $0.28^{+0.09}_{-0.10}$ & $-1.7 \pm 1.6$ \non
$B^+ \to \pi^0 K^+$  &$2.28^{+1.53+0.50}_{-1.65-0.57}$&$\cdots $    &$\cdots$ & $4.9^{+6.3}_{-5.8}$ & $3.0 \pm 2.1$ \non
$B^+ \to \eta^\prime K^+$  &$-1.83^{+0.40+0.77}_{-0.40-1.03}$& $-1\pm 1$    &$7\pm 1$  & $0.45^{+1.4}_{-1.1}$ & $0.4 \pm 1.1$ \non
$B^+ \to \eta K^+$  &$-7.75^{+1.06+0.81}_{-0.99-0.43}$& $33\pm 31$    &$-33\pm 40$  & $-14.5^{+18.6}_{-28.1}$ & $-37 \pm 8$ \non
$B^+ \to K^+ K^0_S$ &$1.83^{+1.93+0.14}_{-1.87-0.18}$& $\cdots$   &$\cdots$ & $-6.4 \pm 2.0$ & $4\pm 14$ \non
$B^+ \to \pi^0 \pi^+$  &$0.08^{+0.06+0.07}_{-0.06-0.04}$&$\cdots$   &$\cdots$ & $-0.11^{+0.06}_{-0.03}$ & $3 \pm 4$ \non
$B^+ \to \pi^+\eta^\prime$  &$68.9^{+2.4+1.0}_{-2.4-0.9}$&$21\pm 21$     &$2\pm 18$ & $1.6^{+10.6}_{-13.8}$ & $6 \pm 16$ \non
$B^+ \to \pi^+\eta$  &$24.8^{+3.6+0.8}_{-3.3-0.7}$&$5\pm 29$    &$37\pm 29$ & $-5.0^{+8.7}_{-10.8}$ & $-14 \pm 7$ \non
\hline
$B^0 \to \pi^- K^+$  &$-5.43^{+1.86+1.26}_{-1.92-1.34}$&$\cdots$    &$\cdots$ & $-7.4^{+4.6}_{-5.0}$ & $-8.3 \pm 0.4$ \non
$B^0 \to \pi^0 K^0_S$  &$-7.70^{+0.17+0.12}_{-0.09-0.09}$&$\cdots$    &$\cdots$ & $-10.6^{+6.2}_{-5.7}$ & $C_{\pi^0 K^0}=0 \pm 13$ \non
 &$71.9^{+0.3+0.5}_{-0.3-0.5}$&$\cdots$   &$\cdots$ & $79.0^{+7.2}_{-5.7}$ & $S_{\pi^0 K^0}=58 \pm 17$ \non
$B^0 \to \eta^\prime K^0_S$  &$-2.65^{+0.10+0.07}_{-0.10-0.11}$& $1.1\pm 1.4$    &$-2.7\pm 1.2$  & $3.0^{+1.0}_{-0.9}$ & $C_{\eta^\prime K^0}=-6 \pm 4$ \non
 &$69.8^{+0.1+0.1}_{-0.1-0.1}$& $70.6$   &$71.5$  & $67.0 \pm 1.4$ & $S_{\eta^\prime K^0}=63 \pm 6$ \non
$B^0 \to \eta K^0_S$  &$-7.88^{+0.14+0.06}_{-0.10-0.02}$& $21\pm 21$   &$-18\pm 23.2$  & $-23.6^{+16.0}_{-29.0}$ & $\cdots$ \non
 &$70.0^{+0.2+0.2}_{-0.3-0.1}$& $69$    &$79$  & $79.0^{+8.9}_{-8.5}$ & $\cdots$ \non
$B^0 \to K_S^0 K_S^0$  &$-17.3^{+0.6+0.4}_{-0.4-0.3}$&$\cdots$    &$\cdots$  & $-10.0^{+1.2}_{-2.0}$ & $C_{K_S^0 K_S^0}=0 \pm 40$ \non
 &$5.34^{+1.05+0.53}_{-1.06-0.49}$&$\cdots$& $\cdots$  &$\cdots$  & $S_{K_S^0 K_S^0}=-80 \pm 50$ \non
$B^0 \to \pi^+ \pi^-$  &$-12.8^{+3.3+1.1}_{-3.1-1.1}$&$\cdots$   &$\cdots$  & $17.0^{+4.5}_{-8.8}$ & $C_{\pi^+ \pi^-}= -32 \pm 4$ \non
 &$-36.4^{+0.5+1.4}_{-0.4-1.4}$&$\cdots$    &$\cdots$  & $-69^{+20.6}_{-13.5}$ & $S_{\pi^+ \pi^-}=-65 \pm 4$ \non
$B^0 \to \pi^0 \pi^0$  &$-80.2^{+5.2+0.4}_{-6.7-0.2}$&$\cdots$     &$\cdots$  & $57.2^{+33.7}_{-40.4}$ & $C_{\pi^0 \pi^0}= -33 \pm 22$ \non
 &$53.5^{+8.7+3.1}_{-8.4-3.0}$&$\cdots$   &$\cdots$  & $\cdots$ & $\cdots $ \non
\toprule
\end{tabular}\end{center}
\end{table}%

In Table \ref{tab:BR-B2PP} we show the updated NLO PQCD predictions of branching ratios of $B \to PP$ decays, together with the experiment data. 
As comparisons, the results from the SCET \cite{Williamson:2006hb} and from the QCDF approach \cite{Cheng:2014rfa,Cheng:2009cn} are also listed.
The two dominant uncertainties shown in this table and the following tables come from the $B$ meson distribution amplitude parameter $\omega_B$ 
and Gegenbauer moments in the LCDAs of light meson. 
Other sources of small theoretical errors, such as the combined uncertainty in the CKM matrix elements, are not included. 
For most $B \to PP$ decays, one can see that the PQCD predictions for the branching ratios do agree well with those SCET and QCDF predictions, 
and also agree with measured ones within errors.
For $B^0 \to \pi^0\pi^0$ decay, however, the NLO enhancements as listed in Table \ref{tab:B2pipi-Kpi-NLO} are still not large enough to interpret the measured one.
It is worth of mentioning that the authors of Ref. \cite{Liu:2015sra}  have studied the Glauber-gluon effect to $B^0 \to \pi^0\pi^0$ decay and 
found that such effect could provide a factor of $2.1$ enhancement to the decay rate, reaching $Br(B^0 \to \pi^0\pi^0)=(0.61^{+0.21}_{-0.17}) \times 10^{-6}$.
The NLO corrections play an important role to explain the data in penguin dominated channels,
particularly in $B^+ \to \pi^0 K^+, \pi^+ K^0, \eta^\prime K^+$ and $B^0 \to \pi^0K^0, \pi^-K^+, \eta^\prime K^0$ \cite{Fan:2012kn}.
For the pure annihilation decay $B^0 \to K^+ K^-$, the NLO effect is also found sizable comparing to the former PQCD prediction at LO \cite{Xiao:2011tx}.

In Table \ref{tab:CPV-B2PP} we show the updated ${\bf CP}$ asymmetry parameters of those $B \to PP$ decays that have been measured by experiments. 
As expected, the ${\bf CP}$ asymmetry is large compared with the $K$ and $D$ meson decays, 
since the weak phases arose from tree and penguin amplitudes are comparable in size for $B$ meson decays. 
For the neutral $B$ meson decays we considered, the ${\bf CP}$ asymmetries are not sensitive to the new power corrections being taken into account in this work.
For the charged $B$ meson decays, however, the direct ${\bf CP}$ asymmetries are affected apparently by the inclusion of these corrections,
especially for the channels with at least one $\eta$ or $\eta^\prime$ in the final states. 
For comparison, we also include those results from the SCET and QCDF approach. 
Despite the good agreement of branching ratios between these approaches and PQCD approach, 
there is a large difference for CP asymmetry parameters between these approaches, 
even a different sign for some decay channels, such as $B^+ \to K^+ K^0_S$, $B^0 \to \pi^+ \pi^-$, $B^0 \to \pi^0 \pi^0$ and $B^0 \to \eta^\prime K^0_S$ etc.  
It is easy to see that the PQCD results have a better agreement with the current experimental data than the QCDF one. 
This verifies that the dominant strong phase in the two body charmless B decays comes from the penguin annihilation type diagrams rather than the NLO QCD corrections.
 
\subsection{\texorpdfstring{$B \to PV$}{} decay modes}\label{sec:b2pv}

\begin{table}[t]
\vspace{-2mm}
\caption{The updated PQCD results for the branching ratios of $B^+\to PV$ decays (in unit of $10^{-6}$).} 
\label{tab:BR-B2PV}.
\label{tab:BR-B+2PV}
\vspace{-4mm}
\begin{center}
\begin{tabular}{l | r | r  r | r | r}
\toprule
Mode     &    PQCD \quad\quad & \quad SCET1 \cite{Wang:2008rk} & \quad SCET2 \cite{Wang:2008rk} &
\quad QCDF \cite{Cheng:2009cn} \quad &    PDG \cite{Workman:2022ynf} \non
\hline
$B^+ \to \pi^+ K^{\ast 0}$  &$5.52^{+1.93+0.38}_{-1.36-0.41}$&$8.5^{+5.0}_{-3.9}$     & $9.9^{+3.7}_{-3.2}$  & $10.4^{+4.5}_{-4.2}$ & $10.1 \pm 0.8$ \non
$B^+ \to \pi^0 K^{\ast +}$  &$3.58^{+1.19+0.18}_{-0.82-0.15}$&$4.2^{+2.3}_{-1.8}$    & $6.5^{+2.0}_{-1.8}$ & $6.7^{+2.5}_{-2.3}$ & $6.8 \pm 0.9$ \non
$B^+ \to \eta^\prime K^{\ast +}$  &$1.54^{+0.51+0.17}_{-0.34-0.08}$  &$4.5^{+6.7}_{-4.0}$   & $4.8^{+5.4}_{-3.7}$ & $1.7^{+4.9}_{-1.6}$ & $4.8^{+1.8}_{-1.6}$ \non
$B^+ \to \eta K^{\ast +}$  &$6.08^{+0.41+2.02}_{-0.30-1.45}$  &$17.9^{+6.5}_{-6.1}$   & $18.6^{+5.1}_{-5.3}$  & $15.7^{+12.7}_{-8.3}$ & $19.3 \pm 1.6$ \non
$B^+ \to K^{+} \omega$  &$6.17^{+1.25+1.59}_{-0.90-1.33}$&$5.1^{+2.6}_{-2.1}$     & $5.9^{+2.2}_{-1.8}$   & $4.8^{+5.6}_{-3.0}$ & $6.5 \pm 0.4$ \non
$B^+ \to K^{+} \phi$  &$4.61^{+1.41+2.29}_{-0.82-0.63}$&$9.7^{+5.2}_{-4.2}$     & $8.6^{+3.4}_{-2.9}$   & $8.8^{+5.5}_{-4.5}$ & $8.8^{+0.7}_{-0.6}$ \non
$B^+ \to K^+\rho^0$  &$3.28^{+0.25+0.50}_{-0.21-0.48}$&$6.7^{+2.9}_{-2.4}$    & $4.6^{+1.9}_{-1.6}$  & $3.5^{+4.1}_{-4.5}$ & $3.7\pm 0.5$ \non
$B^+ \to K^0\rho^+$  &$6.11^{+0.45+0.96}_{-0.34-0.86}$&$9.3^{+5.0}_{-4.0}$     & $10.1^{+4.3}_{-3.5}$   & $7.8^{+9.6}_{-5.3}$ & $7.3^{+1.0}_{-1.2}$ \non
$B^+ \to K^+\bar{K}^{\ast 0}$  &$0.47^{+0.15+0.02}_{-0.10-0.03}$&$0.49^{+0.28}_{-0.22}$    & $0.51^{+0.2}_{-0.17}$   & $0.80^{+0.36}_{-0.33}$ & $0.59 \pm 0.08$ \non
$B^+ \to \bar K^0 K^{\ast +}$  &$0.31^{+0.02+0.10}_{-0.02-0.09}$&$0.54^{+0.28}_{-0.22}$     & $0.51^{+0.22}_{-0.18}$   & $0.46^{+0.56}_{-0.31}$ & $\cdots$ \non
$B^+ \to \pi^+ \rho^0$  &$4.96^{+1.34+0.13}_{-1.01-0.14}$&$10.7^{+1.2}_{-1.1}$    & $7.9^{+0.8}_{-0.8}$   & $8.7^{+3.2}_{-1.9}$ & $8.3 \pm 1.2$ \non
$B^+ \to \pi^0 \rho^+$  &$10.9^{+3.4+0.6}_{-2.4-0.6}$&$8.9^{+1.0}_{-1.0}$      & $11.4^{+1.3}_{-1.1}$  & $11.8^{+2.3}_{-1.8}$ & $10.9 \pm 1.4$ \non
$B^+ \to \eta^\prime \rho^+$  &$4.06^{+1.22+0.84}_{-0.89-0.77}$& $0.37^{+2.5}_{-0.23}$     & $0.44^{+3.2}_{-0.20}$  & $5.6^{+1.2}_{-0.9}$ & $9.7 \pm 2.2$ \non
$B^+ \to \eta \rho^+$ &$5.59^{+1.68+1.17}_{-1.22-1.06}$& $3.9^{+2.0}_{-1.7}$   & $3.3^{+1.9}_{-1.6}$  & $8.3^{+1.3}_{-1.1}$ & $7.0 \pm 2.9$ \non
$B^+ \to \pi^+ \omega$  &$5.42^{+1.44+0.47}_{-1.10-0.45}$&$6.7^{+0.80}_{-0.70}$    & $8.5^{+0.9}_{-0.9}$   & $6.7^{+2.5}_{-1.5}$ & $6.9 \pm 0.5$  \non
$B^+ \to \pi^+ \phi$  &$0.042^{+0.014+0.002}_{-0.010-0.002}$ & $\sim 0.003$  &$\sim 0.003$   & $\sim 0.043$ & $0.032 \pm 0.015$  \non
\toprule
\end{tabular}\end{center} \end{table}%
\begin{table}[t]
\vspace{-2mm}
\caption{The updated PQCD results for the branching ratios of $B^0\to PV$ decays (in unit of $10^{-6}$).}
\label{tab:BR-B02PV}       
\begin{center}
\begin{tabular}{l| r|r  r |r|r}
\toprule  
Mode     &    PQCD \quad\quad & \quad SCET1 \cite{Wang:2008rk} & \quad SCET2 \cite{Wang:2008rk} &
\quad QCDF \cite{Cheng:2009cn} \quad &  \quad  PDG \cite{Workman:2022ynf} \non
\hline
$B^0 \to \eta^\prime K^{\ast 0}$  &$1.60^{+0.46+0.04}_{-0.32-0.03}$  &$4.1^{+6.3}_{-3.7}$   &$4.0^{+4.8}_{-3.5}$  & $1.5^{+4.6}_{-1.7}$ & $2.8 \pm 0.6$ \non
$B^0 \to \eta K^{\ast 0}$  &$5.37^{+0.35+1.69}_{-0.25-1.25}$ &$16.6^{+6.0}_{-5.7}$   &$16.5^{+4.7}_{-4.7}$  & $15.6^{+12.3}_{-8.2}$ & $15.9 \pm 1.0$ \non
$B^0 \to K^0 \omega$  &$5.60^{+0.99+1.55}_{-0.77-1.35}$&$4.1^{+2.2}_{-1.8}$    &$4.9^{+2.0}_{-1.7}$  & $4.1^{+5.3}_{-2.8}$ & $4.8 \pm 0.4$ \non
$B^0 \to K^0 \phi$  &$4.25^{+1.25+0.66}_{-0.71-0.55}$&$9.1^{+4.9}_{-3.9}$     &$8.0^{+3.2}_{-2.7}$  & $8.1^{+5.1}_{-4.1}$ & $7.3 \pm 0.7$ \non
$B^0 \to K^{+} \rho^-$  &$6.05^{+0.57+0.84}_{-0.42-0.75}$&$9.8^{+4.9}_{-4.0}$   &$10.2^{+4.1}_{-3.4}$  & $8.6^{+9.3}_{-5.3}$ & $7.0 \pm 0.9$ \non
$B^0 \to K^{0} \rho^0$  &$3.64^{+0.49+0.47}_{-0.38-0.45}$&$3.5^{+2.11}_{-1.6}$    &$5.8^{+2.2}_{-1.9}$  & $5.4^{+5.5}_{-3.3}$ & $3.4 \pm 1.1$ \non
$B^0 \to \pi^-K^{\ast +}$  &$4.67^{+1.54+0.24}_{-1.09-0.27}$&$8.4^{+4.7}_{-3.6}$    &$9.5^{+3.41}_{-3.0}$  & $9.2^{+3.8}_{-3.4}$ & $7.5\pm 0.4$ \non
$B^0 \to \pi^0K^{\ast 0}$  &$1.57^{+0.55+0.14}_{-0.38-0.16}$&$4.6^{+2.5}_{-1.9}$    &$3.7^{+1.5}_{-1.3}$  & $3.5^{+1.6}_{-1.5}$ & $3.3\pm 0.6$ \non
$B^0 \to K^+ K^{\ast -}$ &&&&&  \non
$\quad\quad + K^- K^{\ast +}$  &$0.39^{+0.03+0.04}_{-0.03-0.03}$&$0.10^{+0.10}_{-0.08}$    &$\cdots$  &$0.15^{+0.05}_{-0.04}$ & $ < 0.4$ \non
$B^0 \to K^0 \bar{K}^{\ast 0}$ &&&&&  \non
$\quad\quad + \bar{K}^0 K^{\ast 0} $  &$0.79^{+0.17+0.11}_{-0.12-0.09}$&$0.96^{+0.38}_{-0.30}$    &$0.95^{+0.30}_{-0.25}$ & $1.37^{+0.65}_{-0.27}$ & $ < 0.96$  \non
$B^0 \to \pi^0 \rho^0$  &$0.32^{+0.10+0.04}_{-0.07-0.04}$&$2.5^{+0.3}_{-0.2}$    &$1.5^{+0.1}_{-0.1}$  & $1.3^{+2.1}_{-0.8}$ & $2.0 \pm 0.5$  \non
$B^0 \to \pi^+ \rho^-$ &&&&&   \non
$\quad\quad + \pi^- \rho^+$  &$25.2^{+5.7+1.1}_{-7.8-1.1}$&$13.4^{+1.3}_{-1.3}$   &$16.8^{+1.7}_{-1.6}$  & $25.1^{+2.1}_{-2.8}$ & $23.0 \pm 2.3$  \non
$B^0 \to \eta^\prime\rho^0$  &$0.12^{+0.03+0.02}_{-0.02-0.02}$&$0.43^{+2.51}_{-0.13}$    &$1.0^{+3.5}_{-0.9}$  & $0.09^{+0.12}_{-0.05}$ & $< 1.3$ \non
$B^0 \to \eta \rho^0$  &$0.13^{+0.02+0.03}_{-0.02-0.02}$&$0.04^{+0.20}_{-0.01}$    &$0.14^{+0.33}_{-0.13}$  & $0.10^{+0.04}_{-0.03}$ & $< 1.5$ \non
$B^0 \to \pi^0 \phi$  &$0.02^{+0.01+0.00}_{-0.01-0.00}$&$\sim 0.001$  &$\sim 0.001$  & $0.01^{+0.03}_{-0.01}$ & $< 0.15$ \non
$B^0 \to \eta^\prime \phi$  &$0.022^{+0.003+0.009}_{-0.002-0.008}$&$\sim 0.0001$   &$\sim 0.0007$  & $\sim 0.004$ & $< 0.5$ \non
$B^0 \to \eta \phi$  &$0.006^{+0.002+0.001}_{-0.001-0.001}$&$\sim 0.0004$   &$\sim 0.0008$  & $\sim 0.005$ & $< 0.5$ \non
$B^0 \to \pi^0 \omega$  &$0.10^{+0.02+0.01}_{-0.01-0.01}$&$0.0003^{+0.0299}_{-0}$    &$0.015^{+0.024}_{-0.002}$  & $0.01^{+0.04}_{-0.01}$ & $< 0.5$ \non
$B^0 \to \eta^\prime \omega$  &$0.53^{+0.13+0.12}_{-0.10-0.10}$&$0.18^{+1.31}_{-0.10}$  &$3.1^{+4.9}_{-2.6}$  & $0.59^{+0.60}_{-0.27}$ & $\cdots$ \non
$B^0 \to \eta \omega$  &$0.73^{+0.16+0.17}_{-0.13-0.14}$&$0.91^{+0.66}_{-0.50}$   &$1.4^{+0.8}_{-0.6}$  & $0.85^{+0.76}_{-0.35}$ & $\cdots$ \non
\toprule
\end{tabular} \end{center}
\end{table}%

In Tables \ref{tab:BR-B+2PV} and \ref{tab:BR-B02PV}, we list the NLO PQCD predictions for the branching ratios of $B\to PV $ decays. 
There are three special cases of decay modes $B^0/\bar B^0 \to K^+ K^{\ast -}/ K^- K^{\ast +}$, $B^0/\bar B^0 \to K^0 \bar{K}^{\ast 0}/\bar{K}^0 K^{\ast 0}$ 
and $B^0 /\bar B^0\to \pi^+ \rho^-/ \pi^- \rho^+$ in Table \ref{tab:BR-B02PV}, each containing four individual  decay channels. 
The experiments can not distinguish them unless a time dependent analysis is required, because of the $B^0-\bar B^0$ mixing \cite{Rui:2011dr}. 
Therefore, the branching ratios shown in the table are the sum of the four corresponding decay channels divided by 2.
We mark that the branching ratios for the channels decaying to final states $\pi K^\ast, K \rho, KK^\ast$ and  $\pi \rho$ are comparable to
the previous PQCD predictions \cite{Li:2006jv,Zhang:2008by,Rui:2011dr} with no much change for the parameters of these meson states,
which also agree with the experimental measurements within errors. 
For $B^+ \to \eta^{(\prime)} \rho^+, \eta^{(\prime)} K^{\ast +}$ and $B^0 \to \eta^{(\prime)} K^{\ast 0}$ decays, the NLO PQCD results are smaller than the measured ones. 
The contribution from the considered NLO corrections can provide some enhancements to these modes 
than the previous PQCD predictions at LO \cite{Liu:2005mm}, but still not large enough to explain the data.  
More studies are required in these decays, especially the $U(1)$ anomaly problem and mixing uncertainty of flavour SU(3) singlet mesons $\eta^{(\prime)} $.
In the $\omega$ and $\phi$ involved channels, the NLO corrections play important role to explain the data.
Note that the mixing between $\omega$ and $\phi$ can  also change the result of $B \to K \phi$ and $B \to K \omega$ decay channels.

\begin{table}[t]
\vspace{-2mm}
\caption{The updated PQCD results for the ${\bf CP}$ asymmetries of $B\to PV$ decays (in unit of $10^{-2}$). }  
\label{tab:CPV-B2PV}           
\begin{center}
\begin{tabular}{l| r|r  r |r|r}
\toprule  
Mode     &    PQCD \quad\quad & \quad SCET1 \cite{Wang:2008rk} & \quad SCET2 \cite{Wang:2008rk} &
\quad QCDF \cite{Cheng:2009cn} \quad &   \quad\quad PDG \cite{Workman:2022ynf} \non
\hline
$B^+ \to \eta^\prime K^{\ast +}$  &$1.54^{+9.05+14.9}_{-8.16-9.74}$&$2.7^{+27.4}_{-19.5}$  &$2.6^{+26.7}_{-32.9}$ & $65.5^{+35.7}_{-63.9}$ & $-26 \pm 27$ \non
$B^+ \to \eta K^{\ast +}$  &$-34.5^{+2.5+0.9}_{-2.4-0.8}$&$-2.6^{+5.4}_{-5.5}$   &$-1.9^{+3.4}_{-3.6}$ & $-9.7^{+7.3}_{-8.0}$ & $2 \pm 6$ \non
$B^+ \to K^+ \omega$  &$31.5^{+0.6+0.1}_{-1.1-0.7}$&$11.6^{+18.2}_{-20.4}$    &$12.3^{+16.6}_{-17.3}$  & $22.1^{+19.6}_{-18.2}$ & $-2 \pm 4$ \non
$B^+ \to \pi^+ K^{\ast 0}$  &$-0.94^{+0.26+0.04}_{-0.29-0.03}$&$0$     &$0$  & $0.4^{+4.5}_{-4.2}$ & $-4 \pm 9$ \non
$B^+ \to \pi^0 K^{\ast +}$  &$-0.01^{+4.40+1.12}_{-4.87-1.26}$&$-17.8^{+30.4}_{-24.7}$   &$-12.9^{+12.0}_{-12.2}$  & $1.6^{+11.5}_{-4.2}$ & $-39 \pm 21$ \non
$B^+ \to K^+ \rho^0$  &$58.7^{+4.3+3.2}_{-4.0-2.8}$&$9.2^{+15.2}_{-16.1}$    &$16.0^{+20.5}_{-22.5}$  & $45.4^{+36.1}_{-30.2}$ & $37 \pm 10$ \non
$B^+ \to K^0 \rho^+$  
&$0.99^{+0.01+0.13}_{-0.01-0.18}$&$0$   &$0$  & $0.3^{+0.5}_{-0.3}$ & $-3 \pm 15$ \non
$B^+ \to K^+ \bar{K}^{\ast 0}$  &$21.3^{+6.2+1.2}_{-5.7-1.4}$&$-3.6^{+6.1}_{-5.3}$     &$-4.4^{+4.1}_{-4.1}$  & $-8.9^{+3.0}_{-2.6}$ & $12 \pm 10$ \non
$B^+ \to K^+ \phi$  &$-1.93^{+0.66+0.66}_{-0.60-0.42}$&$0$     &$0$  & $0.6^{+0.1}_{-0.1}$ & $2.4 \pm 2.8$ \\ 
\hline
$B^+ \to \pi^+ \phi$  &$0.0$&$\cdots$   &$\cdots$  & $0.0$ & $1 \pm 5$ \non
$B^+ \to \pi^+ \omega$  &$-29.8^{+0.5+1.1}_{-0.4-0.8}$&$0.5^{+19.1}_{-19.6}$   &$2.3^{+13.4}_{-13.2}$  & $-13.2^{+12.4}_{-10.9}$ & $-4 \pm 5$ \non
$B^+ \to \pi^+ \rho^0$  
&$14.9^{+0.4+0.5}_{-0.4-0.6}$&$-10.8^{+13.1}_{-12.7}$   &$-19.2^{+15.6}_{-13.5}$  & $-9.8^{+11.9}_{-10.5}$ & $0.9 \pm 1.9$ \non
$B^+ \to \pi^0 \rho^+$  &$-7.31^{+0.06+0.07}_{-0.02-0.03}$&$15.5^{+17.0}_{-19.0}$    &$12.3^{+9.4}_{-10}$  & $9.7^{+8.3}_{-10.8}$ & $2 \pm 11$ \non
$B^+ \to \eta^\prime \rho^+$  &$29.0^{+0.4+0.0}_{-0.4-0.1}$&$-19.8^{+66.6}_{-37.6}$   &$-21.7^{+135.9}_{-24.3}$  & $1.4^{+14.0}_{-11.9}$ & $26 \pm 17$ \non
$B^+ \to \eta \rho^+$  
&$-13.0^{+0.1+0.1}_{-0.1-1.5}$&$-6.6^{+21.5}_{-21.3}$   &$-9.1^{+16.7}_{-15.8}$  & $-8.5^{+6.5}_{-5.3}$ & $11 \pm 11$ \\
\hline
$B^0 \to \eta^\prime K^{\ast 0}$  &$12.4^{+0.1+0.5}_{-0.3-1.7}$&$9.6^{+9.0}_{-11}$    &$9.9^{+6.3}_{-4.4}$  & $6.8^{+34.9}_{-51.0}$ & $-7 \pm 18$ \non
$B^0 \to \eta K^{\ast 0}$  &$2.10^{+0.71+0.18}_{-0.55-0.21}$&$-1.1^{+2.3}_{-2.4}$   &$-0.7^{+1.2}_{-1.3}$  & $3.5^{+2.7}_{-2.5}$ & $19 \pm 5$ \non
$B^0 \to K^+ \rho^-$  &$54.3^{+0.6+0.8}_{-0.4-0.7}$&$7.1^{+11.2}_{-12.4}$     &$9.6^{+13.0}_{-13.5}$  & $31.9^{+22.7}_{-16.8}$ & $20 \pm 11$ \non
$B^0 \to \pi^- K^{\ast +}$  &$-14.1^{+6.0+2.9}_{-6.4-3.1}$&$-11.2^{+19.0}_{-16.3}$    &$-12.2^{+11.4}_{-11.3}$  & $12.1^{+12.6}_{-16.0}$ & $-27 \pm 4$ \non
$B^0 \to \pi^0 K^{\ast 0}$  &$-14.8^{+0.4+1.2}_{-0.1-1.5}$&$5.0^{+7.5}_{-8.4}$     &$5.4^{+4.8}_{-5.1}$  & $-10.8^{+9.3}_{-6.9}$ & $-15 \pm 13$ \non
$B^0 \to \pi^- \rho^+$  &$-0.59^{+0.18+0.72}_{-0.17-0.68}$&$11.8^{+17.5}_{-20}$   &$10.8^{+9.4}_{-10.2}$  & $4.4^{+5.8}_{-6.8}$ & $13 \pm 6$ \non
$B^0 \to \pi^+ \rho^-$  &$-30.9^{+0.1+1.7}_{-0.1-1.6}$&$-9.9^{+17.2}_{-16.7}$   &$-12.4^{+17.6}_{-15.3}$  & $-22.7^{+8.2}_{-4.5}$ & $-8 \pm 8$ \\ 
\hline
$B^0 \to K_S^0 \omega$  &$-5.29^{+0.83+0.21}_{-0.99-0.40}$&$5.2^{+8.0}_{-9.2}$      &$3.8^{+5.2}_{-5.4}$  & $-4.7^{+5.8}_{-6.0}$ & $C_{K_S^0 \omega}=0 \pm 40$ \non
 &$79.2^{+0.1+0.3}_{-0.2-0.2}$&$\cdots$     &$\cdots$  & $\cdots$ & $S_{K_S^0 \omega}=70 \pm 21$ \non
$B^0 \to K_S^0 \phi$  &$2.67^{+0.05+0.28}_{-0.18-0.26}$&$0$      &$0$  & $0.9^{+0.3}_{-0.1}$ & $C_{K_S^0 \phi}=1 \pm 14$ \non
 &$70.6^{+0.3+0.8}_{-0.3-0.1}$&$71\pm 1$  & $\cdots$    &$\cdots$  & $S_{K_S^0 \phi}=59 \pm 14$ \non
$B^0 \to K_S^0 \rho^0$  &$8.96^{+1.81+0.05}_{-1.48-0.01}$&$-6.6^{+11.6}_{-9.7}$     &$3.5^{+4.8}_{-4.8}$  & $8.7^{+8.8}_{-6.0}$ & $C_{K_S^0 \rho^0}=-4 \pm 20$ \non
 &$57.9^{+0.5+0.1}_{-0.4-0.0}$&$50^{+10}_{-6}$    &$\cdots$  & $\cdots$ & $S_{K_S^0 \rho^0}=50^{+17}_{-21}$ \non
$B^0 \to \pi^0 \rho^0$  &$66.1^{+2.1+3.9}_{-1.7-3.6}$&$-0.6^{+21.4}_{-21.9}$      &$-3.5^{+21.4}_{-20.3}$  & $\cdots$ & $C_{\pi^0 \rho^0}=-27 \pm 24$ \non
 &$-46.8^{+0.1+2.7}_{-0.8-3.5}$&$24^{+27}_{-22}$     &$\cdots$  & $\cdots$ & $S_{\pi^0 \rho^0}=-23 \pm 34$ \non
\toprule
\end{tabular} \end{center}
\end{table}%

In Table \ref{tab:CPV-B2PV}, we list the PQCD predictions for ${\bf CP}$ asymmetry parameters of $B \to PV$ decays. 
We do not show explicitly the result of  channels that have not been measured.
The NLO PQCD predictions agree with the available measured values, although there are still large experimental uncertainties. 
For comparison, we also show results of SCET and QCDF approach. 
It is easy to see that there are large differences between these approaches for many of the decay channels due to different source of strong phases in these approaches, 
which need further precision experimental data to resolve. 
With comparing the table with the table \ref{tab:CPV-B2PP}, 
we can see that the measured direct CP asymmetry of the $B\to PV$ decays are much larger than the corresponding $B\to PP$ decays. 
\begin{itemize}
\item
For the direct ${\bf CP}$ asymmetry of $B^+ \to K^+ \rho^0$ and $B^0 \to K^+ \rho^-$ decays, 
the PQCD predictions are larger than the measurements by $50 \%$, 
and the newly LHCb measurement $(15 \pm 2.2) \%$ \cite{LHCb:2022fxf} is more agree with the SCET result. 
Similar results also happen for the $B^+ \to K^+ \omega$ decay with a $30 \%$ ${\bf CP}$ asymmetry from PQCD calculation. 
\item
These large direct ${\bf CP}$ asymmetry prediction in theories agree with the current experimental data within uncertainties.
Unfortunately, there is difficulty for the experiments to measure these kinds of decays precisely, 
since the vector meson (immediately decayed two pseudoscalar mesons) in the final state is not directly measurable by experiments. 
The measured three body $B$ meson decays contains direct three-body $B$ decays together with the intermediate $B\to PV$ decays, which are difficult to resolve. 
Recently, the experimenters give large direct ${\bf CP}$ asymmetry in the three-body $B$ decays in some regions of the Dalitz plot, 
which provides strong evidence of large CP asymmetry in $B\to PV$ decays.
\item
There is large direct CP asymmetry for the rare decay channels with branching ratios one order magnitude smaller than other channels, 
like $B^+ \to K^+{\bar K}^{\ast 0}$ and $B^0 \to \pi^0\rho^0$ decays, in which many of those suppressed contributions also play important role. 
The theoretical calculation of ${\it CP}$ asymmetry is significantly affected by the NLO corrections and 
newly added power corrections (${\cal O}(x_1)$ and ${\cal O}(m_b/m_B)$) in this work. 
Note that they may change significantly again if a complete NLO correction is included \cite{Zhang:2008by,Rui:2011dr}. 
\item
For the decays $B^0 /\bar B^0\to \pi^+ \rho^-/ \pi^- \rho^+$ as discussed in the beginning of this subsection, 
they require the experimental measurements of the time dependence of all the four decay channels. 
The CP asymmetry analysis is quite complicated as discussed in Ref \cite{Rui:2011dr}. 
We show only two of the six ${\bf CP}$ asymmetry parameters of these decays.
\end{itemize}

\subsection{\texorpdfstring{$B \to VV$}{} decay modes}\label{sec:b2vv}

The two body $B$ decays with two vector mesons in the final states are more complicated than the $B\to PP$ and $B\to PV$ decays. 
As shown in the last section, only the vertex correction, quark loop and magnetic penguin contributions for $B\to VV$ decays are available \cite{Li:2006cva}. 
Although all other NLO corrections are still missing, we include here results of all $B\to VV$ decays with the current known NLO corrections. 

\begin{table}[t]
\vspace{-2mm}
\caption{The updated PQCD results for the branching ratios of $B \to VV$ decays (in unit of $10^{-6}$).} 
\label{tab:BR-B2VV}   
\begin{center}
\begin{tabular}{l|r|r|r|r}
\toprule
Mode    & \quad\quad\quad\quad  PQCD \quad\quad   & \quad\quad SCET \cite{Wang:2017rmh} &
\quad QCDF \cite{Cheng:2009cn,Beneke:2006hg}  & \quad PDG \cite{Workman:2022ynf} \non
\hline
$B^+ \to \rho^+ K^{\ast 0}$  &$9.40^{+1.43+1.05}_{-1.34-0.95}$&$ 8.93\pm 3.18 $& $9.2^{+3.8}_{-5.5}$ & $9.2 \pm 1.5$ \non
$B^+ \to \rho^0 K^{\ast +}$  &$6.25^{+1.12+0.59}_{-0.84-0.53}$&$4.64 \pm 1.37 $& $5.5^{+1.4}_{-2.5}$ & $4.6 \pm 1.1$ \non
$B^+ \to \omega K^{\ast +}$  &$5.48^{+1.52+0.81}_{-1.36-0.66}$&$5.56\pm 1.60 $& $3.0^{+2.5}_{-1.5}$ & $ < 7.4$ \non
$B^+ \to \phi K^{\ast +}$  &$12.3^{+1.7+1.5}_{-1.4-1.4}$&$9.86 \pm 3.39 $& $10.0^{+12.4}_{-3.5}$ & $10.2 \pm 2.0$ \non
$B^+ \to K^{\ast +} \bar{K}^{\ast 0}$  &$0.66^{+0.12+0.09}_{-0.09-0.08}$&$0.52\pm 0.18 $& $0.6^{+0.3}_{-0.3}$ & $0.91 \pm 0.29$  \non
$B^+ \to \rho^0 \rho^+$  &$14.0^{+4.1+0.4}_{-3.0-0.4}$&$ 22.1\pm 3.7  $& $20.06^{+4.5}_{-2.1}$ & $24.0 \pm 1.9$ \non
$B^+ \to \rho^+\omega$  &$10.9^{+2.8+1.0}_{-2.1-0.9}$&$ 19.2 \pm 3.1 $& $16.9^{+3.6}_{-1.8}$ & $15.9 \pm 2.1$ \non
$B^+ \to \rho^+\phi$  &$0.042^{+0.011+0.004}_{-0.008-0.003}$&$0.005\pm 0.001 $& $\cdots$ & $< 3.0$ \non
\hline
$B^0 \to \rho^- K^{\ast +}$  &$8.72^{+1.27+0.97}_{-0.96-0.87}$&$ 10.6\pm 3.2 $& $8.9^{+4.9}_{-5.6}$ & $10.3 \pm 2.6$ \non
$B^0 \to \rho^0 K^{\ast 0}$  &$3.37^{+0.38+0.43}_{-0.29-0.39}$&$5.87\pm 1.87 $& $4.6^{+3.6}_{-3.6}$ & $3.9 \pm 1.3$ \non
$B^0 \to \omega K^{\ast 0}$  &$5.93^{+0.89+1.74}_{-0.73-1.55}$&$3.82\pm 1.39 $& $2.5^{+2.5}_{-1.5}$ & $2.0 \pm 0.5$ \non
$B^0 \to \phi K^{\ast 0}$  &$11.8^{+1.6+1.5}_{-1.3-1.5}$&$9.14\pm 3.14  $ & $10.0 \pm 0.5$   &$\cdots$  \non
$B^0 \to K^{\ast 0} \bar{K}^{\ast 0}$  &$0.38^{+0.09+0.02}_{-0.06-0.01}$&$0.48\pm 0.16 $& $0.6^{+0.2}_{-0.3}$ & $0.83 \pm 0.24$ \non
$B^0 \to K^{\ast +}K^{\ast -}$  &$0.17^{+0.02+0.05}_{-0.02-0.03}$&$\cdots $& $0.16^{+0.1}_{-0.1}$ & $< 2.0$ \non
$B^0 \to \rho^+ \rho^-$  &$22.7^{+6.3+0.6}_{-4.8-0.6}$&$ 27.7\pm 4.1 $  & $25.5^{+2.8}_{-3.0}$& $27.7 \pm 1.9$  \non
$B^0 \to \rho^0 \rho^0$  &$0.54^{+0.16+0.04}_{-0.11-0.04}$&$1.00\pm 0.29$& $0.9^{+1.9}_{-0.5}$ & $0.96 \pm 0.15$ \non
$B^0 \to \rho^0 \omega $  &$0.76^{+0.13+0.14}_{-0.11-0.12}$&$0.59\pm 0.19 $& $0.08^{+0.36}_{-0.02}$ & $< 1.6$ \non
$B^0 \to \rho^0 \phi$  &$0.019^{+0.005+0.002}_{-0.004-0.001}$&$\sim 0.002 $& $\cdots$ & $< 3.3$ \non
$B^0 \to \omega\omega$  &$1.21^{+0.24+0.31}_{-0.19-0.24}$&$0.39\pm 0.13 $& $0.7^{+1.1}_{-0.4}$ & $1.2 \pm 0.4$ \non
$B^0 \to \omega \phi$  &$0.018^{+0.005+0.005}_{-0.004-0.005}$&$\sim 0.002 $& $\cdots$ & $< 0.7$\non
$B^0 \to \phi\phi$  &$0.029^{+0.002+0.006}_{-0.002-0.006}$&$\cdots $& $\cdots$ & $< 0.027$ \\
\toprule
\end{tabular} \end{center}
\end{table}

In Tables \ref{tab:BR-B2VV}, we list the updated PQCD predictions for the branching ratios, together with the SCET and QCDF results, as well as the experiment data. 
For most $B\to VV$ decays, the NLO PQCD predictions agree well with the measured values within errors. 
Since the vector meson is not directly measurable by experiments, the $B\to VV$ decays are measured as four or five-body decays in experiments. 
Again, there is much larger systematic uncertainty in these decays than the corresponding $B\to PP$ decays for experiments due to the interference between different resonances and resonance with continuum.  
Theoretically, it is expected that the branching ratios of $B\to VV$ decays should be much larger than the corresponding $B\to PP$ decays, 
since the decay constant of vector meson is larger than that of the pseudoscalar meson and the transverse polarization provides additional contribution. 
This argument stands for most of the $B$ meson decays, except for the $B\to \pi\pi$ and $B\to \rho \rho$ decays. 
It is easy to see from tables \ref{tab:BR-B2PP} and \ref{tab:BR-B2VV} that the PQCD branching ratio of $B^0\to \rho^0 \rho^0$ decay 
is two times larger than that of $B^0\to \pi^0\pi^0$ decay, while experimentally it is the inverse case, which is the long-standing $\pi\pi$ puzzle. 
In fact, the experimental data for the $B^0\to \rho^0 \rho^0$ decay are not consistent between the two B factories \cite{Zou:2015iwa}.
Because of the isospin symmetry between the decay amplitudes for $B \to \rho^0\rho^0$, $\rho^0\rho^+$ and $\rho^+\rho^-$, 
and the smallness of $B^0 \to \rho^0\rho^0$ decay branching ratio, 
one generally expect a relation of ${\cal B}(B^0 \to \rho^+ \rho^-) \approx 2 {\cal B}(B^+ \to \rho^0 \rho^+)$.  
From table \ref{tab:BR-B2VV} one can easily see that theoretically the isospin triangle is hold at a very good precision 
while the experimental data shows ${\cal B}(B^0 \to \rho^+ \rho^-)\approx 1.5 {\cal B}(B^+ \to \rho^0 \rho^+)$. 
There should be a serious problem in these experimental measurements, since no one expects that a new physics violates the QCD isospin symmetry.
 The considered power corrections of ${\cal O}(x_1)$ and ${\cal O}(m_b/m_B)$ in this work 
do not bring measurable change to the previous PQCD predictions of the branching ratios at LO \cite{Zou:2015iwa}. 
For most $B\to VV$ decays, the variations induced by the inclusion of the known partial NLO contributions are small or moderate. 
For the rare decays $B^+ \to \rho^+\phi$, $B^0\to \rho^0\rho^0, \rho^0\omega, \rho^0\phi, \omega\omega, \omega\phi$, however, 
the enhancements can be as large as 100 percents.

\begin{table}
\vspace{-2mm}
\caption{The updated PQCD results for the ${\bf CP}$ asymmetries of $B\to VV$ decays (in unit of $10^{-2}$).}
\label{tab:ACP-B2VV}       
\begin{center}
\begin{tabular}{l| r|r|r|r} \toprule
Mode \quad & \quad\quad\quad PQCD \quad\quad & \quad\quad SCET  \cite{Wang:2017rmh} &
\quad QCDF \cite{Cheng:2009cn,Beneke:2006hg} & \quad PDG \cite{Workman:2022ynf} \non 
\hline
$B^+ \to \rho^+ K^{\ast 0}$  &$0.58^{+0.13+0.16}_{-0.12-0.18}$&$-0.56 \pm 0.61 $& $-0.3^{+2}_{-1}$ & $ -1 \pm 16$ \non
$B^+ \to \rho^0 K^{\ast +}$  &$30.6^{+0.5+0.1}_{-0.7-0.2}$&$29.3\pm 31.0 $& $43^{+13}_{-28}$ & $ 31 \pm 13$ \non
$B^+ \to \omega K^{\ast +}$  &$43.0^{+1.7+3.8}_{-2.0-3.2}$&$24.3\pm 27.1 $ & $ 29 \pm 35$   &  $\cdots$ \non
$B^+ \to \phi K^{\ast +}$  &$2.40^{+0.14+0.13}_{-0.14-0.10}$&$-0.39\pm 0.44 $& $0.05$ & $-1 \pm 8$ \non
$B^+ \to K^{\ast +} \bar{K}^{\ast 0}$  &$-26.8^{+2.3+1.0}_{-2.4-2.0}$&$9.5 \pm 10.6 $&$\cdots$ & $\cdots$ \non
$B^+ \to \rho^0 \rho^+$  &$0.03^{+0.00+0.00}_{-0.01-0.00}$&$0.0 $& $0.06$ & $ -5 \pm 5$ \non
$B^+ \to \rho^+\omega$  &$-25.9^{+1.8+1.3}_{-1.9-1.2}$&$-13.6 \pm 16.1 $& $-8^{+3}_{-4}$ & $ -20 \pm 9$ \non
$B^+ \to \rho^+\phi$  &$0.0$&$0.0$&$\cdots$ &  $\cdots$ \non
\hline
$B^0 \to \rho^- K^{\ast +}$  &$32.4^{+0.1+0.1}_{-0.1-0.2}$&$20.6\pm 23.3 $& $32^{+2}_{-14}$ & $ 21 \pm 15$ \non
$B^0 \to \rho^0 K^{\ast 0}$  &$-14.4^{+1.2+0.9}_{-1.4-1.0}$&$-3.30\pm 3.91 $& $-15 \pm 16$ & $-6 \pm 9$ \non
$B^0 \to \omega K^{\ast 0}$  &$9.89^{+0.96+1.59}_{-0.80-1.12}$&$3.66 \pm 4.05 $&$23^{+10}_{-18}$& $45 \pm 25$  \non
$B^0 \to \phi K^{\ast 0}$  &$0.86^{+0.06+0.07}_{-0.06-0.06}$&$-0.39 \pm 0.44$ & $0.8^{+0.4}_{-0.5}$ & $0 \pm 4$ \\ 
\hline
$B^0 \to \rho^+ \rho^-$  &$-1.85^{+0.20+0.01}_{-0.11-0.00}$&$-7.68\pm 9.19 $& $11^{+11}_{-4}$ & $C_{\rho^+\rho^-} = 0 \pm 9 $ \non
 &$-12.7^{+0.1+0.4}_{-0.1-0.3}$&$\cdots$& $-19^{+9}_{-10}$ & $S_{\rho^+\rho^-} = -14 \pm 13 $ \non
$B^0 \to \rho^0 \rho^0$  &$74.6^{+1.3+1.9}_{-1.9-2.3}$&$19.5\pm 23.5 $& $-53^{+26}_{-54}$ & $C_{\rho^0\rho^0} = 20 \pm 90$ \non
 &$1.38^{+0.74+2.15}_{-0.03-1.93}$&$\cdots$& $16^{+50}_{-49}$ & $S_{\rho^0\rho^0} = 30 \pm 70$ \non
\toprule
\end{tabular} \end{center} \end{table}
 
In Table \ref{tab:ACP-B2VV}, we list the updated PQCD predictions for the ${\bf CP}$ asymmetry parameters ${\cal A}_{{\rm CP}}$, $C$, $S$ 
of two-body charmless $B \to VV$ decays. 
Although some of the $B\to VV$ decays are predicted with large CP asymmetry, 
there is not any experimental measurement with signal significance more than $3\sigma$. 
The reason for this difficulty in experiment is that there are at least four mesons in the final states to measure in these decays \cite{Rui:2021kbn,Li:2021qiw}, 
and hence the interference between different resonances and interference between resonance and continuum are more complicated than the $B\to PV$ decays. 
Therefore more experimental data and more precision measurements of angular distributions are needed for the $B \to VV$ decays.
At the theoretical side, the inclusion of the known NLO contribution can change the CP asymmetry parameters more significantly than the branching ratios. 
For example, the sign of the  ${\cal A}_{\rm CP} (\rho^+ K^{\ast 0}), {\cal A}_{\rm CP} (K^{\ast +} \bar{K}^{\ast 0})$ and $S_{\rho^0\rho^0}$ 
is changed by the NLO corrections in PQCD approach. 
In fact, as discussed in the previous section, less NLO corrections are finished for vector meson final states than pseudo-scalar meson final states. 
More efforts are needed in the $B\to VV$ decays in the theoretical improvement of NLO corrections.

\begin{table}[t]
\vspace{-2mm}
\caption{The updated PQCD results for the longitudinal polarisation fractions $f_L$ of $B\to VV$ decays (in unit of percentage).} \
\label{tab:PFractions-B2VV} 
\begin{center}
\begin{tabular}{l | r | r | r | r | r}
\toprule
Mode \quad &  \quad PQCD$_{\rm LO}$ \cite{Zou:2015iwa} & \quad\quad  PQCD \quad & \quad\quad SCET  \cite{Wang:2017rmh} \quad &
\quad QCDF \cite{Cheng:2009cn,Beneke:2006hg} & \quad  HFLAV \cite{Amhis:2019ckw} \non \hline  
$B^+ \to \rho^+ K^{\ast 0}$   \quad &$70.0\pm 5.0$  &$76.6^{+1.5}_{-1.4}$ &$45.0\pm 18.0$ & $48.0^{+52.0}_{-40.0}$  & $48 \pm 8$ \non
$B^+ \to \rho^0 K^{\ast +}$    \quad &$75.0^{+4.0}_{-5.0}$  &$80.0^{+1.5}_{-1.5}$ &$42.0\pm 14.0$ & $67.0^{+31.0}_{-48.0}$ & $78 \pm 12$ \non
$B^+ \to \omega K^{\ast +}$    \quad &$64.0\pm 7.0$  &$77.4^{+0.5}_{-0.9}$ &$53.0\pm 14.0$  & $67.0^{+32.0}_{-39.0}$ & $41\pm 19$ \non
$B^+ \to \phi K^{\ast +}$   \quad &$57.0^{+6.3}_{-5.9}$  &$68.7^{+1.3}_{-1.5}$ &$51.0\pm 16.4 $ & $49.0^{+51.0}_{-43.0}$ & $50 \pm 5$ \non
$B^+ \to K^{\ast +} \bar{K}^{\ast 0}$    \quad &$74.0 \pm 7.0$  &$82.4^{+1.1}_{-1.1}$ &$50.0 \pm 16.0$ & $45.0^{+55.0}_{-38.0}$ & $82^{+15}_{-21}$\non
$B^+ \to \rho^0 \rho^+$    \quad &$98.0 \pm 1.0$  &$96.9^{+0.1}_{-0.1}$ &$\sim 100$ & $96.0 \pm 2.0$ & $95 \pm 1.6$ \non
$B^+ \to \rho^+\omega$   \quad &$97.0 \pm 1.0$ &$96.3^{+0.3}_{-0.4}$ &$97.0 \pm 1.0$ & $96.0^{+2.0}_{-3.0}$ & $90 \pm 6$ \non
$B^+ \to \rho^+\phi$    \quad &$95.0 \pm 1.0$ &$81.3^{+1.9}_{-1.8}$ &$\sim 100$ & $\cdots$ & $\cdots$ \non
\hline
$B^0 \to \rho^- K^{\ast +}$   \quad &$68.0^{+5.0}_{-4.0}$  &$75.7^{+1.5}_{-1.4}$  &$55\pm 14$ & $53.0^{+45.0}_{-32.0}$ & $38 \pm 13$ \non
$B^0 \to \rho^0 K^{\ast 0}$   \quad &$65.0^{+4.0}_{-5.0}$  &$71.0^{+1.5}_{-1.3}$ &$61.0\pm 13.0$ & $39.0^{+60.0}_{-31.0}$ & $17.3 \pm 2.6$ \non
$B^0 \to \omega K^{\ast 0}$   \quad &$65.0 \pm 5.0$ &$77.7^{+0.4}_{-0.9}$ &$40.0\pm 20.0 $ & $58.0^{+44.0}_{-17.0}$ & $69 \pm 11$  \non
$B^0 \to \phi K^{\ast 0}$    \quad &$56.5^{+5.8}_{-5.9}$  &$69.5^{+1.2}_{-1.5}$ &$51.0 \pm 16.4 $ & $50.0^{+51.0}_{-44.0}$ & $49.7 \pm 1.7$ \non
$B^0 \to K^{\ast 0} \bar{K}^{\ast 0}$    \quad &$58.0\pm 8.0$ &$68.8^{+5.3}_{-5.3}$ &$50.0\pm 16.0 $ & $52.0^{+48.0}_{-49.0}$  & $74 \pm 5$ \non
$B^0 \to K^{\ast +}K^{\ast -}$   \quad &$\sim 100.0$  &$\sim 100.0$ &$\cdots $ & $\sim 100.0$  & $\cdots$ \non
$B^0 \to \rho^+ \rho^-$   \quad  &$95.0 \pm 1.0$ &$93.8^{+0.1}_{-0.1}$ &$99.1\pm 0.3$ & $92.0^{+1.0}_{-3.0}$ & $99.0^{+2.1}_{-1.9}$ \non
$B^0 \to \rho^0 \rho^0$   \quad &$12.0^{+16.0}_{-2.0}$  &$80.9^{+1.9}_{-1.9}$ &$87.0\pm 5.0 $ & $92.0^{+7.0}_{-37.0}$  & $71^{+8}_{-9}$ \non
$B^0 \to \rho^0 \omega $   \quad &$67.0^{+8.0}_{-9.0}$  &$74.2^{+0.1}_{-0.1}$  &$58.0\pm 14.0 $ & $52.0^{+12.0}_{-44.0}$  & $\cdots$ \non
$B^0 \to \rho^0 \phi$   \quad &$95.0 \pm 1.0$ \quad  &$81.3^{+1.9}_{-1.8}$ &$\sim 100$ & $\cdots$ & $\cdots$\non
$B^0 \to \omega\omega$   \quad &$66.0^{+10.0}_{-11.0}$  &$88.4^{+0.9}_{-0.8}$  &$64.0\pm 15.0 $  & $94.0^{+4.0}_{-20.0}$  & $\cdots$\non
$B^0 \to \omega \phi$   \quad &$94.0^{+2.0}_{-3.0}$   &$80.8^{+0.8}_{-1.4}$  &$\sim 100$ & $\cdots$ & $\cdots$\non
$B^0 \to \phi\phi$    \quad &$97.0 \pm 1.0$ &$99.9^{+0.0}_{-0.0}$ &$\cdots $ & $\cdots$ & $\cdots$ \\
\toprule
\end{tabular}
\end{center}
\end{table}%

Besides the decay width, the fraction of a given polarisation state of the hadronic $B \to VV$ decays is also an interesting observable, 
because the studies of these physical quantities offer more opportunities for our understanding of the mechanism for hadronic weak decays and ${\bf CP}$ asymmetries. 
In Table \ref{tab:PFractions-B2VV}, we list the updated PQCD predictions for the longitudinal polarisation fractions $f_L$ of two-body charmless $B \to VV$ decays, 
together with results calculated in SCET and QCDF approach, as well as the experiment data.
In fact, the systematic uncertainty of these experimental studies cancel in the ratios, 
thus it makes the experimental measurements more reliable than the branching ratios and the ${\bf CP}$ asymmetry parameter in the $B\to VV$ decays, 
which can be seen from the comparison of the experimental data in Tables \ref{tab:BR-B2VV}, \ref{tab:ACP-B2VV} and \ref{tab:PFractions-B2VV}. 
From the tables one see that the NLO QCD corrections and the newly added two power suppressed terms can help us to explain the longitudinal polarisation fraction
as well as the branching fraction of the color suppressed $B^0 \to \rho^0\rho^0$ decay.
For the penguin dominant decays, we note that the LO PQCD results of the polarization fractions have better agreement with the current experimental data 
than the partly NLO corrected PQCD results. 
The reason is very simple, the included NLO corrections only enhance the longitudinal polarisation in $B\to VV$ decays, 
making the branching ratios larger than the LO ones, which can be seen explicitly in Table \ref{tab:PFractions-B2VV} and Ref. \cite{Zou:2015iwa}.  
A more complete NLO result together with an update of LCDAs of vector mesons are needed.

It is expected that the size of transverse polarization is suppressed by $m_V/m_B$ 
in naive power counting comparing with the longitudinal polarization in $B\to VV$ decays \cite{Kagan:2004uw}. 
The large transversal polarisation fractions observed in the penguin-dominated $B \to VV$ modes, 
such as in $B \to \rho K^\ast, \omega K^\ast, \phi K^\ast$ decays, are therefore challenging for the the QCDF approach \cite{Kagan:2004uw}. 
In the framework of PQCD, the large transverse polarization fraction is interpreted on the basis of the chirality enhanced annihilation diagrams, 
particularly on the ${\rm (S-P) \times (S+P)}$ QCD penguin operator, 
where the light quarks in the final states are not produced through chiral currents 
and hence the transversal polarisation state is not suppressed by helicity flip \cite{Zou:2015iwa}.  
In contrast to the PQCD approach, where the annihilation amplitudes can be perturbative calculated, later QCDF calculation introduces plural annihilation parameters
to mitigate the troublesome endpoint divergence, and fits to the existing data of branching ratios \cite{Cheng:2009cn}. 
More mechanisms such as the possible larger electroweak penguin contributions are added in the QCDF approach 
to make the transverse polarizations larger to match the experimental data \cite{Beneke:2006hg}.

More physical observables, such as the perpendicular polarization fraction ($f_\perp$), 
the relative phase ($\varphi_\parallel, \, \varphi_\perp, \, \delta_0$) and the helicity ${\bf CP}$ asymmetry parameters ($\acp^0, \, \acp^\perp$) etc. in $B
\to  VV$ decays have been discussed in Ref.  \cite{Zou:2015iwa}. 
Since the complete NLO corrections for this kinds of decays are not available, we do no upgrade these variables in this review. 

\section{Conclusion}\label{sec:conclusion}

Encouraged by the upgrade of LHCb and the physical running of Belle-II 
which would improve greatly the accuracy of experiment measurements of most $B$ meson decay modes,
more precision theoretical investigation of the high order QCD corrections are also in progress. 
Regulating the endpoint divergence by the transverse momentum of quarks in the propagators, 
one can do the perturbation calculation for kinds of diagrams in the PQCD factorization approach based on $k_T$ factorization, including the annihilation type diagrams. 
We summarize the comprehensive study of the two-body charmless $B \to PP, PV, VV$ decays in the PQCD factorization approach up to the NLO.
The vertex corrections, quark loop diagrams and the chromo-magnetic operator contributions have been considered in all of these decays. 
The NLO QCD corrections to the naive factorization type diagrams of $B\to P$ transition and corrections to the timelike pion form factors are also included. 
In short, most of the NLO QCD corrections for the naive factorization diagrams have been finished for the $B\to PP$ decays, 
leaving only the NLO corrections for hard scattering diagrams. 
On the other hand, the NLO corrections for the $B\to VV$ decay channels are least to know, leaving a bad precision.  

Because of the complication of the NLO calculations, different $B$ decay channels are calculated by different authors at a different time. 
Unavoidably, slightly different non-perturbative parameters are used by different papers for different decay channels. 
In this review, we take into account the latest determination of nonperturbative parameters in the hadron distribution amplitudes, 
the currently known NLO QCD corrections, as well as the newly two power suppressed terms in decaying amplitude 
which are proportional to the momentum fraction of light (anti-) quark in $B$ meson and to the ratio $m_b/m_B$ respectively. 
We also use the most recent values of those relevant CKM matrix elements and weak phases in the numerical calculations.
As a result, the branching ratios and the ${\bf CP}$ asymmetry parameters of all considered $B \to PP, PV, VV$ decays, 
summarized in this review are a little different from the existing NLO PQCD predictions for these decays.
For the $B \to VV$ decays, the PQCD predictions of the longitudinal polarisation fractions $f_L$ are also shown.
The major theoretical errors come from the input hadronic parameters, 
such as those from the uncertainties of $\omega_B$ and the Gegenbauer moments.
Other sources of small theoretical errors, such as the combined uncertainty in the CKM matrix elements, are not included.

For most considered B meson decay channels, 
the PQCD results of branching ratios agree well with other approaches and the experimental measurements within errors 
after the inclusion of all currently known NLO contributions and the power suppressed terms. 
The PQCD predictions for ${\bf CP}$ asymmetry parameters disagree with the SCET and QCDF predictions for many of the decay channels, 
but have a better agreement with the measured ones.
The longstanding $K \pi $ puzzle about the pattern of the direct ${\bf CP}$ asymmetries of the penguin-dominated $B \to K \pi$ decays 
can be understood directly after the inclusion of the NLO contributions.
The newly added two power suppressed terms play an important role in the $B \to \eta P$ decay modes,
and also affect significantly the PQCD predictions of the ${\bf CP}$ asymmetries in rare decay channels like $B^+ \to K^+ {\bar K}^{\ast 0}$ and $B^0 \to \pi^0\rho^0$.
Nevertheless, their contributions to other decays are negligible.

In the PQCD factorization approach, the NLO contributions are only partially known at present. 
For the $B\to VV$ decays, least NLO QCD corrections are calculated, 
making many of the longitudinal polarization fraction of these decays do not agree well with the experimental data.
The still missing NLO contributions, such as those related with $B \to V$ transition form factors, 
hard scattering diagrams and the annihilation diagrams, should be calculated as soon as possible,
in order to meet the requirement for precision test of the standard model and to find possible signal or evidence of the new physics. 
The meson LCDAs as the main nonperturbative parameters in this approach, absorb the infrared divergences in the factorization of the QCD corrections order by order. 
They should be changed after each NLO corrections included. 
The precise parameters should be determined by a global fit of the experimental data after a complete NLO result available. 
 
\section*{Acknowledgements}

This work is partly supported by the National Science Foundation of China  under Grants No. 11805060, 11975112, 12105028, 12070131001 and 11775117,
and the Joint Large-Scale Scientific Facility Funds of the NSFC and CAS under Contract No. U1932110.
This work is also partly supported by the Natural Science Foundation of Jiangsu Province under Grants No. BK20200980 and the National Key Research
and Development Program of China under Contract No. 2020YFA0406400.

\begin{appendix}

\section{The hard functions and Sudakov factors}\label{app-hardfunction}

At leading order, the hard functions $h_i$ in decay amplitudes are 
\beq
h_i(x_1,x_2,(x_3),b_1,b_2) = h_{1}(\beta_i,b_2)\times h_{2}(\alpha_i,b_1,b_2) \,, \quad i \in \{e,a,ne,na,{na}^\prime\} 
\eeq
with the multiplicative functions 
\beq
&&h_{1}(\beta,b_2) = \left\{\begin{array}{ll}
K_0( \beta b_2), & \quad  \quad \beta^2 >0\\
\frac{i\pi}{2}H_0^{(1)}(\sqrt{-\beta^2}b_2),& \quad  \quad \beta^2<0
\end{array} \right.\\
&&h_{2}(\alpha,b_1,b_2) = \left\{\begin{array}{ll}
\theta(b_2-b_1)I_0(\alpha b_1)K_0( \alpha b_2)+(b_1\leftrightarrow b_2), & \quad   \alpha^2 >0\\
\frac{i\pi}{2}\theta(b_2-b_1)J_0(\sqrt{-\alpha^2}b_1)H_0^{(1)}(\sqrt{-\alpha^2}b_2)+(b_1\leftrightarrow b_2),& \quad   \alpha^2<0
\end{array} \right.
\eeq
Here $J_0$ is the Bessel function, $K_0$ and $I_0$ are the modified Bessel functions, $N_0$ is the Neumann function, 
and $H_0$ is the Hankel function of the first kind with relation $H_0^{(1)}(x)=J_0(x)+iN_0(x)$.
In different types of decay amplitudes, the $x_i$ dependent integrated variables $\alpha_i, \beta_i$ are different, which are arranged as 
\beq
&&\left(\alpha_e\right)^2 = x_3 m_B^2 \,, \quad \left(\alpha_{e^\prime}\right)^2 = x_1 m_B^2 \,, \quad
\left(\beta_e \right)^2=\left(\beta_{e^\prime} \right)^2= x_1 x_3 m_B^2 \,,\\
&&\left(\alpha_{ne}\right)^2 =\left(\alpha_{ne^\prime}\right)^2= x_1 x_3 m_B^2 \,, \quad
\left(\beta_{ne}\right)^2 = x_3 \left( x_1 - \bar{x}_2 \right) m_B^2 \,, \quad \left(\beta_{ne^\prime}\right)^2 = x_3 \left( x_1 - x_2 \right) m_B^2 \,,\non
&&\left(\alpha_a\right)^2 = - \bar{x}_3 m_B^2 \,, \quad \left(\alpha_{a^\prime}\right)^2 = -x_2 m_B^2 \,, \quad
\left(\beta_a\right)^2 =\left(\beta_{a^\prime}\right)^2= - \bar{x}_3 x_2 m_B^2 \,,\non
&&\left(\alpha_{na}\right)^2 = \left(\alpha_{na^\prime}\right)^2  = - \bar{x}_3 x_2 m_B^2 \,, \quad 
\left(\beta_{na}\right)^2 = [1+x_3\left( x_2-\bar{x}_1 \right)] m_B^2 \,,   \quad \left(\beta_{na^\prime}\right)^2 = \bar{x}_3 \left( x_1 -x_2 \right) m_B^2 \,. \nonumber
\label{eq:alpha-beta}
\eeq

The hard function in the decay amplitude of the chromomagnetic penguin correction is
\beq
&&h_g(\alpha_{g^{(\prime)}},\beta_{g^{(\prime)}}, \gamma_{g^{(\prime)}},b_1,b_2,b_3) = -K_0(\beta_{g^{(\prime)}} b_1) K_0(\gamma_{g^{(\prime)}} b_2) \non
&& \hspace{3cm} \int_0^{\pi/2} d \theta \, \tan\theta \, J_0(\alpha_{g^{(\prime)}}b_1\tan\theta)J_0(\alpha_{g^{(\prime)}} b_2\tan\theta)J_0(\alpha_{g^{(\prime)}} b_3\tan\theta)  \,
\label{eq:hg}
\eeq
with the inner virtualities
\beq
&~&\left(\alpha_q \right)^2 = x_3 m_B^2, \,\quad
\left(\beta_q \right)^2  = x_1x_3m_B^2, \, \quad
\left(\gamma_q\right)^2 = \bar{x}_3 x_2 m_B^2 \,, \non
&~& \left(\alpha_{q^\prime} \right)^2 = x_1m_B^2, \,\quad     \left(\beta_{q^\prime} \right)^2 = \left(\beta_g\right)^2 , \,\quad 
\left(\gamma_{q^\prime} \right)^2 = \left( x_2-x_1 \right) m_B^2 \,.
\label{eq:hg-scale}
\eeq

The Sudakov factors as well as the strong coupling in the $E$ functions inside the integration of various decay amplitudes are
\beq
&&E_e(\mu) = \alpha_s(\mu){\rm exp}[-S_B(\mu)-S_3(\mu)]S_t(x_3)\,, \non
&&E_{e^{\prime}}(\mu)=\alpha_s(\mu){\rm exp}[-S_B(\mu)-S_3(\mu)]S_t(x_1) \,,   \non
&&E_{ne}(\mu)=\alpha_s(\mu){\rm exp}[-S_B(\mu)-S_2(\mu)-S_3(\mu)]\vert_{b_3=b_1}  \,,   \non
&&E_a(\mu)=\alpha_s(\mu){\rm exp}[-S_2(\mu)-S_3(\mu)]S_t(x_3)\,, \non
&&E_{a^{\prime}}(\mu)=\alpha_s(\mu){\rm exp}[-S_2(\mu)-S_3(\mu)]S_t(x_2)\,,   \non
&&E_{na}(\mu)=\alpha_s(\mu){\rm exp}[-S_B(\mu)-S_2(\mu)-S_3(\mu)]|_{b_3=b_2}  \,, \non
&&E_{q}(\mu)=\alpha_s(\mu)^2 {\cal C}^{(q)}(\mu, l^2) {\rm exp}[-S_B(\mu) -S_3(\mu)]S_t(x_3)   \,, \non
&&E_{q^\prime}(\mu)=\alpha_s(\mu)^2 {\cal C}^{(q)}(\mu, l^2) {\rm exp}[-S_B(\mu) -S_3(\mu)]S_t(x_1)   \,, \non
&&E_{8g}(\mu)=\alpha_s(\mu)^2 {\cal C}_{\rm 8g}^{eff}(\mu) {\rm exp}[-S_B(\mu)-S_2(\mu)-S_3(\mu)]S_t(x_3)   \,,\non
&&E_{8g^\prime}(\mu)=\alpha_s(\mu)^2 {\cal C}_{\rm 8g}^{eff}(\mu) {\rm exp}[-S_B(\mu)-S_2(\mu)-S_3(\mu)]S_t(x_1)  \,.
\eeq
The renormalization scale $\mu$ in the above functions are taken at the largest virtualities in each decay amplitude in order to suppress the high order contributions, 
which are chosen as
\beq
&&\mu_e = {\rm Max}\{|\alpha_e|,|\beta_e|,1/b_1,1/b_3\}\,,\quad
\mu_{e^{\prime}} = {\rm Max}\{|\alpha_{e^{\prime}}|,|\beta_{e^{\prime}}|,1/b_3,1/b_1\}\,,  \\
&&\mu_{ne} = {\rm Max}\{|\alpha_{ne}|,|\beta_{ne}|,1/b_1,1/b_2\}\,,\quad
\mu_{ne^{\prime}} = {\rm Max}\{|\alpha_{ne^{\prime}}|,|\beta_{ne^{\prime}}|,1/b_1,1/b_2\}\,, \non
&&\mu_a = {\rm Max}\{|\alpha_a|,|\beta_a|,1/b_2,1/b_3\}\,,\quad
\mu_{a^{\prime}} = {\rm Max}\{|\alpha_{a^{\prime}}|,|\beta_{a^{\prime}}|,1/b_3,1/b_2\}\,,  \non
&&\mu_{na} = {\rm Max}\{|\alpha_{na}|,|\beta_{na}|,1/b_1,1/b_2\}\,,\quad
\mu_{na^{\prime}} = {\rm Max}\{|\alpha_{na^{\prime}}|,|\beta_{na^{\prime}}|,1/b_1,1/b_2\} \,, \non
&&\mu_{q} = {\rm Max}\{|\alpha_{q}|,|\beta_{q}|,|\gamma_{q}|,1/b_1,1/b_3\}\,,\quad
\mu_{q^{\prime}} = {\rm Max}\{|\alpha_{q^{\prime}}|,|\beta_{q^{\prime}}|,|\gamma_{q^{\prime}}|,1/b_3,1/b_1\}\,  \non
&&\mu_{g} = {\rm Max} \{|\alpha_{g}|,|\beta_{g}|,|\gamma_{g}|,1/b_1,1/b_2,1/b_3\}\,,\quad
\mu_{g^{\prime}} = {\rm Max} \{|\alpha_{g^{\prime}}|,|\beta_{g^{\prime}}|,|\gamma_{g^{\prime}}|,1/b_1,1/b_2,1/b_3\} \,.\nonumber
\eeq

The Sudakov factor associated in $B$ and light meson wave functions read respectively as 
\beq
&&S_{B}(x_1, b_1, \mu)= s\left(x_1 \frac{m_B}{\sqrt{2}}, b_1\right) + s_q(b_1, \mu) \,, 
\label{eq:sudakov-B}\\
&&S_{M_i}(x_i, \bar{x}_i, b_i, \mu)= s\left(x_i \frac{m_B}{\sqrt{2}}, b_i\right) + s\left(\bar{x}_i \frac{m_B}{\sqrt{2}}, b_i\right) +  s_q(b_i, \mu) \,.
\label{eq:sudakov-M}
\eeq
The factor $s(Q,b)$ is gauge invariant with the gauge cancelation between the reducible and irreducible soft gluon corrections \cite{Catani:1989ne,Catani:1990rp,Li:1996gi}, 
which collects the resummation of the double logarithms in the $k_T$ factorization \cite{Li:1992nu},
\beq
s(\xi Q,b) &=& \frac{A^{(1)}}{2\, \beta_1} \, \hat{q} \, \ln \left( \frac{\hat{q}}{\hat{b}} \right) 
+ \frac{A^{(2)}}{4 \, \beta_1^2} \left( \frac{\hat{q}}{\hat{b}} - 1 \right) - \frac{A^{(1)}}{2 \, \beta_1} \left( \hat{q} - \hat{b} \right) \non
&-& \frac{A^{(1)} \, \beta_2 \, \hat{q}}{4\, \beta_1^3} \left[ \frac{\ln(2\, \hat{b}) + 1}{\hat{b}} - \frac{\ln(2 \, \hat{q}) + 1}{\hat{q}} \right] 
- \left[ \frac{A^{(2)}}{4\,\beta_1^2} - \frac{A^{(1)}}{4 \,\beta_1} \, \ln\left(\frac{e^{2\,\gamma_E - 1}}{2} \right) \right] \ln\left(\frac{\hat{q}}{\hat{b}}\right) \non
&-&\frac{A^{(1)} \, \beta_2}{8 \, \beta_1^3} \left[ \ln^2(2 \, \hat{b}) - \ln^2(2\, \hat{q}) \right]  
- \frac{A^{(1)} \, \beta_2}{8\,\beta_1^3} \, \ln\left(\frac{e^{2\,\gamma_E - 1}}{2}\right) 
\left[\frac{\ln(2\, \hat{b}) + 1}{\hat{b}} - \frac{\ln(2 \, \hat{q}) + 1}{\hat{q}} \right] \non
&-&\frac{A^{(2)} \, \beta_2}{16 \, \beta_1^4} \left[ \frac{2 \ln(2 \, \hat{q}) + 3}{\hat{q}} - \frac{2 \ln(2 \, \hat{b}) + 3}{\hat{b}}\right] 
- \frac{A^{(2)} \, \beta_2}{16 \, \beta_1^4}  \, \frac{\hat{q} - \hat{b}}{\hat{b}^2} \left[ 2\, \ln(2 \, \hat{b}) + 1 \right] \non
&-& \frac{A^{(2)} \, \beta_2^2}{432 \, \beta_1^6} \, \frac{\hat{b} - \hat{q}}{\hat{b}^3} \left[ 9 \, \ln^2(2\,\hat{b}) + 6 \, \ln(2\hat{b}) + 2 \right]  \non
&-& \frac{A^{(2)} \, \beta_2^2}{1728 \, \beta_1^6} 
\left[ \frac{18 \, \ln^2(2\,\hat{b}) + 30 \, \ln(2\hat{b}) + 19}{\hat{b}^2} - \frac{18 \, \ln^2(2\,\hat{q}) + 30 \, \ln(2\hat{q}) + 19}{\hat{b}^2} \right] \,.
\label{sudakov}
\eeq
The abbreviated variables are 
\beq
\hat{q} \equiv \ln\left(\frac{\xi Q}{\sqrt{2}\Lambda}\right) \,, \,\,\,\,\,\, \hat{b} \equiv \ln\left(\frac{1}{b \Lambda}\right) \,, 
\eeq
and the coefficients $A^{(i)}$ and $\beta_i$ are
\beq
&&A^{(1)} = \frac{4}{3} \,, \,\,\,\,\,\, 
A^{(2)} = \frac{67}{9} - \frac{\pi^2}{3} - \frac{10}{27} \, n_f + \frac{8}{3} \, \beta_1 \, \ln\left(\frac{e^{\gamma_E}}{2}\right) \,, \non
&&\beta_1 = \frac{33 - 2 \, n_f}{12} \,, \,\,\,\,\,\, \beta_2 = \frac{152 - 19 \, n_f}{24} \,.
\eeq
The factor $s_q(b, \mu)$ in Eqs. (\ref{eq:sudakov-B},\ref{eq:sudakov-M}) comes from the resummation of the single logarithms in the quark self-energy correction
\beq
s_q(b_1, \mu) = \frac{5}{3} \int_{1/b_1}^\mu \frac{d \bar{\mu}}{\bar{\mu}} \, \gamma_q(g(\bar{\mu})) \,, \quad
s_q(b_i, \mu) = 2 \int_{1/b_i}^\mu \frac{d \bar{\mu}}{\bar{\mu}} \, \gamma_q(g(\bar{\mu})),
\label{eq:sudakov-sq}
\eeq
with the quark anomaly dimension $\gamma_q = - \alpha_s(\mu)/\pi$. 

Beside the end-point singularity introduced by the external quark lines,
the longitudinal momentum fraction also generate large logarithm when the momentum distribution of internal quark line is shrined to on shell,
which is resummed into a universal jet function regarding as a part of the hard kernel in the decay amplitudes. 
Under the covariant gauge $\partial \cdot A = 0$ \cite{Sterman:1986aj,Li:1998is}, the solution is  
\beq 
J(x) = - \mathrm{Exp} \left(\frac{\pi \alpha_s C_F }{4}\right) \int_\infty^\infty \frac{dt}{\pi} \, (1-x)^{\mathrm{Exp(t)}} \, \sin \left( \frac{\alpha_s C_F t}{2}\right) \,
\mathrm{Exp} \left( - \frac{\alpha_s}{4\pi} \, C_F \, t^2 \right) \,.
\label{eq:jet-function}
\eeq
This threshold Sudakov factor \cite{Botts:1989kf} is usually parameterized in the form 
\beq
S_t(x) = \frac{2^{1+2c} \, \Gamma(3/2+c)}{\sqrt{\pi}\Gamma(1+c)} \, \Big[ x(1-x)\Big]^{c} \, ,
\label{eq:St}
\eeq
and in our evaluation the parameter $c$ is chosen at $0.3$.

\end{appendix}


\end{document}